\documentclass[12pt]{article}

\pdfoutput=1 
\usepackage{graphicx} 
\usepackage{color}                                                                                                                                                                                                                                                                                                                                                                                                                                                                                                                                                                                                                                                                                                                                                                                                                                                                                                                                                                                                                        
\usepackage{cancel,tabularx,moreverb,fancybox,amsmath,float,bm,braket,slashbox,txfonts,amssymb,bm,accents}
\usepackage[top=30truemm,bottom=30truemm,left=25truemm,right=25truemm]{geometry}
\usepackage{latexsym}
\usepackage{here}
\usepackage{cite}
\usepackage{hyperref}                                                                                                                                                                                                                                                                                                                                                                                                                                                                                                                                                                                                                                                                                                                                                                                                                                                                                                                                                                                                                                                                                                                                                                                                                                                                                                                                                                                                                                                                                                                                                                                                                                                                                                                                                                                                                                                                                                                                                                                                                                                                                                                                                                                                                                                                                                                                                                                                                                                                                                                                                                                                                                                                                                                                                                                                                                                                                                                                                                                                                                                                                                                                                                                                                                                                                                                                                                                                                                                                                                                                                                                                                                                                                                                                                                                                                                                                                                                                                                                                                                                                                                                                                                                                                                          
\usepackage[utf8]{inputenc}

\newcommand{\ex}[1]{\mathrm{e}^{#1}}

\newcommand{\pa}[1]{\left(#1 \right)}

\newcommand{\BR}[1]{\Biggl[#1 \Biggr]}

\newcommand{\bb}[1]{\mathbb{#1}}

\newcommand{\ca}[1]{\mathcal{#1}}

\newcommand{\abs}[1]{\left|#1\right|}

\newcommand{\ar}[1]{\xrightarrow[#1]{}}

\newcommand{\circled}[1]{\textcircled{\scriptsize#1}}

\newcommand{\red}[1]{\color{red}{#1}}

\newcommand{\ti}[1]{\tilde{#1}}
\newcommand{\dg}[1]{#1^{\dagger}}
\newcommand{\fr}{\frac}
\newcommand{\s}[1]{\sqrt{#1}}

\def\be{\begin{equation}}
\def\ee{\end{equation}}
\def\ba{\begin{eqnarray}}
\def\ea{\end{eqnarray}}

\def\m{{\mu}}
\def\la{{\lambda}}

 \def\ep{{\epsilon}}
 \def\d{{\delta}}

 \def\a{{\alpha}}
 \def\ba{{\bar{\alpha}}}
 
 \def\l{{\lambda}}
 \def\G{{\Gamma}}
 \def\D{{\Delta}}
 \def\g{{\gamma}}
 
 \def\b{{\beta}}
 \def\e{{\epsilon}}

 \def\p{\partial}

\def\tr{{\text{tr}}}

\def\dd{{\mathrm{d}}}

  \makeatletter
    
    \@addtoreset{equation}{section}
  \makeatother

\begin{document}
\begin{titlepage}
\thispagestyle{empty}

\begin{flushright}
YITP-19-82
\\

\end{flushright}

\bigskip

\begin{center}
\noindent{{\large \textbf{
Entanglement Wedge Cross Section from CFT: \\
Dynamics of Local Operator Quench
}}}\\
\vspace{2cm}
Yuya Kusuki, Kotaro Tamaoka
\vspace{1cm}

{\it
Center for Gravitational Physics, \\
Yukawa Institute for Theoretical Physics (YITP), Kyoto University, \\
Kitashirakawa Oiwakecho, Sakyo-ku, Kyoto 606-8502, Japan.
}
\vskip 2em
\end{center}

\begin{abstract}
We derive dynamics of the entanglement wedge cross section from the reflected entropy for local operator quench states in the holographic CFT. By comparing between the reflected entropy and the mutual information in this dynamical setup, we argue that (1) the reflected entropy can diagnose a new perspective of the chaotic nature for given mixed states and (2) it can also characterize classical correlations in the subregion/subregion duality. Moreover, we point out that we must improve the bulk interpretation of a heavy state even in the case of well-studied entanglement entropy. Finally, we show that we can derive the same results from the odd entanglement entropy. 
The present paper is an extended version of our earlier report arXiv:1907.06646 and includes many new results: non-perturbative quantum correction to the reflected/odd entropy, detailed analysis in both CFT and bulk sides, many technical aspects of replica trick for reflected entropy which turn out to be important for general setup, and explicit forms of multi-point semi-classical conformal blocks under consideration. 
 \end{abstract}

\end{titlepage}

\restoregeometry

\tableofcontents

\clearpage

\section{Introduction \& Summary} 

\subsection{Introduction} 

The non-equilibrium dynamics in a given strongly coupled system attracts a lot of attention in the physics community. One useful tool to capture this dynamical process is {\it entanglement entropy}, which measures entanglement between subsystem $A$ and its complement $\bar{A}$. This quantity is defined by
\begin{equation}
S(A)=-\tr \rho_A \log \rho_A,
\end{equation}
where $\rho_A$ is a reduced density matrix for a subsystem $A$, obtained by tracing out its complement. The Renyi entropy is a generalization of the entanglement entropy, which is defined as
\begin{equation}
S^{(n)}(A)=\fr{1}{1-n} \log \tr \rho_A^n,
\end{equation}
and the limit $n\to1$ of the Renyi entropy defines the entanglement entropy $S(A)$.
For this measure, a large number of works have been done to characterize the dynamics, for example, after joining quench \cite{Calabrese2007}, global quench \cite{Calabrese2006a,Calabrese2005},  splitting quench \cite{Shimaji2018} and double quench  \cite{Guo2018, Caputa2019, Kusuki2019c}.
In particular, our interest in this paper is to study a local operator quench state \cite{Nozaki2014,Nozaki2014a}, which is created by acting a local operator $O(x)$ on the vacuum in a given CFT at $t=0$,
\be\label{eq:defO}
\ket{\Psi(t)}=\s{\ca{N}}\ex{-\ep H-iHt} O(x)\ket{0}, 
\ee
where $x$ represents the position of insertion of the operator, $\e$ is an UV regularization of
the local operator and $\ca{N}$ is a normalization factor so that $\braket{\Psi(t)|\Psi(t)}=1$.

One main goal of this paper is to understand {\it dynamics} of correlations between two disjoint intervals.
A natural challenge for this purpose is to investigate the dynamics of some quench state by utilizing correlation measures. One progress in this direction had already done in \cite{Calabrese2005a,Calabrese2009a}, which studied universal features of dynamics after a global quench by using the entanglement entropy for two disjoint intervals, or equivalently, the {\it mutual information} and showed that entanglement spreads as if correlations were carried by free quasiparticles after a global quench. And also it was shown that this quasiparticle picture breaks down in the holographic CFT \cite{Asplund2015a}. It suggests that the mutual information is very useful to probe the universal feature of correlation dynamics in a given CFT class. (see also \cite{Asplund2014}, which studied the dynamics of the mutual information after a joining quench.)
However, what we have to mention is that our interest is the correlation between two disjoint intervals, which are not necessarily complementary to each other, therefore, the state cannot be described by pure state. 
For mixed states, we do not have the unique measure for the bi-partite correlation.
For this reason, we are also interested in other correlation measures. For example, one of other interesting correlation measures is  {\it negativity} \cite{Calabrese2012,Calabrese2013a} and in \cite{Coser2014, Wen2015}, the time-dependence of the correlation between two disjoint intervals is studied by using the negativity.

In this paper, we will make use of {\it reflected entropy} \cite{Dutta2019} as a tool to probe dynamics of correlations between two intervals. The definition is as follows. We consider the following mixed state,
\begin{equation}
\rho_{AB} = \sum_ {n}   p_n \rho^{(n)}_{AB},
\end{equation}
where each $\rho^{(n)}_{AB}$ represents a pure state as
\begin{equation}
\rho^{(n)}_{AB}=\sum_{i,j} \s{\l^i_n \l^j_n} \ket{i_n}_A  \ket{i_n}_B  \bra{j_n}_A  \bra{j_n}_B  ,
\end{equation}
where $\ket{i_n}_A \in \ca{H}_A$, $\ket{i_n}_B \in \ca{H}_B$ and $\l_n^i$ is a positive number such that $\sum_i \l_n^i=1$.
The real number $p_n$ is the corresponding probability associated with its appearance in the ensemble.
For this mixed state, we can provide the simplest purification for this mixed state as
\begin{equation}
\ket{\s{\rho_{AB}} }= \sum_{i,j,n} \s{p_n  \l^i_n \l^j_n}  \ket{i_n}_A  \ket{i_n}_B  \ket{j_n}_{A^*}  \ket{j_n}_{B^*}   ,
\end{equation}
where $\ket{i_n}_{A^*} \in \ca{H}^*_A$ and $\ket{i_n}_{B^*} \in \ca{H}^*_B$ are just copies of $\ca{H}_A$ and $\ca{H}_B$.  Then, the reflected entropy is defined by
\begin{equation}
S_R(A:B) \equiv -\tr \rho_{AA^*} \log \rho_{AA^*},
\end{equation}
where $\rho_{AA^*}$ is the reduced density matrix of $\rho_{AA^*BB^*}=\ket{\s{\rho_{AB}}} \bra{\s{\rho_{AB}}}$  after tracing over $\ca{H}_B \otimes \ca{H}^*_B$.
We have to emphasize that this quantity measures not only quantum correlations but also classical correlations, like mutual information. Actually, these two quantities for the vacuum are very similar, however, we will give quite differences by considering dynamical setups.
\footnote{
Here, we mean the {\it vacuum} by the mixed state $\rho_{AB}$ which comes form the vacuum for the whole system by tracing over $\ca{H}_{\overline{AB}}$. }
Interestingly, if we restrict ourselves to two-dimensional CFTs, we can analytically evaluate this quantity in the path integral formalism, like entanglement entropy. For this reason, we consider a 2D CFT in this paper.

An important point is that this quantity has a simple holographic dual interpretation, so-called {\it entanglement wedge cross section},
\begin{equation}
S_R(A:B)=2E_W(A:B),
\end{equation}
where $E_W(A:B)$ is entanglement wedge cross section defined as the area of the minimal surface bipartitioning the entanglement wedge region, first introduced in \cite{Takayanagi2018a, Nguyen2018}. (See also  \cite{Bao2018, Umemoto2018a, Hirai2018, Bao2019c, Espindola2018, Bao2018b, Guo2019a, Bao2019b, Yang2019, Kudler-Flam2019a, BabaeiVelni2019, Prudenziati2019, Du2019, Liu2019, Jokela2019, Caputa2019a, Tamaoka2019, Guo2019, Bao2019, Kudler-Flam2019,Harper2019, Kusuki2019a, Kusuki2019b, Umemoto2019, Jeong2019, Bao2019a,Levin2019} for further developments in this direction.)
That is, the reflected entropy is computable both in bulk side and CFT side and also meaningful in quantum information theory, in a similar manner to the RT and HRT formula \cite{Ryu2006a, Ryu2006, Hubeny2007}. Thus this is a very good useful to investigate the quantum gravity in the context of the AdS/CFT, however, there is little understanding of its property for now. 
In particular, there is no understanding on the non-equilibrium properties of the reflected entropy even in the holographic CFT.
This naturally motivates us to study the dynamics of the reflected entropy.
This study might give new insights into the relation of dynamics of correlations between in the holographic CFT and in the quantum gravity.

On this background, in this paper, we will study the time-dependence of the reflected entropy after a local quench as a first step to understand the dynamics of the reflected entropy. We would like to point out the advantage of considering the local operator quench.
Technically, the reflected entropy after a local quench is calculated by the {\it Regge limit} of $n$-point conformal blocks.
Fortunately, the method to calculate the Regge limit was recently invented in \cite{Kusuki2019}, therefore, it is now possible to easily calculate what we need. This is one of reasons to focus on the local operator quench.
Another advantage comes from a physical reason. Local quenches have a richer structure than global quenches, because they are inhomogeneous. Thus, we can extract more information about dynamics from local quenches than global quenches.

It is worth paying attention to another proposal for the entanglement wedge cross section in \cite{Tamaoka2019}, which is so-called {\it odd entanglement entropy}. The odd entanglement entropy is defined by
\begin{equation}\label{eq:OEE}
S_O(A:B)\equiv\lim_{n_O\to1} \fr{1}{1-n_O}\BR{\tr \pa{\rho_{AB}^{T_B}}^{n_O}-1},
\end{equation}
where $\rho_{AB}$ is a reduced density matrix for subsystems $A$ and $B$, obtained by tracing out its complement. The limit $n_O \to 1$ is the analytic continuation of an odd integer and $T_B$ is the partial transposition with respect to the subsystem $B$.
(Note that it is equivalent to act $T_A$ instead of $T_B$.)  Interestingly, it is conjectured that this quantity has a simple bulk interpretation as
\begin{equation}\label{eq:OEEdef}
S_O(A:B)-S(A:B)=E_W(A:B),
\end{equation}
where $S(A:B)$ is the entanglement entropy for the subsystems $A$ and $B$.
This is verified for the vacuum and thermal state in the 2D holographic CFT \cite{Tamaoka2019}, however, it is nontrivial that this relation also holds in other setups. For this reason, we will also study this quantity in the same setup and investigate whether the relation can also be applied to nontrivial states or not. We would like to mention that this quantity can be calculated in the same way as negativity \cite{Calabrese2012,Calabrese2013a}. More precisely,  this is given by the analytic continuation of an {\it odd} integer of the same replica partition function as negativity.

\subsection{Summary} 
Here we briefly summarize our results.

\begin{itemize}
\item CFT vs. Gravity (in Section \ref{sec:localCFT} and \ref{sec:bulk})

It has been argued that the entanglement entropy after a local quench state is realized by geometries with a falling particle \cite{Nozaki2013}.
From this observation, it is naturally expected that the reflected entropy for a locally excited state would be also the dual to the entanglement wedge cross section in that geometry. In this paper, we calculate the reflected entropy for such a dynamical state and compare it to the dynamics of the entanglement wedge cross section. As a result, we find the perfect agreement.
This is a new support of the dynamical generalization of the reflected entropy/entanglement wedge cross section conjecture.

\item Technical aspects of replica trick for reflected entropy (in Section \ref{sec:localCFT})

When we use the replica trick, we should use the conformal blocks not for original theory (Virasoro conformal blocks) but for orbifold theory. In the case of the entanglement entropy (and odd one) we can justify the use of former blocks. However, this turns out to be not the case for the reflected entropy. We clarify many technical aspects of the replica trick for reflected entropy which were not described in the literature. 
We hope that our description will be useful and technically important to study further the reflected entropy for QFTs in more general setup.

\item Dynamics of reflected entropy (and entanglement of purification) vs. mutual information (in Section \ref{sec:dynamics})

One motivation is to understand the dynamics of the reflected entropy.
In this paper, we will consider three patterns of a local operator quench as shown in Figure \ref{fig:setup}.
First observation for the reflected entropy is that the time-dependence is captured by the {\it quasi-particle picture}  \cite{Calabrese2005a, Calabrese2006a} as seen in the mutual information and the negativity. For example, if we consider a setup $\circled{3}$ in Figure \ref{fig:setup}, we find that the reflected entropy becomes non-zero only in the time region $t\in[u_2,v_2]$.
However, the time-dependence in the non-zero region is very complicated, therefore, it cannot be completely explained by the quasi-particle picture.

\begin{figure}[t]
 \begin{center}
  \includegraphics[width=12.0cm,clip]{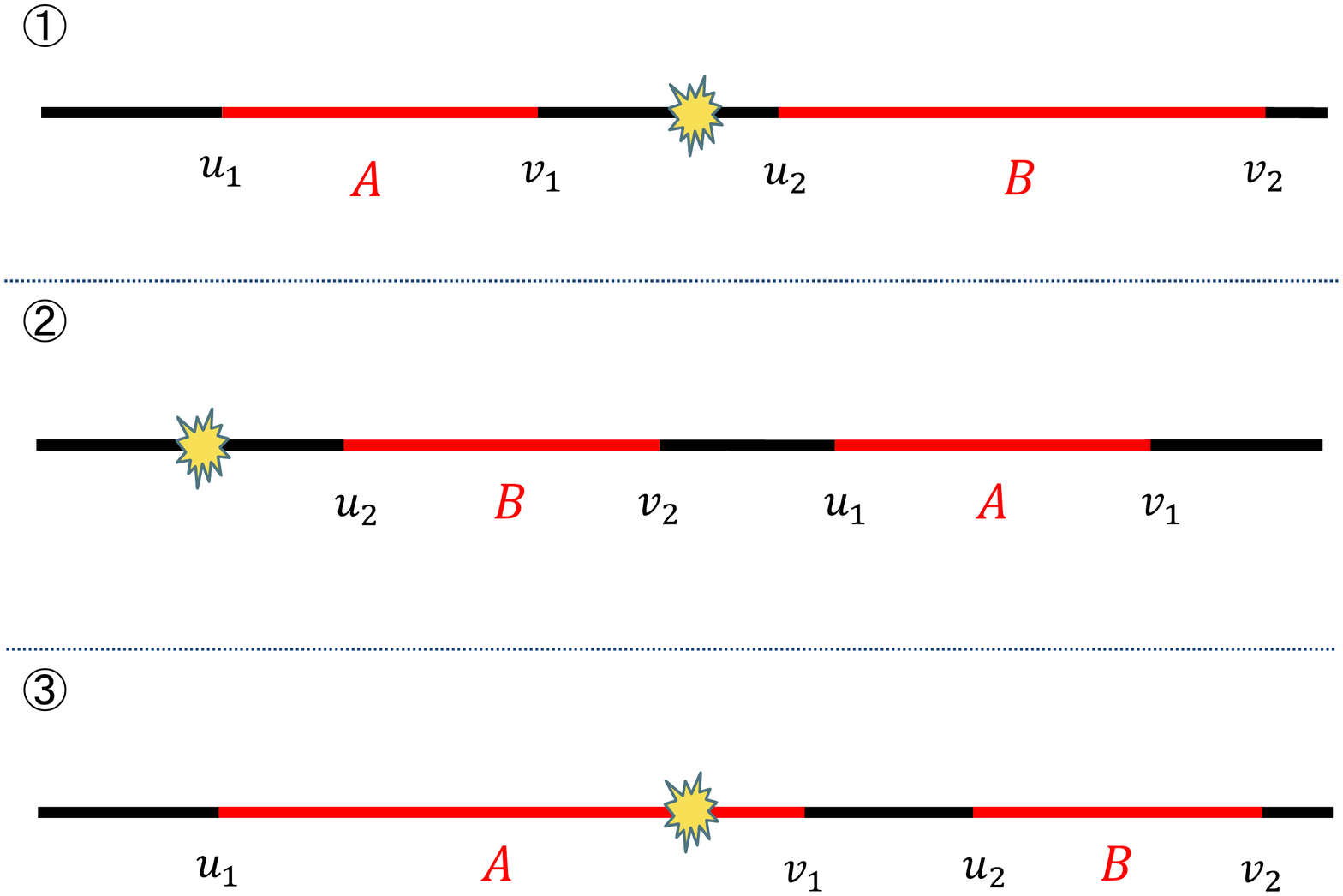}
 \end{center}
 \caption{Three setups considered in this paper. We fist study the setup  $(0< u_2<-v_1<-u_1<v_2)$, second, $(0< u_2<v_2<u_1<v_1)$ , and finally,  $( 0< v_1<u_2<v_2<-u_1  )$. In any setups, we excite the vacuum by acting an local operator on $x=0$ at $t=0$.}
 \label{fig:setup}
\end{figure}

We compare our results for the reflected entropy to the dynamics of the mutual information in the same setup and find both similarities and differences. For example, the time dependence of the reflected entropy is discontinuous, unlike the mutual information.
Moreover, we give a natural explanation that the reflected entropy probes more classical correlations than the mutual information from our dynamical setup\footnote{This implies the quantities dual to the entanglement wedge cross section can not be any axiomatic measures of the quantum entanglement. This conjecture has been proven recently by \cite{Umemoto2019}.}.

As a comment, our physical interpretation in this section can be also applied to entanglement of purification because in the holographic CFT, the reflected entropy reduces to the entanglement of purification.

\item What is dual to a heavy state? (in Section \ref{sec:Heavy})

Another interest is to understand what is the holographic dual to a heavy state in CFT.
The first study has been done in \cite{Asplund2015} by making use of the entanglement entropy. The result suggests that the entanglement entropy for a heavy state can be approximated the holographic entanglement entropy in the BTZ background. 
In this paper, we consider reflect entropy for a heavy state to make it clear. This approach is quite natural because reflected entropy is more refined tool than entanglement entropy. Consequently, we find a contradiction between their bulk interpretation and the entanglement entropy which comes from the {\it pure state limit} of the reflected entropy. To resolve this problem, we give an improved bulk interpretation of the heavy state and then we obtain the perfect agreement between our bulk interpretation and the reflected entropy in the heavy state.

\item Quantum correction (in Section \ref{sec:quantum})

We can evaluate some quantum corrections to the reflected entropy, which is consistent with a naive expectation from the physical viewpoint.
And also the reflected entropy with some quantum corrections also satisfies some important inequalities of the holographic reflected entropy.

\item Dynamics in other CFTs (in Section \ref{sec:other})

If one wants to characterize the holographic CFT by the reflected entropy, it is necessary to find out a unique feature of the holographic reflected entropy.  For this purpose, we first tried to compare the holographic result to that in rational CFTs (RCFTs).
As a result, we show that the time-dependence for these two CFTs are quite different.
From this observation, we could argue that the dynamics of the reflected entropy is very sensitive to whether a given CFT is chaotic or not. In other words, we can make use of the reflected entropy as a probe of the chaotic nature of a given CFT (see also \cite{Wang2019}).

\item Agreement with odd entanglement entropy (in Section \ref{sec:odd})

We can show that the odd entanglement entropy also reproduces the entanglement wedge cross section in our dynamical setup.
Actually, the similarity between the holographic odd entanglement entropy and the holographic reflected entropy can be explained by a special property of the linearized conformal block.
Therefore, instead of providing the detailed calculations, we show how the odd entanglement entropy reduces to the reflected entropy in the holographic CFT.

\end{itemize}

\section{Reflected Entropy of Local Operator from CFT} \label{sec:localCFT}

\begin{figure}[H]
 \begin{center}
  \includegraphics[width=16cm,clip]{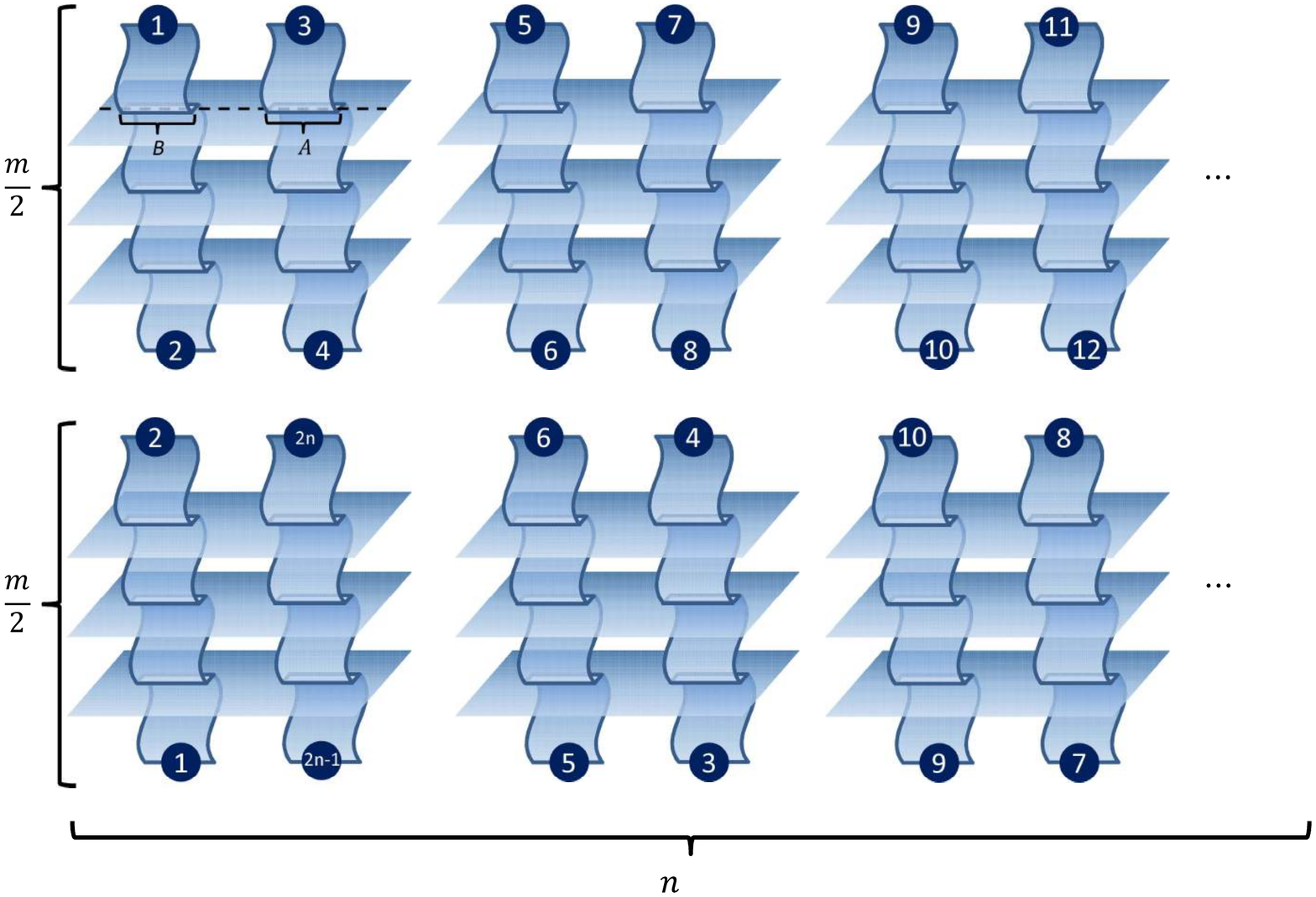}
  \includegraphics[width=6cm,clip]{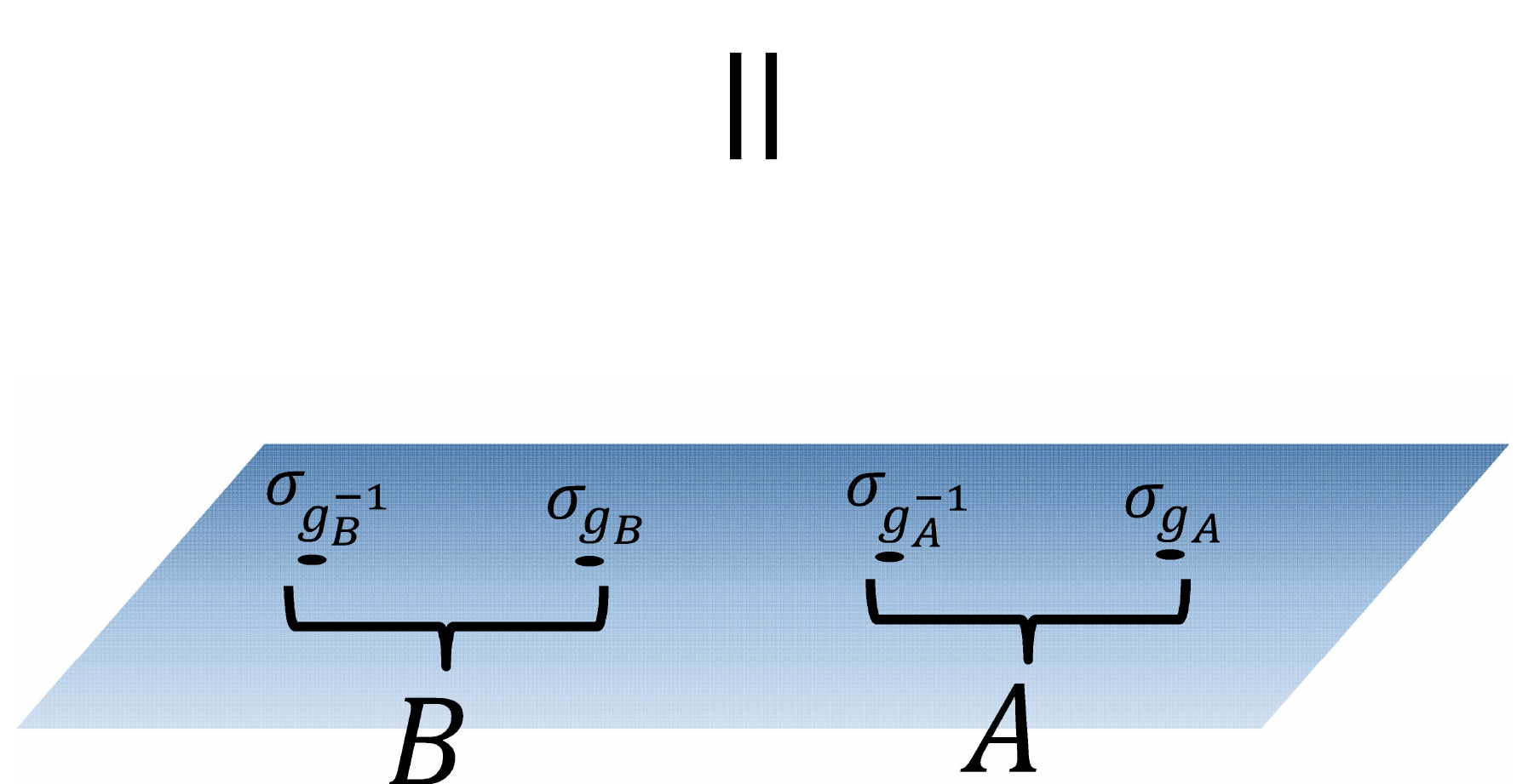}
 \end{center}
 \caption{The path integral representation of the Renyi reflected entropy. Edges labeled with the same number get glued together.
We can instead view it as a correlator with four twist operators $\Braket{\sigma_{g_A}(u_1)\sigma_{g_A^{-1}}(v_1)   \sigma_{g_B}(u_2) \sigma_{g_B^{-1}}(v_2) }_{\text{CFT}^{\otimes mn}}$. 
 }
 \label{fig:replica}
\end{figure}

The reflected entropy can be evaluated in the path integral formalism \cite{Dutta2019}.
For example, the Renyi reflected entropy in the vacuum can be computed by a path integral on $m \times n$ copies as shown in Figure \ref{fig:replica}. Here, we would view this manifold as a correlator with twist operators as in the lower of Figure \ref{fig:replica}, where we define the twist operators $\sigma_{g_A}$ and $\sigma_{g_B}$. 
Here, we focus on the following mixed state,
\begin{equation}
\rho_{AB}=\tr_{\overline{AB}} \ket{\Psi(t)} \bra{\Psi(t)},
\end{equation}
where $\Psi(t)$ is a time-dependent pure state as $\ket{\Psi(t)}=\s{\ca{N}}\ex{-\ep H-iHt} O(0)\ket{0} $.
Then, in a similar manner to the method in \cite{Nozaki2014}, the replica partition function in this state can be obtained by a correlator as
\begin{equation}\label{eq:Renyi}
\fr{1}{1-n} \log
	\fr{Z_{n,m}}
	{\pa{ Z_{1,m}}^n },
\end{equation}
and
\begin{equation}\label{eq:dRenyi}
Z_{n,m}\equiv \Braket{\sigma_{g_A}(u_1)\sigma_{g_A^{-1}}(v_1)  {O^{\otimes mn}}(w_1,\bar{w}_1)  \dg{{O^{\otimes mn}}}(w_2,\bar{w}_2)   \sigma_{g_B}(u_2) \sigma_{g_B^{-1}}(v_2) }_{\text{CFT}^{\otimes mn}},
\end{equation}
where we abbreviate $V(z,\bar{z})\equiv V(z)$ if $z\in\bb{R}$ and the operators $O$ are inserted at
\begin{equation}
w_1=t+ i \e, \ \ \ 
\bar{w}_1=-t+ i \e, \ \ \ 
w_2=t- i \e, \ \ \ 
\bar{w}_2=-t- i \e. \ \ \ 
\end{equation}
Here $O^{\otimes N}\equiv O\otimes O\otimes\cdots \otimes O$ is an abbreviation of the operator on $N$ copies of CFT ($\text{CFT}^{\otimes N}$)\footnote{For simplicity, we always omit the transposition of operators on the reflected sheets.}. 
To avoid unnecessary technicalities, here we do not show the precise definition of the twist operators $\sigma_{g_A}$ and $\sigma_{g_B}$ (which can be found in \cite{Dutta2019}) because in many parts of this paper, we only use the following properties of the twist operators,
\begin{equation}
\begin{aligned}
&h_{\sigma_{g_A}}=h_{\sigma_{g_A^{-1}}}=h_{\sigma_{g_B}}=h_{\sigma_{g_B^{-1}}}=\fr{cn}{24} \pa{m-\fr{1}{m}}  (= n h_m)   ,\\
& \sigma_{g_A^{-1} g_B}=\sigma_{g_n} \otimes \sigma_{g_n^{-1}},
\end{aligned}
\end{equation}
where the twist operator $\sigma_{g_n}$ is just the usual twist operator $\sigma_n$ based on the $n$-cyclic permutation group, which has the conformal dimension $h_{\sigma_{g_n}}=\fr{c}{24}\pa{n-\fr{1}{n}} ( \equiv h_n ) $.
Note that the second property is a naive expression, which will be explained more explicitly in Section \ref{subsubsec:orbifold}.
 
The reflected entropy is defined by the von-Neumann limit of this partition function,
\begin{equation}
\lim_{n,m \to 1} \fr{1}{1-n} \log
	\fr{Z_{n,m}}
	{\pa{ Z_{1,m}}^n },
\end{equation}
where the analytic continuation $m \to 1$ is taken for ``even'' integer $m$.

\noindent

It is hard to calculate the numerator in (\ref{eq:Renyi}) in general. Fortunately, in the case of interest, i.e., holographic CFTs, this 6-point function can be approximated by a single conformal block as in \cite{Asplund2015}, for example, if we set $0<\e\ll t< u_2<-v_1<-u_1<v_2$ then the correlation function is approximated by
\newsavebox{\boxpa}
\sbox{\boxpa}{\includegraphics[width=190pt]{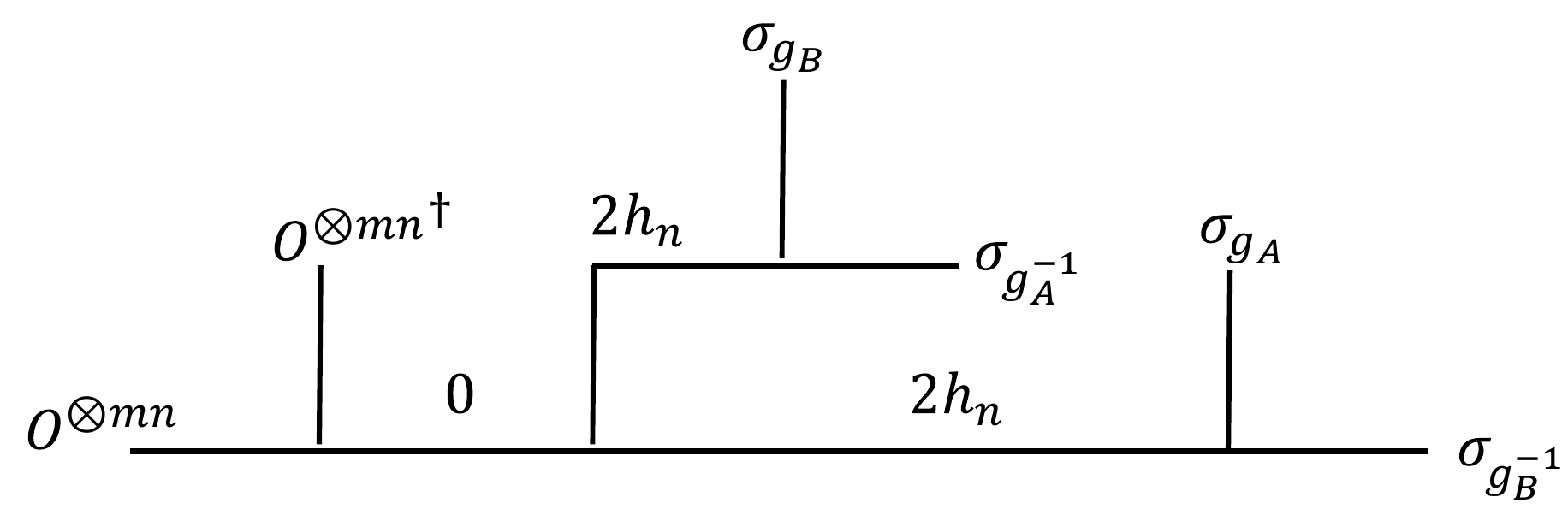}}
\newlength{\paw}
\settowidth{\paw}{\usebox{\boxpa}} 

\begin{equation}\label{eq:localchannel}
{(C_{n,m})}^2 \parbox{\paw}{\usebox{\boxpa}} \times (\text{anti-holomorphic part}),
\end{equation}
where $C_{n,m}$ is the OPE coefficient $\braket{\sigma_{g_A^{-1}}| \sigma_{g_B} (1) |\sigma_{g_B g_A^{-1}}}$. This coefficient can be calculated by the method developed in \cite{Lunin2001}, and the result is
\begin{equation}\label{eq:Mathur}
C_{n,m}=(2m)^{-4h_n}.
\end{equation}
The details of this derivation can be found in Appendix C of \cite{Dutta2019}.
As explained in \cite{Asplund2015}, we have many choices of the single block approximation aside from (\ref{eq:localchannel}) because we can also decompose the 6-point correlator in terms of the conformal block transformed by the {\it monodromy transformation}.
The correct result is obtained by the maximal single conformal block approximation. More detailed explanations and its explicit calculations are shown in the following subsections.

\subsection{Technical remarks on the replica trick}

Before moving on to the calculation, we discuss two technical complications due to the unusual replica trick for the reflected entropy: (1) We have an analytic continuation of an even integer $m$ related to preparing a canonically purified state, but eventually take $m\rightarrow1$ limit. We should properly treat this tricky manipulation. (2) We also have a replica number $n$ related to the Renyi index. Since we finally take the $m,n\rightarrow1$ limit, we should pay attention to the order of these limits. In contrast to the vacuum case, these two issues become relevant to final results in the present analysis. 

\subsubsection{Orbifold block and an even integer analytic continuation}\label{subsubsec:orbifold}
In general, we cannot approximate the conformal block of the orbifold theory (``orbifold block'') appeared in (\ref{eq:localchannel}) by the Virasoro conformal block. This is because there is the current associated with the replica symmetry. However, in the limits $n,m \to 1$ and $c \to \infty$, these two blocks can be related. Indeed, the orbifold block in this limit can be related to a ``square'' of the Virasoro conformal block. This ``squaring'' (or ``doubling'') essentially comes from the doubling of the purified Hilbert space. Interestingly, this doubling also explains the origin of the double of entanglement wedge cross section for reflected entropy in holographic CFTs. Therefore, let us first explain why this works in our case. 

Since we analytically continue an even integer $m$ to the real number, replica sheets labelled by $m=0,\dots,\fr{m}{2}-1$ and ones for $\fr{m}{2},\dots,m-1$ should decouple. A similar decoupling of the replica sheets is well-known in the context of the logarithmic negativity because we also have to consider the analytic continuation of an even integer to evaluate the negativity \cite{Calabrese2012,Calabrese2013a} (see also \cite{Kusuki2019b}).
To make it clear, we introduce the following notations:

\begin{table}[H]
  \begin{tabular}{|c|l|} 
\hline
     $O_{(k,l)}$
				&
		Operator on $(k,l)$-sheet. ($k=0,\dots, m-1$ and $l=0,\dots, n-1$.)
\\  \hline
     $O^{\otimes n}_{(k)}$
				& $\bigotimes^{n-1}_{l=0}O_{(k,l)} $ 
\\  \hline

     $\sigma_n^{(0)}$
				&
		$O_{(k,l)}(\ex{2\pi i}z)\sigma_n^{(0)}(0) = O_{(k,l+1)}(z)\sigma_n^{(0)}(0) $, \ \ \ \ \ \ (if $k=0$),  \\
		&$O_{(k,l)}(\ex{2\pi i}z)\sigma_n^{(0)}(0) = O_{(k,l)}(z)\sigma_n^{(0)}(0) $, \ \ \ \ \ \ \ \ \  (otherwise). 
\\  \hline
     $\sigma_n^{(m/2)}$
				&
		$O_{(k,l)}(\ex{2\pi i}z)\sigma_n^{(m/2)}(0) = O_{(k,l+1)}(z)\sigma_n^{(m/2)}(0) $, \ \ \ \ \ \   (if $k=\fr{m}{2} $),  \\
		&$O_{(k,l)}(\ex{2\pi i}z)\sigma_n^{(m/2)}(0) = O_{(k,l)}(z)\sigma_n^{(m/2)}(0) $, \ \ \ \ \ \ \ \ \ (otherwise). 
\\  \hline
     $\sigma_m^{\otimes n}$
				&
		$O_{(k,l)}(\ex{2\pi i}z)\sigma_m^{\otimes n}(0) = O_{(k+1,l)}(z)\sigma_m^{\otimes n}(0) $.
\\  \hline
     $\bar{\sigma}_m^{\otimes n}$
				&
		$O_{(k,l)}(\ex{2\pi i}z)\bar{\sigma}_m^{\otimes n}(0) = O_{(k-1,l)}(z)\bar{\sigma}_m^{\otimes n}(0) $.
\\  \hline
     ${\sigma'}_m^{\otimes n}$
				&
		$O_{(k,l)}(\ex{2\pi i}z){\sigma'}_m^{\otimes n}(0) = O_{(k+1,l+1)}(z){\sigma'}_m^{\otimes n}(0) $,  \ \ \ \ \ (if $k=0,\fr{m}{2}$),   \\
		&$O_{(k,l)}(\ex{2\pi i}z){\sigma'}_m^{\otimes n}(0) = O_{(k+1,l)}(z){\sigma'}_m^{\otimes n}(0) $, \ \ \ \ \ \ \ \ (otherwise) .
\\  \hline
     $\bar{\sigma'}_m^{\otimes n}$
				&
		$O_{(k,l)}(\ex{2\pi i}z)\bar{\sigma'}_m^{\otimes n}(0) = O_{(k-1,l-1)}(z)\bar{\sigma'}_m^{\otimes n}(0) $,  \ \ \ \ \ (if $k=0,\fr{m}{2}$),   \\
		&$O_{(k,l)}(\ex{2\pi i}z)\bar{\sigma'}_m^{\otimes n}(0) = O_{(k-1,l)}(z)\bar{\sigma'}_m^{\otimes n}(0) $, \ \ \ \ \ \ \ \ (otherwise) .
\\  \hline
  \end{tabular}
\end{table}
\ \\

Then, the operator $O^{\otimes mn}$ can be written as
\be
O^{\otimes mn}=O^{\otimes n}_{(0)} \otimes \cdots \otimes O^{\otimes n}_{(m/2)} \otimes \cdots.
\ee
Throughout this paper, we suppress the transposition acting on the operators on second half sheets concerning to $m$. 
We have to emphasize that in the analytic continuation of even $m$, the operator $O^{\otimes mn}$ does NOT reduce to $O$ but
the ``square'' of $O$ as
\begin{equation}\label{eq:squaring}
\lim_{m \in \text{even} \to 1} O^{\otimes mn} \to  O^{\otimes n}_{(0)} \otimes O^{\otimes n}_{(1/2)}.
\end{equation}
One can also find the same decoupling in the original paper \cite{Dutta2019}, where the analytic continuation leads to
\begin{equation}
\lim_{m \in \text{even} \to 1} \sigma_{g_A^{-1} g_B}\to \sigma_n^{(0)} \otimes \bar{\sigma}_n^{(1/2)}.
\end{equation}
It means that this tricky analytic continuation provides {\it two decoupled sheets} labeled by $0$ and $1/2$. 

The relation between the above notations and the twist operators in \eqref{eq:dRenyi} is given by
\begin{equation}
\begin{aligned}
\sigma_{g_B}={\sigma}_m^{\otimes n}, \ \ \ \ \ 
\sigma_{g_B^{-1}}=\bar{\sigma}_m^{\otimes n}, \ \ \ \ \ 
\sigma_{g_A}={\sigma'}_m^{\otimes n}, \ \ \ \ \ 
\sigma_{g_A^{-1}}=\bar{\sigma'}_m^{\otimes n}, \ \ \ \ \ 
\sigma_{g_A^{-1} g_B}=\sigma_n^{(0)} \otimes \bar{\sigma}_n^{(m/2)}, \ \ \ \ \ 
\end{aligned}
\end{equation}
and the conformal block can be re-expressed by
\newsavebox{\boxpCa}
\sbox{\boxpCa}{\includegraphics[width=250pt]{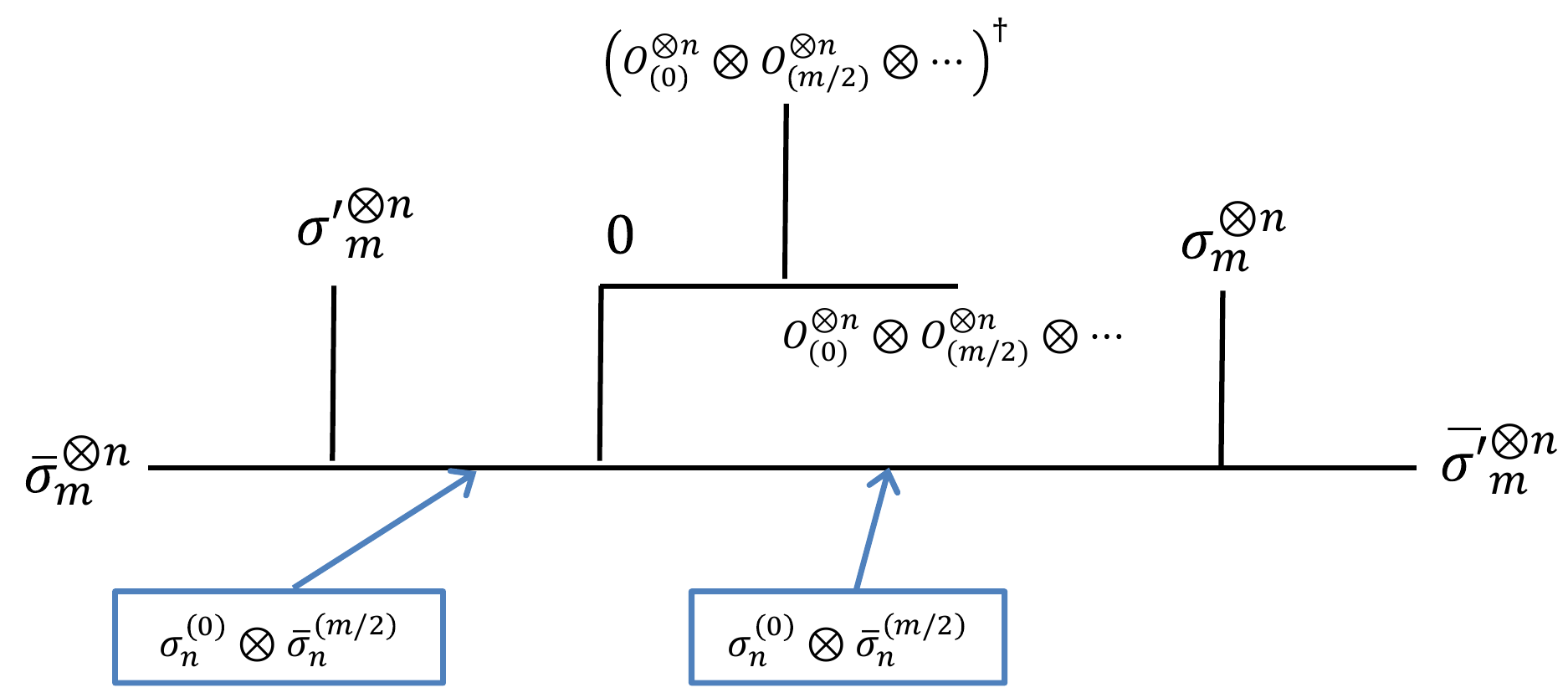}}
\newlength{\pCaw}
\settowidth{\pCaw}{\usebox{\boxpCa}} 

\begin{equation}
 \parbox{\pCaw}{\usebox{\boxpCa}},
\end{equation}
where $\cdots$ means the rest of $O^{\otimes mn}$, that is, $\bigotimes_{l=\hat{0},1,2,\dots, \hat{\fr{m}{2}}, \dots, n-1}  O^{\otimes n}_{(l)}$, which is not important because it disappears in the limit $m \to 1$.

The point is that $\{ O^{\otimes n}_{(0)}, \sigma_n^{(0)} \}$ do not interact with $\{ O^{\otimes n}_{(m/2)}, \sigma_n^{(m/2)} \}$, therefore, the component of the conformal block (i.e., three point block) is decoupled into two parts, for example,
\begin{equation}\label{eq:square2}
\braket{  \sigma_n^{(0)}  \otimes  \bar{\sigma}_n^{(m/2)}  |  O^{\otimes n}_{(0)} \otimes   O^{\otimes n}_{(m/2)}   | \sigma_n^{(0)}  \otimes \bar{\sigma}_n^{(m/2)}  } = 
\braket{  \sigma_n^{(0)}   |  O^{\otimes n}_{(0)} |   \sigma_n^{(0)}  } 
\braket{  \sigma_n^{(m/2)}   |  O^{\otimes n}_{(m/2)} |   \sigma_n^{(m/2)}  }.
\end{equation}
Let us highlight this decoupling by
\newsavebox{\boxpCb}
\sbox{\boxpCb}{\includegraphics[width=250pt]{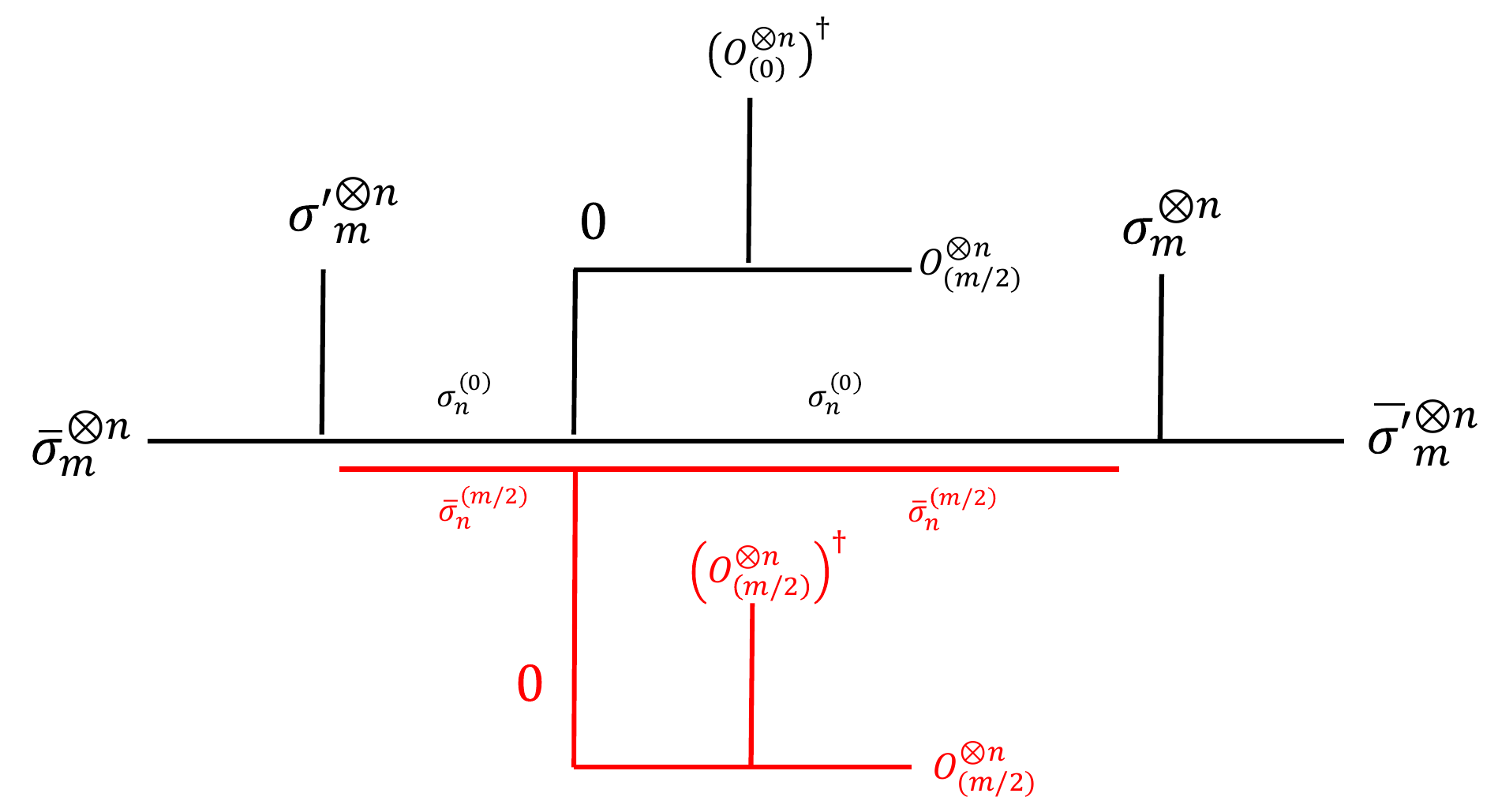}}
\newlength{\pCbw}
\settowidth{\pCbw}{\usebox{\boxpCb}} 

\begin{equation}\label{eq:decoupleblock}
 \parbox{\pCbw}{\usebox{\boxpCb}}.
\end{equation}
Roughly, each decoupled contribution can be regarded as the independent Virasoro conformal blocks up to the universal contributions from external operators\footnote{To be precise, the decoupled conformal block in (\ref{eq:decoupleblock}) is still not the Virasoro block because there is the current associated with the $\bb{Z}_n$ symmetry (see \cite{Asplund2015a}, which discusses this problem). However, in the large $c$ limit, this type of blocks with twist operators can be related to the Virasoro block. In fact, this assumption is often used in the calculation of the entanglement entropy and it is verified in the holographic CFT by comparing with the gravity calculation \cite{Asplund2015}. Moreover, this is also verified by comparing with a completely independent calculation without relying on twist operators \cite{Kusuki2019}.
}. We will see each decoupled block provides the entanglement wedge cross section, thus we obtain the double of the entanglement wedge cross section in total. Note that these blocks are quite similar to the one for the odd entanglement entropy. This is the main reason why it also reproduces the cross section. Having this doubling in mind, we will often suppress the above lengthy doubling expression \eqref{eq:decoupleblock} and instead double the conformal dimension for internal operators. 

\if(
\begin{equation}\label{eq:square}
\lim_{m \in \text{even} \to 1} O^{\otimes mn} \to \pa{O^{\otimes n}}^{\otimes 2}.
\end{equation}
This squaring comes from the tricky analytic continuation, ``even'' integer $m \to 1$.
A similar squaring can be found in the calculation of negativity \cite{Calabrese2012,Calabrese2013a}, where a similar analytic continuation leads to squaring of the double trace operator $\sigma_{n}^2$,
\begin{equation}
\lim_{n \in \text{even} \to 1} \sigma_{n}^2 \to \pa{\sigma_{1/2}}^{\otimes 2}.
\end{equation}
To avoid further technicalities, we only mention the rule (\ref{eq:square}) here and we move the detailed explanation to Appendix \ref{app:Vir}.
)\fi
The analytic continuation of the even integer $m$ gives rise to another subtle issue. 
In order to obtain the correct normalization for the density matrix, $Z_{1,m}$ should not be regarded as the naive $\tr\rho_{AB}^m$, namely the Renyi entropy after a local quench,
\begin{equation}
Z_{1,m} \neq \Braket{\sigma_{g_m}(u_1)\sigma_{g_m^{-1}}(v_1)  O^{\otimes m}(w_1,\bar{w}_1)  \dg{{O^{\otimes m}}} (w_2,\bar{w}_2)   \sigma_{g_m}(u_2) \sigma_{g_m^{-1}}(v_2) }_{\text{CFT}^{\otimes m}},
\end{equation}
where the twist operator $\sigma_{g_m}$ is just the usual twist operator $\sigma_m$. 
This is just because the naive $\tr\rho_{AB}^m$ is (strictly speaking) different from the normalization of the purified state. In other words, the naive one cannot take into account the above squaring effect. As a result of this squaring, the analytic continuation of the denominator in (\ref{eq:Renyi}) is given by the square of the two-point function,
\begin{equation}\label{eq:denominator}
\lim_{m\in \text{even} \to 1} \pa{Z_{1,m}}^n=\Braket{O(w_1,\bar{w}_1)\dg{O}(w_2,\bar{w}_2)}^{2n}=(2i\e)^{-8nh_O},
\end{equation}
where $h_O$ is the conformal dimension of the operator $O$. 
It would be worth noting that we can confirm the necessity of this squaring from the pure state limit of $\rho_{AB}$, where our reduced density matrix $\rho_{AA^*}$ becomes ``square'' of $\rho_{A}$. 

\subsubsection{Order of the two limit for replica numbers}
\begin{figure}[H]
 \begin{center}
  \includegraphics[width=15.0cm,clip]{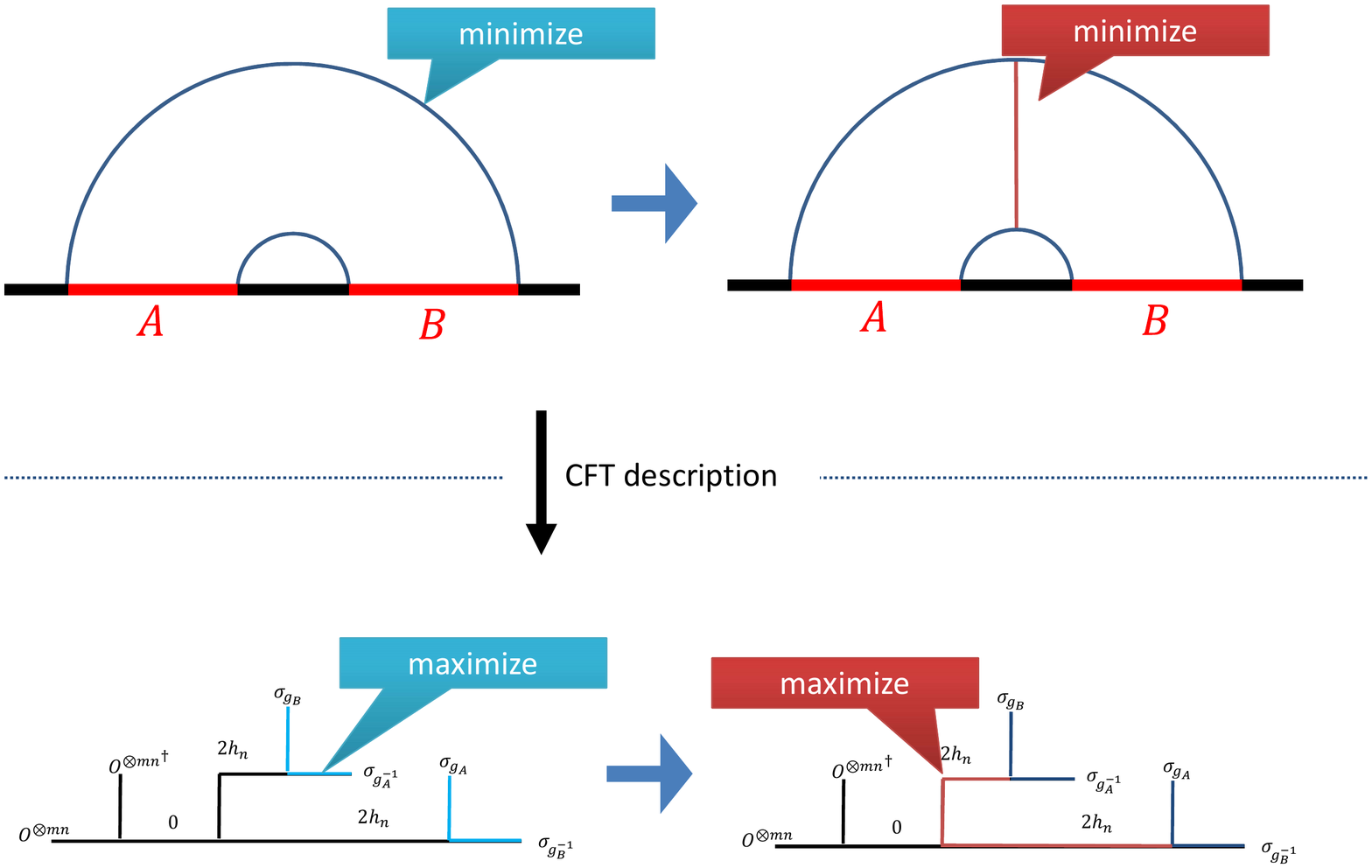}
 \end{center}
 \caption{
To reproduce the entanglement wedge cross section, we first take the large $c$ limit and approximate the correlator by the  maximal single conformal block. However, we have to take care of the fact that this maximization is done by two maximization processes. First, we maximize the propagations between external operators (lines colored by blue) and second, we maximize the internal line (colored by red). This order of processes corresponds to the minimizations in bulk side as shown in the upper of this figure. As mentioned in the main text, this order of maximizations can be accomplished by the large $c$ limit under the assumption $2h_n \ll n h_m$.
}
 \label{fig:minimize}
\end{figure}

Second one is physically more important---the two limits $m \to 1$ and $n \to 1$ do not commute with each other in the large $c$ limit. We should first take the limit $n \to 1$.
There is a physical reason: in order to obtain the correct cross section of the entanglement wedge, we should prepare the precise entanglement wedge at first. 
In terms of the single conformal block approximation, it means that we have to choose the maximal channel in the limit $n \to 1$ with a fixed $m$\footnote{Actually, if we take first the limit $m \to 1$, then the limit $n \to 1$ and choose the maximal single block in these limits to calculate the reflected entropy, we sometimes obtain an incorrect result with a contradiction to the known relation to the mutual information, $S_R(A:B)[O]\geq I(A:B)[O]$.} (see Figure \ref{fig:minimize}).

However, in the following, we calculate the reflected entropy by taking first the limit $m\to 1$ and then $n \to 1$ under $2h_n \ll n h_m$, instead of first taking $n \to 1$ followed by $m \to 1$. Let us stress that this is just for the simplification of calculation and presentation.
Indeed, as we show in the following, our result from this procedure perfectly reproduces the bulk calculation. 
We can also show this validity in another way. The reason why two limits $m\to 1$, $n \to 1$ do not commute with each other is just because the dominant channel in the large $c$ limit could change if the order is reversed. And in fact, we use the assumption $2h_n \ll n h_m$ only to specify the dominant channel. That is, after identifying the dominant channel, the order of the two limits is not important.
Therefore, we can calculate the correct reflected entropy by taking first the limit $m\to 1$ and then $n \to 1$ under $2h_n \ll n h_m$.

It would be interesting to comment that the non-commutativity of $n \to1$ and $m \to 1$ implies that there is a replica transition as the replica number is varied. A similar replica transition can also be found in\cite{Metlitski2009,Belin2013,Belin2015,Belin2017,Dong2018, Kusuki2018c}.
It would be interesting to find this transition from the bulk side in the future.

\subsection{Quench outside Region $A$ and $B$}\label{subsec:outside}

We first consider the setup, $0<\e\ll u_2<-v_1<-u_1<v_2$ and we assume the {\it connected condition}
\begin{equation}
0<\fr{(v_1-u_2)(u_1-v_2)}{(v_1-v_2)(u_1-u_2)}<\fr{1}{2},
\end{equation}
which means that in the bulk side, the entanglement wedge for two intervals $A=[u_1,v_1]$ and $B=[u_2,v_2]$ is connected.
\footnote{
In the CFT side, the transition between connected and disconnected entanglement wedge can be interpreted as a change of the dominant conformal block as shown in \cite{Hartman2013a}. One can show this connected condition from the CFT side by calculating the entanglement entropy for two intervals $A=[u_1,v_1]$ and $B=[u_2,v_2]$ after a local quench. In fact, this calculation cannot be found in previous works but we can calculate it by the method developed in this section (which is explained later in Section \ref{sec:dynamics}).
}
In this article, we only focus on this connected case because the reflected entropy for the disconnected case trivially vanishes, which is not interesting. Note that even if this connected condition is satisfied, the entanglement wedge could become disconnected under the time evolution (which is discussed later above (\ref{eq:disresult})).

In the early time ($0<t<u_2$), the $ \e \to 0$ limit of this block simply reduces as
\footnote{In the following, we will abbreviate $\sigma_n \otimes \bar{\sigma}_n$ by $2h_n$.}

\newsavebox{\boxpb}
\sbox{\boxpb}{\includegraphics[width=160pt]{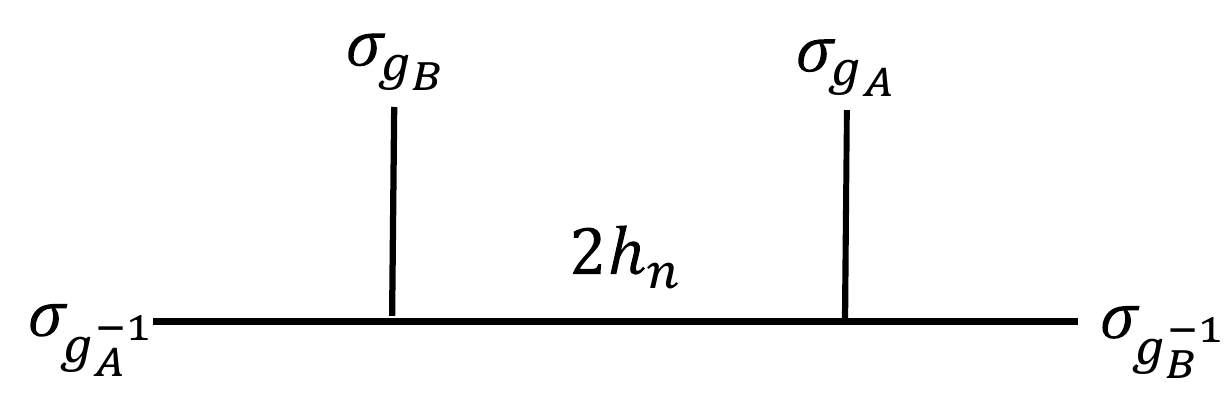}}
\newlength{\pbw}
\settowidth{\pbw}{\usebox{\boxpb}} 

\begin{equation}\label{eq:OPElimit}
  \parbox{\paw}{\usebox{\boxpa}}   \ar{\e \to0} (2i\e)^{-2mnh_O} \times \parbox{\pbw}{\usebox{\boxpb}},
\end{equation}
which means
\begin{equation}\label{eq:vacreduction}
\begin{aligned}
\fr{Z_{n,m}}{\pa{Z_{1,m}}^n}
\ar{ \e \to 0}
	\Braket{\sigma_{g_A}(u_1)\sigma_{g_A^{-1}}(v_1)  \sigma_{g_B}(u_2) \sigma_{g_B^{-1}}(v_2) }_{\text{CFT}^{\otimes mn}},
\end{aligned}
\end{equation}
Therefore, the reflected entropy for the excited state in the early time is just given by that for the vacuum, like the entanglement entropy after a local quench \cite{Asplund2015}.  Note that the explicit form of the vacuum reflected entropy is
\begin{equation}\label{eq:REvacuum}
\fr{c}{3}\log\fr{1+\s{\fr{(-u_1+v_1)(v_2-u_2)}{(-u_1+u_2)(-v_1+v_2)}    }}{1- \s{\fr{(-u_1+v_1)(v_2-u_2)}{(-u_1+u_2)(-v_1+v_2)} }  },
\end{equation}
which exactly matches the entanglement wedge cross section in pure AdS${}_3$ \cite{Takayanagi2018a}.
This can be immediately shown by using the asymptotic form of the Virasoro block (see (\ref{eq:HHLLblock})),
\begin{equation}
\parbox{\pbw}{\usebox{\boxpb}}
\to
2^{4h_n}
\pa{  \fr{1+\s{\fr{(-u_1+v_1)(v_2-u_2)}{(-u_1+u_2)(-v_1+v_2)}    }}{1- \s{\fr{(-u_1+v_1)(v_2-u_2)}{(-u_1+u_2)(-v_1+v_2)} }  }   }^{-2h_n},
\end{equation}
where we take first the limit $c \to \infty$, second $m \to 1$, and finally $n \to 1$.

On the other hand, for $u_2<t<-v_1$, only the holomorphic part of the OPE between $ O^{\otimes mn}$ and $ \dg{{O^{\otimes mn}}}$ crosses a branch cut on the real axis from $u_2$ to $v_2$, which means that the limit $\e \to 0$ is not the usual OPE limit but the {\it Regge limit} \cite{Cornalba2007a,Cornalba2007}.
Before evaluating the 6-point correlator for $u_2<t<-v_1$, we should take care of the fact that there are other choices of the conformal block expansion and a single block approximation besides (\ref{eq:localchannel}).
The point is that the correlator  is invariant under a monodromy transformation, which moves the operators ${O^{\otimes mn}}$, $\dg{{O^{\otimes mn}}}$ around the twist operators. On the other hand, each individual conformal block is not invariant. Thus, we have other choices of the single conformal block approximation and the correct choice is maximal one under the assumption $2h_n \ll nh_m$.

 Fortunately, we find that the correct choice for $u_2<t<\s{-v_1 u_2}$ is just the following channel without monodromy tranformations,
\newsavebox{\boxpg}
\sbox{\boxpg}{\includegraphics[width=190pt]{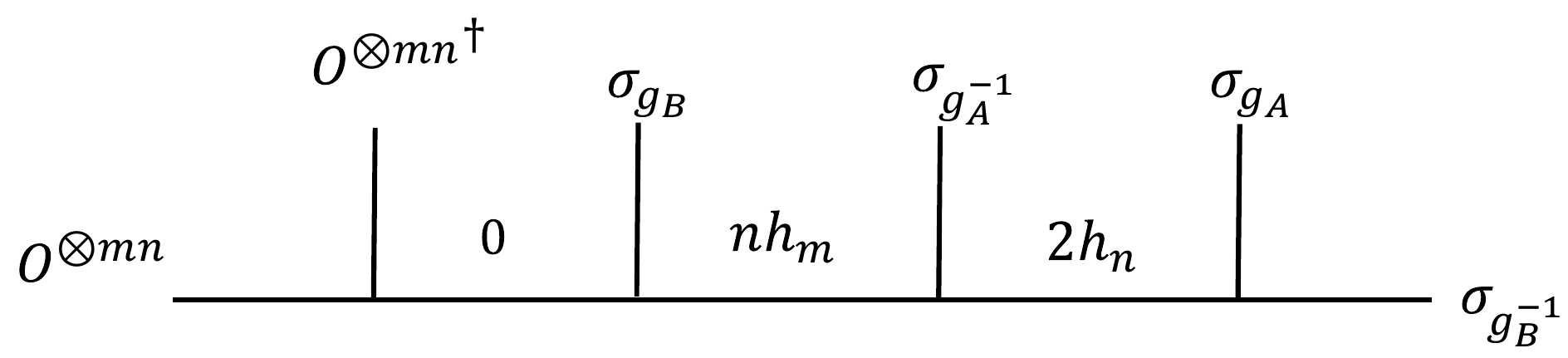}}
\newlength{\pgw}
\settowidth{\pgw}{\usebox{\boxpg}} 

\begin{equation}\label{eq:channel1}
{(C_{n,m})}^2 \parbox{\pgw}{\usebox{\boxpg}} \times (\text{anti-holomorphic part}),
\end{equation}
For  $u_2<t<-v_1$ (in particular, $u_2<t<\s{-v_1 u_2}$) , the effect of crossing the branch cut can be illustrated for the holomorphic part by
\newsavebox{\boxpi}
\sbox{\boxpi}{\includegraphics[width=190pt]{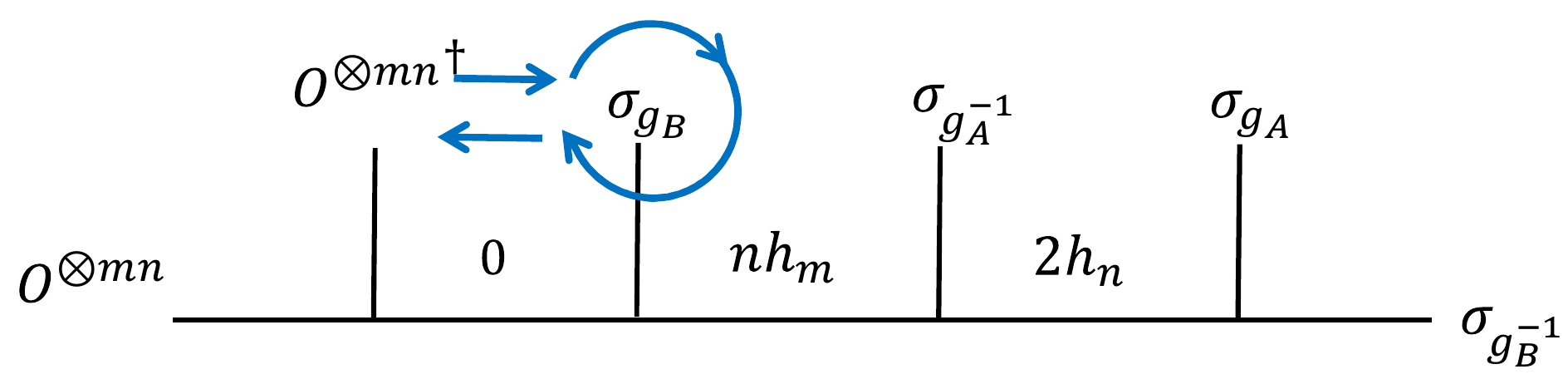}}
\newlength{\piw}
\settowidth{\piw}{\usebox{\boxpi}} 

\begin{equation}\label{eq:effect1}
 \parbox{\piw}{\usebox{\boxpi}},
\end{equation}
which is a conformal block mapped by a monodromy transformation, which moves the operator $\dg{{O^{\otimes mn}}}$ {\it clockwise} around the twist operator $\sigma_{g_B}$ (i.e., $(w_2-u_2) \to \ex{-2\pi i} (w_2-u_2)$).
In general, the effect of the monodromy transformation is encapsulated in the {\it monodromy matrix}, which does not depend on a given CFT data and, therefore, can be evaluated exactly. For the Virasoro block, the monodromy matrix is usually expressed by \cite{Moore1989} (the notation is as in \cite{Kusuki2018c, Kusuki2019})

\newsavebox{\boxpX}
\sbox{\boxpX}{\includegraphics[width=150pt]{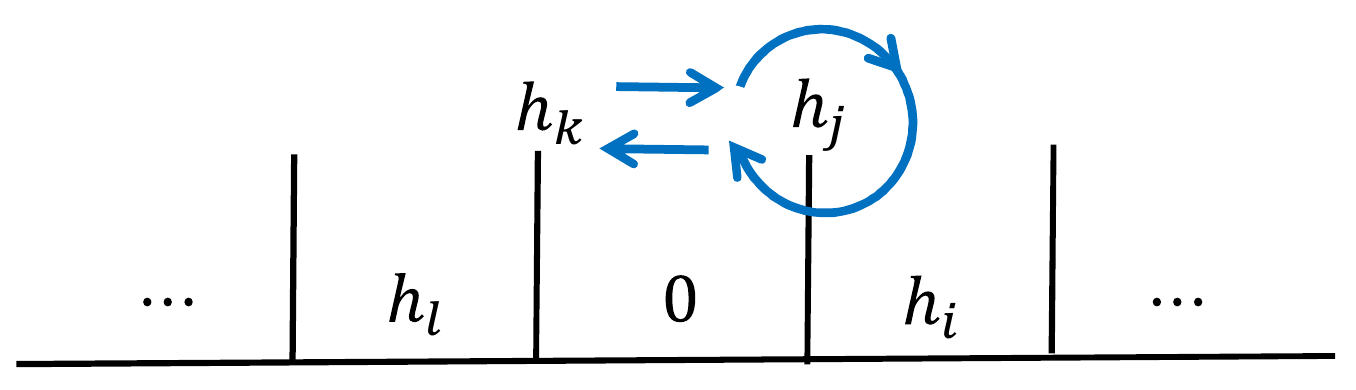}}
\newlength{\pXw}
\settowidth{\pXw}{\usebox{\boxpX}}

\newsavebox{\boxpY}
\sbox{\boxpY}{\includegraphics[width=150pt]{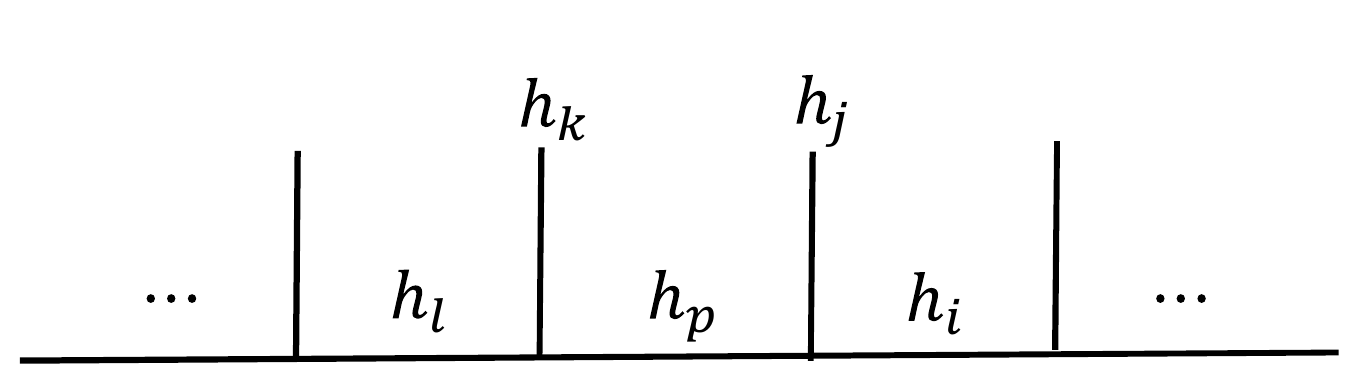}}
\newlength{\pYw}
\settowidth{\pYw}{\usebox{\boxpY}} 

\begin{equation}\label{eq:monodef}
 \parbox{\pXw}{\usebox{\boxpX}}
= \int_{\bb{S}} \dd \a_p 
 {\bold M}_{0, \a_p}^{(-)}
   \left[
    \begin{array}{cc}
    \a_j   & \a_i   \\
     \a_k   &   \a_l \\
    \end{array}
  \right]
	\times \parbox{\pYw}{\usebox{\boxpY}},
\end{equation}
where we introduce the following Liouville notation,
\begin{equation}
c=1+6Q^2, \ \ \ \ \ Q=b+\fr{1}{b},
\end{equation}
and the Liouville momentum,
\begin{equation}
\a_i\pa{Q-\a_i}=h_i.
\end{equation}
The contours run from $\fr{Q}{2}$ to $\fr{Q}{2}+ i\infty$ and also runs clockwise around
$\a_p=\a_i+\a_j+lb<\fr{Q}{2}$ and $\a_p=\a_k+\a_l+lb<\fr{Q}{2}$  \ \ \ ($l \in \bb{Z}_{\geq 0}  $).
\footnote{
Here we choose the convention $0 < b < 1$, which is possible if $c>25$, in particular, large $c$.
}
The superscript $(-)$ of the matrix ${\bold M}$ means the {\it clockwise } monodromy.
It is worth to note that this monodromy matrix only depends on the four external operators $\{ i,j,k,l \}$ and the internal operator $p$, that is, it is independent of other operators described by $\{ \cdots \}$ in (\ref{eq:monodef}).
If one is interested in the details of these transformations, one can refer to \cite{Kusuki2019}.

Like the Virasoro block, the orbifold block (\ref{eq:effect1}) can also be expressed in terms of a certain monodromy matrix as
\newsavebox{\boxph}
\sbox{\boxph}{\includegraphics[width=190pt]{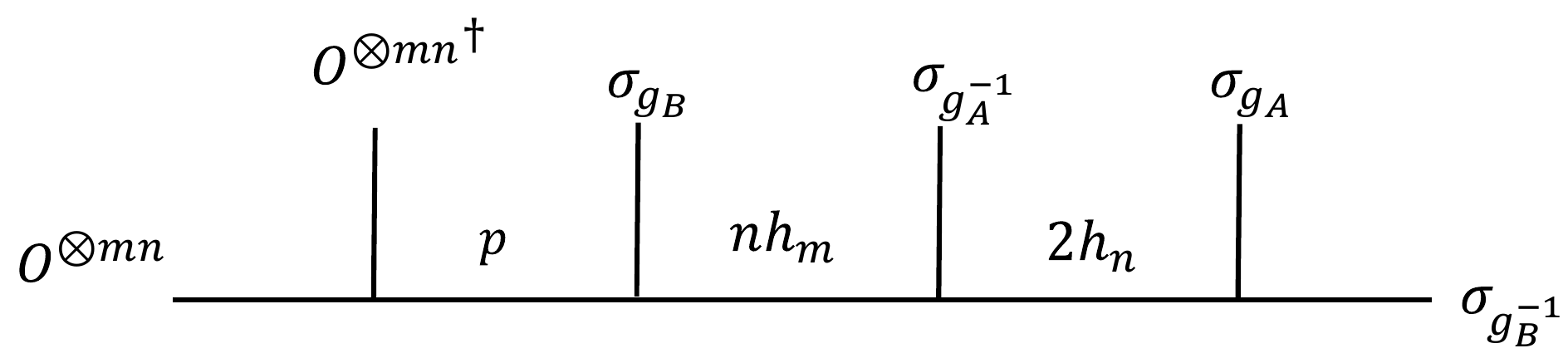}}
\newlength{\phw}
\settowidth{\phw}{\usebox{\boxph}} 

\begin{equation}\label{eq:channel2}
\begin{aligned}
\int_{\bb{S}} \dd \a_p 
 \ti{{\bold M}}_{0, \a_p}^{(-)}
   \left[
    \begin{array}{cc}
    \a_m   & \a_m   \\
     \a_{O}   &   \a_{O} \\
    \end{array}
  \right]
	\times
 \parbox{\phw}{\usebox{\boxph}},
\end{aligned}
\end{equation}
where we denote the monodromy matrix associated with the orbifold block (i.e., not Virasoro block) by $\ti{{\bold M}}$ and define the Liouville momentum,
\begin{equation}
\a_m\pa{Q-\a_m}=h_m\pa{=\fr{c}{24}\pa{m-\fr{1}{m}}}, \ \ \ \ \a_{O}\pa{Q-\a_{O}}=h_O,  \ \ \ \ \bar{\a}_{O}\pa{Q-\bar{\a}_{O}}=\bar{h}_O.
\end{equation}
Although the explicit form of  $\ti{{\bold M}}$ is unknown, that appearing in our calculation can be related to the Virasoto monodromy matrix from the fact (\ref{eq:decoupleblock}). We will explain it in more details when $\ti{{\bold M}}$ appears in the calculation of the reflected entropy.

The anti-holomorphic part does not change in this time region.
As a result, the approximated 6-point function with the monodromy effect (\ref{eq:effect1}) for $u_2<t<\s{-v_1 u_2}$ can be shown as

\begin{equation}
{(C_{n,m})}^2 \parbox{\piw}{\usebox{\boxpi}}  \times \overline{\parbox{\pgw}{\usebox{\boxpg}}},
\end{equation}
where the overline means the anti-holomorphic part.
To proceed further, we consider the Regge limit, which comes from the limit $\e \to 0$.
In fact, the Regge limit of the block is universal 
\cite{Kusuki2019}
\footnote{
The Regge limit of the Virasoro block had first studied in \cite{Kusuki2018b}. And from the observations \cite{Kusuki2018, Kusuki2018a}, it wan shown that the singularity of the Virasoro block is closely related to the fusion matrix \cite{Kusuki2018c, Collier2018} and consequently, the explicit form of the Regge limit is obtained by using the monodromy matrix, which can be rexepressed by the fusion matrix
\cite{Kusuki2019}.
}
because the integral in (\ref{eq:channel2}) is dominated by a Liouville momentum $\a_{\text{min}}$ such that the corresponding conformal dimension $h_{\text{min}}=\a_{\text{min}}( Q-\a_{\text{min}}  )$ is minimal in the set $\{ h| h=\a(Q-\a)  \ \ s.t.  \ \ \a \in\bb{S}    \}$.
In our case, this saddle point contribution comes from the clockwise integral around $\a_{\text{min}}$.
For this reason, we introduce the following notation,
\begin{equation}
 \ca{M}_{0, \a_{\text{min}}}^{(-)} \equiv \text{Res} \pa{-2\pi i  {\bold M}_{0, \a_p}^{(-)}; \a_p=\a_{\text{min}}}.
\end{equation}
We comment on a trivial property of any monodromy matrix,
\begin{equation}
\ti{\ca{M}}_{0, \a_{\text{min}}}^{(-)}
   \left[
    \begin{array}{cc}
    \a   & \a   \\
     \b   &   \b \\
    \end{array}
  \right]
\ar{\a \to 0} 1.
\end{equation}
By using this fact, we obtain (see Appendix \ref{app:5-pt.} in more details)
\begin{equation}\label{eq:F}
\begin{aligned}
 \parbox{\piw}{\usebox{\boxpi}}  
\ar{\e \to0}
	& 2^{4h_n}  (2i\e)^{-4nh_O}\pa{  \fr{1+\s{\fr{(-u_1+v_1)(v_2-t)}{(-u_1+t)(-v_1+v_2)}    }}{1- \s{\fr{(-u_1+v_1)(v_2-t)}{(-u_1+t)(-v_1+v_2)} }  }   }^{-2h_n},
\end{aligned}
\end{equation}
where we take first the limit $c \to \infty$, second $m \to 1$, third $n \to 1$, and finally $\e \to 0$.
Note that the OPE limit between $O^{\otimes mn}$ and $\dg{{O^{\otimes mn}}}$ in the limit $m \in \text{even} \to 1$ is squared by the fact (\ref{eq:squaring}).
Under the limit $m \to 1$, the contribution from the monodromy matrix becomes trivial. In what follows, we will not display the trivial ones under this limit. 
On the other hand, the anti-holomorphic part is just given by the OPE limit,
\begin{equation}\label{eq:antiF}
 \overline{\parbox{\pgw}{\usebox{\boxpg}}}\ar{\e \to 0}
2^{4h_n} (2i\e)^{-4n\bar{h}_O}
\pa{  \fr{1+\s{\fr{(-u_1+v_1)(v_2-u_2)}{(-u_1+u_2)(-v_1+v_2)}    }}{1- \s{\fr{(-u_1+v_1)(v_2-u_2)}{(-u_1+u_2)(-v_1+v_2)} }  }   }^{-2h_n}.
\end{equation}
Substituting these holomorphic part (\ref{eq:F}) and anti-holomorphic part (\ref{eq:antiF}), and (\ref{eq:denominator}), (\ref{eq:Mathur}) into (\ref{eq:Renyi}), we obtain the reflected entropy  at $u_2<t<\s{-v_1 u_2}$ as
\begin{equation}\label{eq:result1}
\fr{c}{6}\log\fr{1+\s{\fr{(-u_1+v_1)(v_2-t)}{(-u_1+t)(-v_1+v_2)}    }}{1- \s{\fr{(-u_1+v_1)(v_2-t)}{(-u_1+t)(-v_1+v_2)} }  }
+
\fr{c}{6}\log\fr{1+\s{\fr{(-u_1+v_1)(v_2-u_2)}{(-u_1+u_2)(-v_1+v_2)}    }}{1- \s{\fr{(-u_1+v_1)(v_2-u_2)}{(-u_1+u_2)(-v_1+v_2)} }  }
.
\end{equation}

For $\s{-v_1 u_2}<t<-v_1$, the 6-point conformal block is NOT dominated by the usual block, but the block illustrated by
\footnote{
In the anti-holomorphic $\bar{z}$ plane, the imaginary direction is flipped, therefore, the arrow of the monodromy transformation is also flipped.
}
\newsavebox{\boxpz}
\sbox{\boxpz}{\includegraphics[width=190pt]{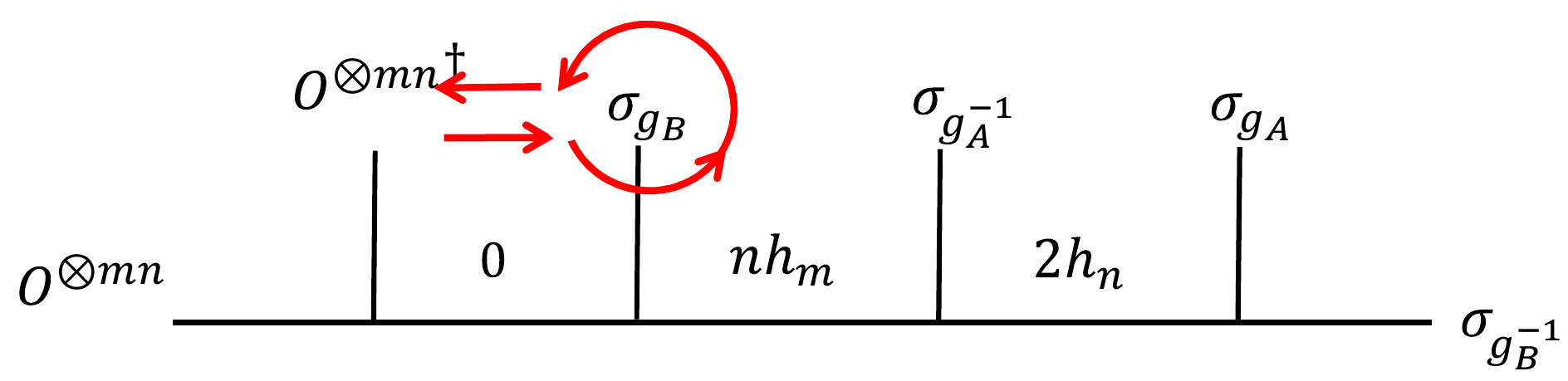}}
\newlength{\pzw}
\settowidth{\pzw}{\usebox{\boxpz}} 

\begin{equation}\label{eq:ch2}
{(C_{n,m})}^2   \parbox{\pzw}{\usebox{\boxpz}}\times     \overline{\parbox{\piw}{\usebox{\boxpi}}}    .
\end{equation}
The effect of crossing the brunch cut is the same as that for  $u_2<t<\s{-v_1 u_2}$. This effect cancels the monodromy illustrated in (\ref{eq:ch2}), therefore, the approximated 6-point function at $\s{-v_1 u_2}<t<-v_1$ results in
\begin{equation}
{(C_{n,m})}^2  \parbox{\pgw}{\usebox{\boxpg}}   \times   \overline{\parbox{\piw}{\usebox{\boxpi}} }.
\end{equation}
Each holomorphic and anti-holomorphic conformal block is the same as (\ref{eq:antiF}) and (\ref{eq:F}), consequently, we obtain 
\begin{equation}\label{eq:result1-2}
\fr{c}{6}\log\fr{1+\s{\fr{(-u_1+v_1)(v_2+t)}{(-u_1-t)(-v_1+v_2)}    }}{1- \s{\fr{(-u_1+v_1)(v_2+t)}{(-u_1-t)(-v_1+v_2)} }  }
+
\fr{c}{6}\log\fr{1+\s{\fr{(-u_1+v_1)(v_2-u_2)}{(-u_1+u_2)(-v_1+v_2)}    }}{1- \s{\fr{(-u_1+v_1)(v_2-u_2)}{(-u_1+u_2)(-v_1+v_2)} }  }
.
\end{equation}

For $-v_1<t<-u_1$, the dominant channel is given by 

\begin{equation}
{(C_{n,m})}^2   \parbox{\pzw}{\usebox{\boxpz}}\times    \overline{\parbox{\piw}{\usebox{\boxpi}}}    .
\end{equation}
In a similar way as (\ref{eq:effect1}), the holomorphic part of the block is affected by crossing the brunch cut as

\begin{equation}\label{eq:effect2}
 \parbox{\piw}{\usebox{\boxpi}},
\end{equation}
In the time region $-v_1<t<-u_1$, the anti-holomorphic part of the OPE between $ O^{\otimes mn}$ and $ \dg{{O^{\otimes mn}}}$ also crosses a branch cut on the real axis from $-u_1$ to $-v_1$. This  affects the anti-holomorphic block as 
\newsavebox{\boxpk}
\sbox{\boxpk}{\includegraphics[width=190pt]{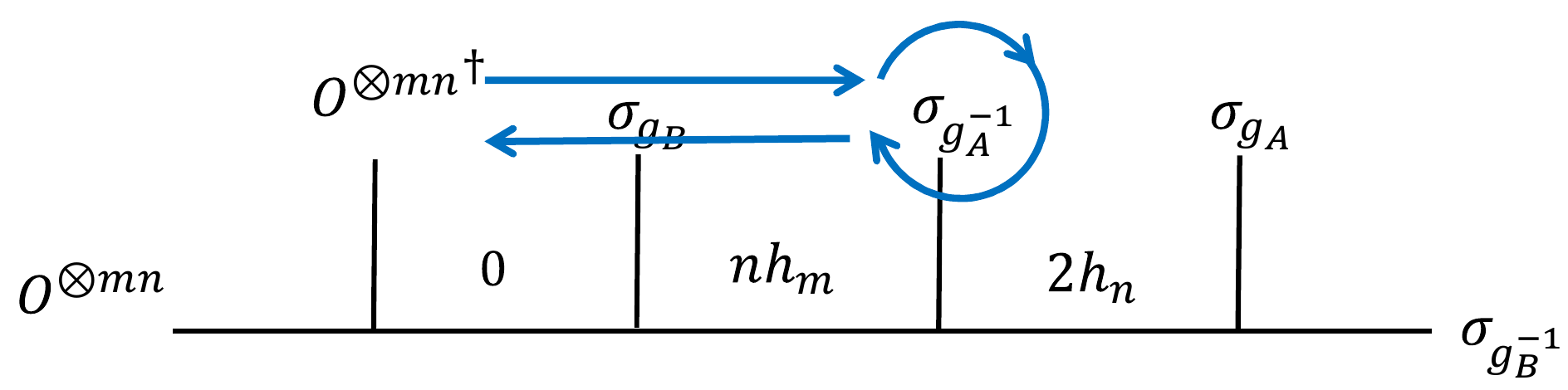}}
\newlength{\pkw}
\settowidth{\pkw}{\usebox{\boxpk}} 

\begin{equation}\label{eq:effect3}
 \overline{\parbox{\pkw}{\usebox{\boxpk}}}.
\end{equation}
Taking account of the effects (\ref{eq:effect2}) and  (\ref{eq:effect3}), the approximated 6-point function for $-v_1<t<-u_1$ can be illustrated by

\newsavebox{\boxpza}
\sbox{\boxpza}{\includegraphics[width=190pt]{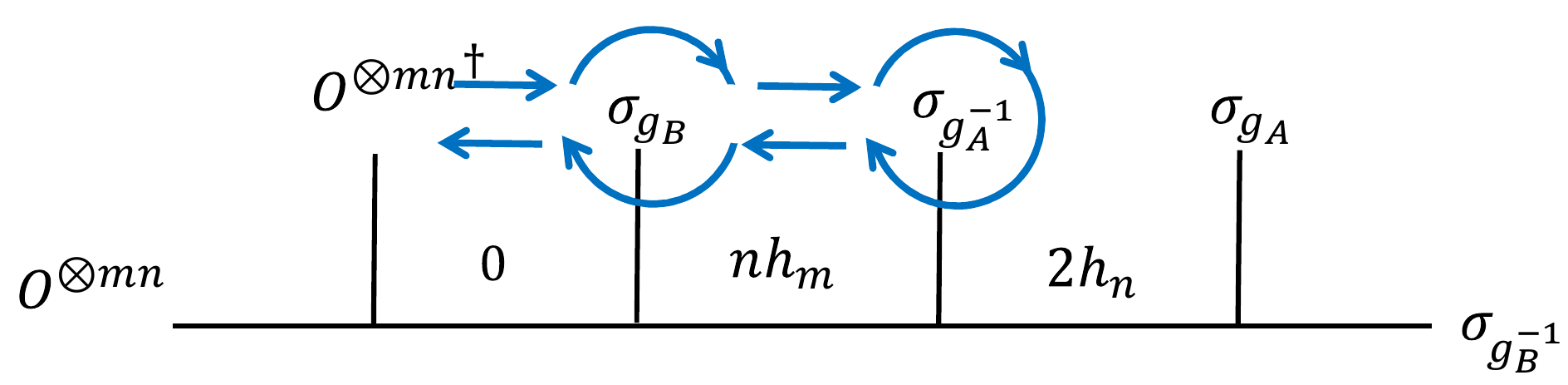}}
\newlength{\pzaw}
\settowidth{\pzaw}{\usebox{\boxpza}} 

\begin{equation}\label{eq:2mono}
{(C_{n,m})}^2   \parbox{\pgw}{\usebox{\boxpg}} \times \overline{ \parbox{\pzaw}{\usebox{\boxpza}}  }.
\end{equation}
Let us evaluate the Regge limit of this approximated 6-point function. By our result \cite{Kusuki2019} again, we obtain
\begin{equation}\label{eq:F2}
\begin{aligned}
&\overline{ \parbox{\pzaw}{\usebox{\boxpza}} } \\
&	\ar{\e \to0} (-2i\e)^{4h_n-4nh_O} 
	\pa{    \fr{(u_2-v_1)(u_1-v_2)}{(t+u_1)(t+u_2)(t+v_1)(t+v_2)}   }^{2h_n}  \ca{M}_n [O],
\end{aligned}
\end{equation}
where we take first the limit $c \to \infty$, second $m \to 1$, third $n \to 1$, and finally $\e \to 0$.
Here $\ca{M}_n[O]$ is a constant, given by the monodromy matrix, and the asymptotic expression in these limits is
\begin{equation}\label{eq:monomono}
 \ca{M}_n[O]\rightarrow\left(\ti{\ca{M}}_{0, 2\a_n}^{(-)}
   \left[
    \begin{array}{cc}
    \a_n   & \a_n   \\
     \a_{O}   &   \a_{O} \\
    \end{array}
  \right]\right)^2
=\pa{ \fr{2}{ i \bar{\g}} \sinh \pi \bar{\g}}^{-4h_n},
\end{equation}
where $\a_n$ is given by $\a_n(Q-\a_n)=h_n$ and we define $\bar{\g}=\s{\fr{24}{c} \bar{h}_O-1}$. 
The square comes from the decoupling of the orbifold block into two Virasoro blocks as explained in (\ref{eq:decoupleblock}).
More detailed calculation can be found in Appendix \ref{app:5-pt.}.
Thus, we obtain the reflected entropy  as
\begin{equation}\label{eq:resultA}
\fr{c}{6} \log \BR{   \fr{4(t+u_1)(t+u_2)(t+v_1)(t+v_2)}{\e^2(u_2-v_1)(u_1-v_2)}   \pa{\fr{\sinh \pi \bar{\g}}{\bar{\g}}}^2   }
+\fr{c}{6}\log\fr{1+\s{\fr{(-u_1+v_1)(v_2-u_2)}{(-u_1+u_2)(-v_1+v_2)}    }}{1- \s{\fr{(-u_1+v_1)(v_2-u_2)}{(-u_1+u_2)(-v_1+v_2)} }  } ,  \ \ \ \ \    \text{if } -v_1<t<-u_1 .
\end{equation}

However, as explained later (in Section \ref{sec:Heavy}), there is another possibility to dominate the 6-point correlater by the following channel,

\newsavebox{\boxpzd}
\sbox{\boxpzd}{\includegraphics[width=190pt]{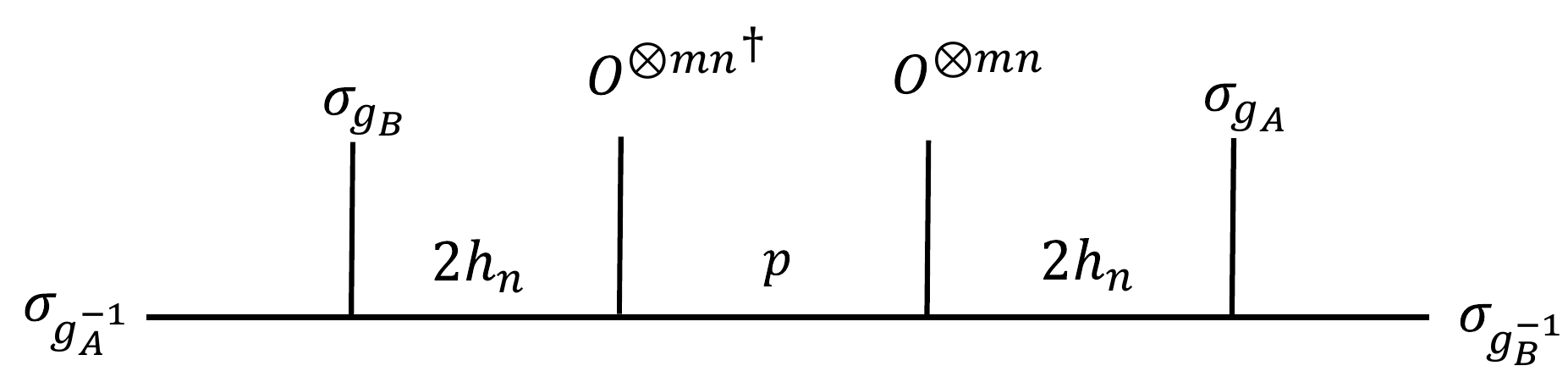}}
\newlength{\pzdw}
\settowidth{\pzdw}{\usebox{\boxpzd}} 

\begin{equation}\label{eq:conchannel}
{(C_{n,m})}^2 {(C_{n,O})}^2   \parbox{\pzdw}{\usebox{\boxpzd}} \times \overline{ \parbox{\pzdw}{\usebox{\boxpzd}}  } .
\end{equation}
The constant $C_{n,O}$ is the OPE coefficient with $O^{\otimes mn}$.
The intermediate state $p$ corresponds to the dominant contribution to the correlator.
In the $m ,n\to 1$ limit, $p$ is given by $O^{\otimes 2}$ in the CFT of interest \cite{Banerjee2016}.
In the bulk side, this channel corresponds to the disconnected entanglement wedge cross section which ends at the block hole horizon (which is discussed more in Section \ref{sec:Heavy}).

The effect of crossing the branch cut is illustrated by
\newsavebox{\boxpx}
\sbox{\boxpx}{\includegraphics[width=190pt]{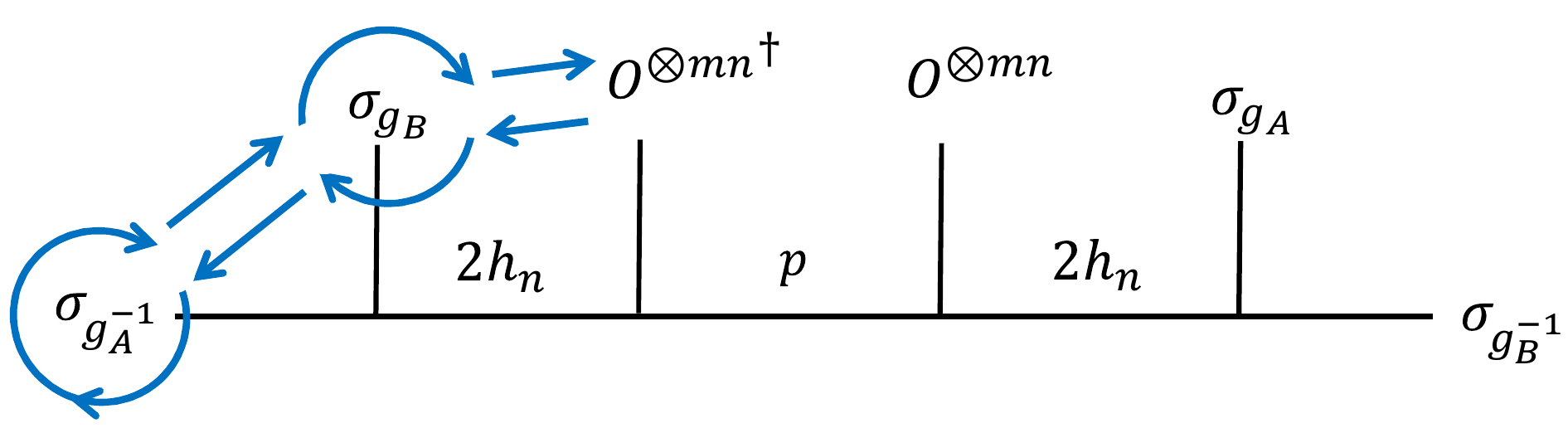}}
\newlength{\pxw}
\settowidth{\pxw}{\usebox{\boxpx}}

\begin{equation}
 \parbox{\pxw}{\usebox{\boxpx}} \times  \overline{\parbox{\pzdw}{\usebox{\boxpzd}} } .
\end{equation}
Each conformal block is approximated in the von-Neumann limit as
\begin{equation}
 \parbox{\pxw}{\usebox{\boxpx}}  \ar{\e \to0} (2i\e)^{4h_n-4nh_O} 
	\pa{    \fr{(u_2-v_1)(u_1-v_2)}{(t-u_1)(t-u_2)(t-v_1)(t-v_2)}   }^{2h_n} ,
\end{equation}
and
\begin{equation}
 \overline{\parbox{\pzdw}{\usebox{\boxpzd}} } \ar{\e \to0} (-2i\e)^{4h_n-4nh_O} 
	\pa{    \fr{(u_2-v_1)(u_1-v_2)}{(t+u_1)(t+u_2)(t+v_1)(t+v_2)}   }^{2h_n} .
\end{equation}
Thus, we obtain the reflected entropy  as
\begin{equation}\label{eq:resultAano}
\fr{c}{6} \log \BR{   \fr{(t-u_1)(t-u_2)(t-v_1)(t-v_2)}{\e^2 \g(u_2-v_1)(u_1-v_2)}   }
+\fr{c}{6} \log \BR{   \fr{(t+u_1)(t+u_2)(t+v_1)(t+v_2)}{\e^2 \bar{\g}(u_2-v_1)(u_1-v_2)}   }
+(\text{const.})
\end{equation}
We can immediately find that the $\e$-singularity of this result is much larger than \ref{eq:resultA}), therefore, we can neglect this possibility.

The calculation of the reflected entropy for $-u_1<t$ is almost the same as the derivation of (\ref{eq:result1}), therefore, we can summarize our results as
\begin{equation}\label{eq:main}
\begin{aligned}
S_R (A:B)[O]&=\left\{
    \begin{array}{ll}
    \fr{c}{3}\log\fr{1+\s{\fr{(-u_1+v_1)(v_2-u_2)}{(-u_1+u_2)(-v_1+v_2)}    }}{1- \s{\fr{(-u_1+v_1)(v_2-u_2)}{(-u_1+u_2)(-v_1+v_2)} }  } ,
			& \text{if } t<u_2 ,\\ \\
    \fr{c}{6}\log\fr{1+\s{\fr{(-u_1+v_1)(v_2-t)}{(-u_1+t)(-v_1+v_2)}    }}{1- \s{\fr{(-u_1+v_1)(v_2-t)}{(-u_1+t)(-v_1+v_2)} }  }
		+
		\fr{c}{6}\log\fr{1+\s{\fr{(-u_1+v_1)(v_2-u_2)}{(-u_1+u_2)(-v_1+v_2)}    }}{1- \s{\fr{(-u_1+v_1)(v_2-u_2)}{(-u_1+u_2)(-v_1+v_2)} }  }
		,  & \text{if } u_2<t<\s{-v_1 u_2} ,\\ \\
    \fr{c}{6}\log\fr{1+\s{\fr{(-u_1+v_1)(v_2+t)}{(-u_1-t)(-v_1+v_2)}    }}{1- \s{\fr{(-u_1+v_1)(v_2+t)}{(-u_1-t)(-v_1+v_2)} }  }
		+
		\fr{c}{6}\log\fr{1+\s{\fr{(-u_1+v_1)(v_2-u_2)}{(-u_1+u_2)(-v_1+v_2)}    }}{1- \s{\fr{(-u_1+v_1)(v_2-u_2)}{(-u_1+u_2)(-v_1+v_2)} }  }
		,  & \text{if } \s{-v_1 u_2}<t<-v_1 ,\\ \\
      \fr{c}{6} \log \BR{   \fr{4(t+u_1)(t+u_2)(t+v_1)(t+v_2)}{\e^2(u_2-v_1)(u_1-v_2)}   \pa{\fr{\sinh \pi \bar{\g}}{\bar{\g}}}^2   }
+\fr{c}{6}\log\fr{1+\s{\fr{(-u_1+v_1)(v_2-u_2)}{(-u_1+u_2)(-v_1+v_2)}    }}{1- \s{\fr{(-u_1+v_1)(v_2-u_2)}{(-u_1+u_2)(-v_1+v_2)} }  } 
	,  & \text{if } -v_1<t<-u_1 ,\\ \\
		\fr{c}{6}\log\fr{1+\s{\fr{(-u_1+v_1)(-t-u_2)}{(-u_1+u_2)(-v_1-t)}    }}{1- \s{\fr{(-u_1+v_1)(-t-u_2)}{(-u_1+u_2)(-v_1-t)} }  }
		+
		\fr{c}{6}\log\fr{1+\s{\fr{(-u_1+v_1)(v_2-u_2)}{(-u_1+u_2)(-v_1+v_2)}    }}{1- \s{\fr{(-u_1+v_1)(v_2-u_2)}{(-u_1+u_2)(-v_1+v_2)} }  }
		,  & \text{if } -u_1<t<\s{-u_1 v_2} ,\\ \\
		\fr{c}{6}\log\fr{1+\s{\fr{(-u_1+v_1)(t-u_2)}{(-u_1+u_2)(-v_1+t)}    }}{1- \s{\fr{(-u_1+v_1)(t-u_2)}{(-u_1+u_2)(-v_1+t)} }  }
		+
		\fr{c}{6}\log\fr{1+\s{\fr{(-u_1+v_1)(v_2-u_2)}{(-u_1+u_2)(-v_1+v_2)}    }}{1- \s{\fr{(-u_1+v_1)(v_2-u_2)}{(-u_1+u_2)(-v_1+v_2)} }  }
		,  & \text{if } \s{-u_1 v_2}<t<v_2 ,\\ \\
     \fr{c}{3}\log\fr{1+\s{\fr{(-u_1+v_1)(v_2-u_2)}{(-u_1+u_2)(-v_1+v_2)}    }}{1- \s{\fr{(-u_1+v_1)(v_2-u_2)}{(-u_1+u_2)(-v_1+v_2)} }  } ,& \text{if } v_2<t .\\ 
    \end{array}
  \right.\\
\end{aligned}
\end{equation}

We can also consider the case $0<\e\ll u_2<v_2<u_1<v_1$. The different monodromy effect from the above case can happen when $t>v_2$. For $v_2<t<u_1$, we find the dominant channel
\newsavebox{\boxpt}
\sbox{\boxpt}{\includegraphics[width=190pt]{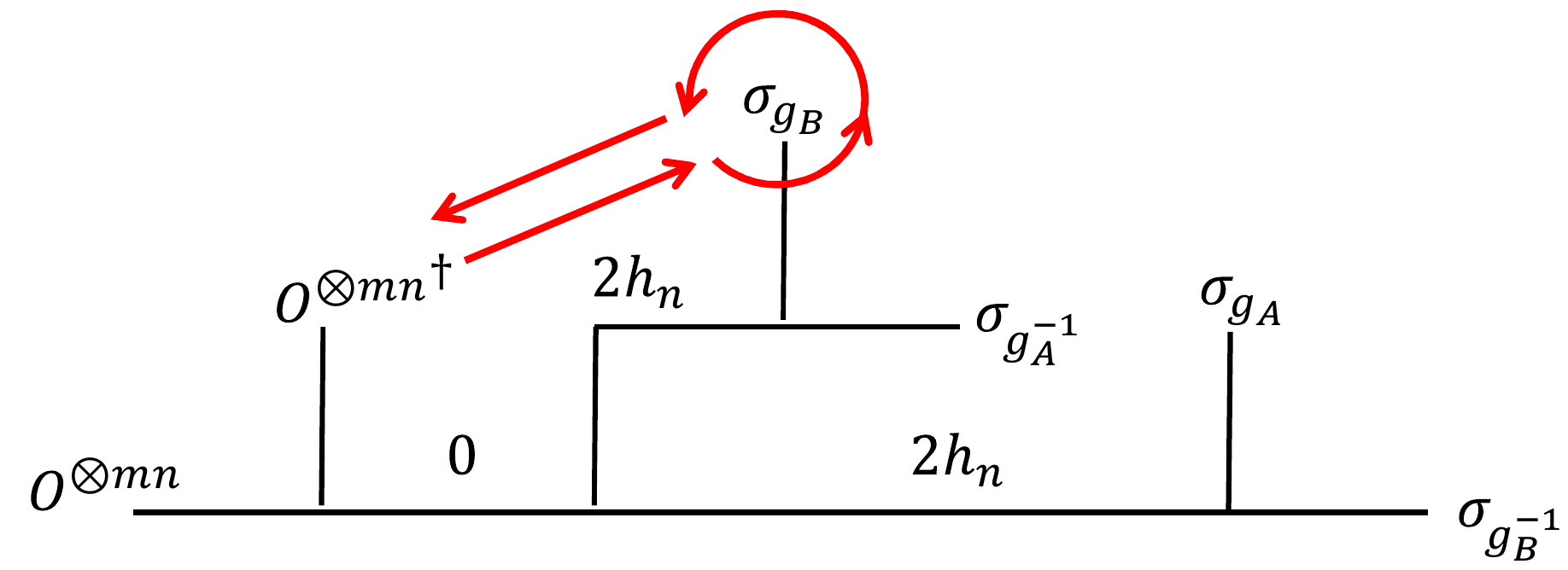}}
\newlength{\ptw}
\settowidth{\ptw}{\usebox{\boxpt}} 

\newsavebox{\boxpzg}
\sbox{\boxpzg}{\includegraphics[width=190pt]{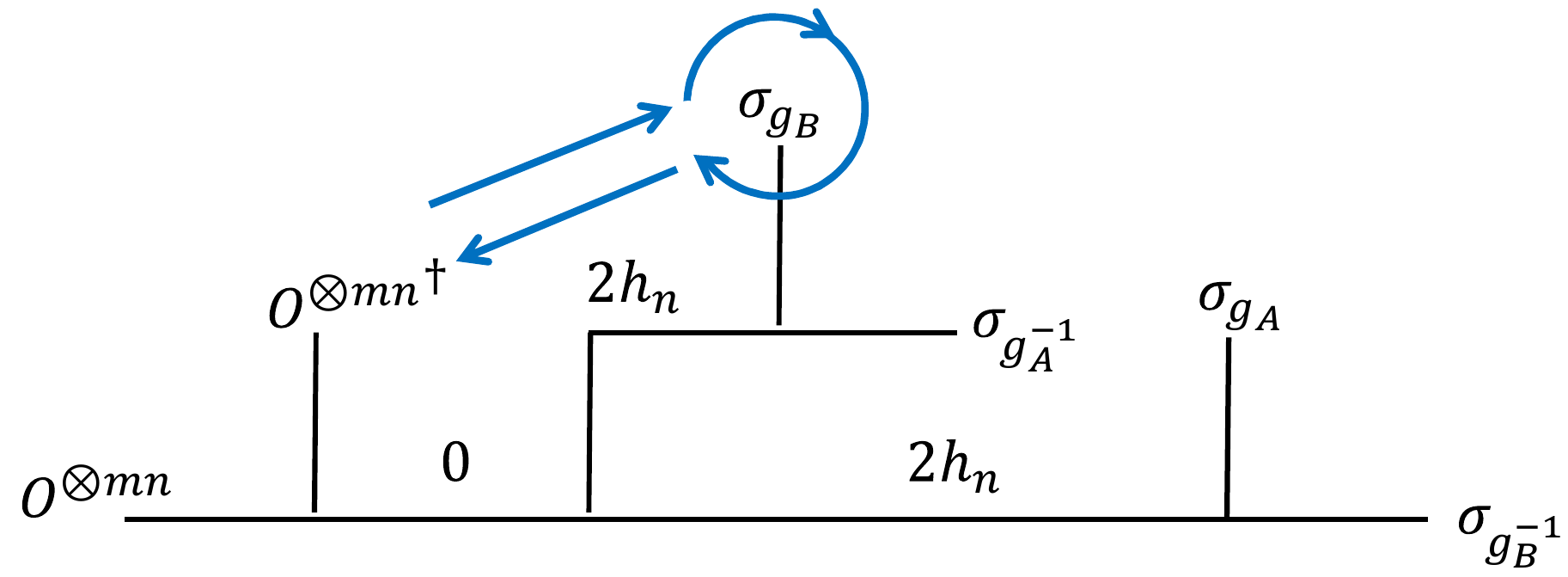}}
\newlength{\pzgw}
\settowidth{\pzgw}{\usebox{\boxpzg}}

\begin{equation}\label{eq:setup2a}
{(C_{n,m})}^2 \parbox{\ptw}{\usebox{\boxpt}} \times \overline{ \parbox{\pzgw}{\usebox{\boxpzg}}}     .
\end{equation}
The monodromy effect can be illustrated as
\newsavebox{\boxpu}
\sbox{\boxpu}{\includegraphics[width=190pt]{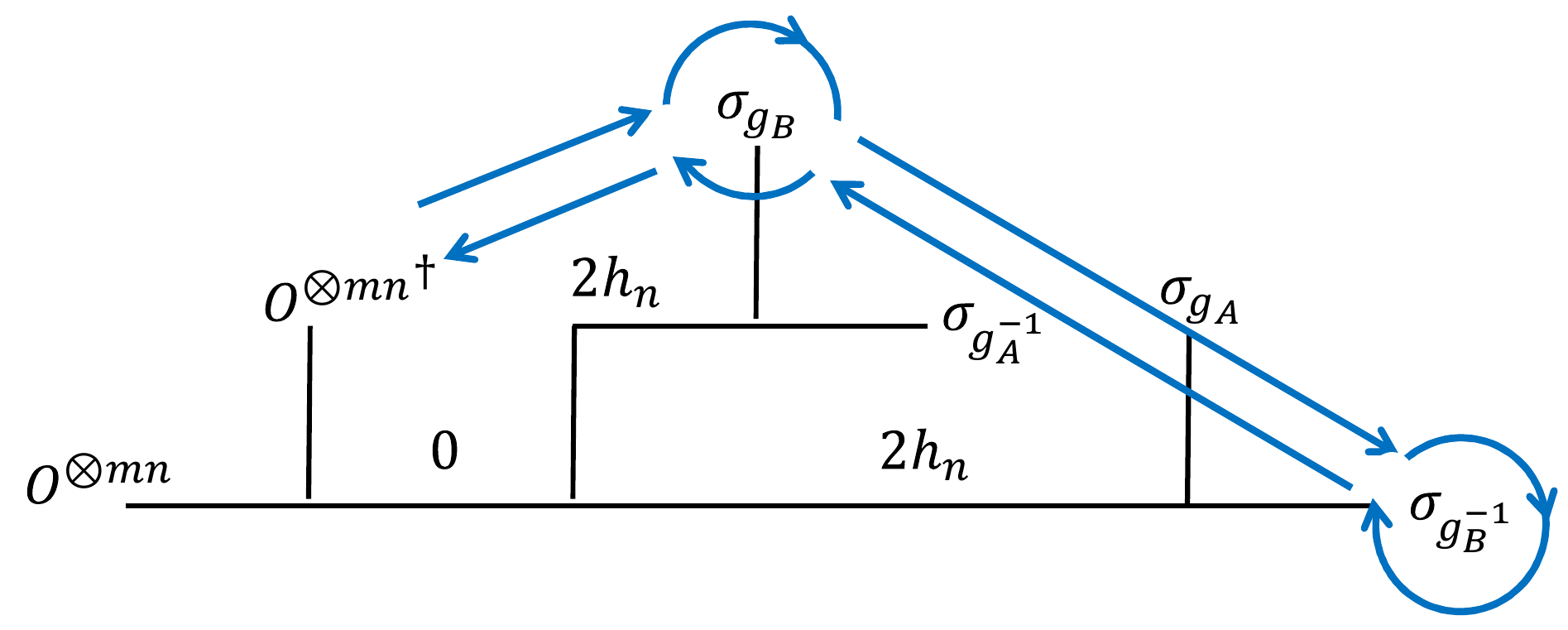}}
\newlength{\puw}
\settowidth{\puw}{\usebox{\boxpu}} 

\begin{equation}\label{eq:setup2b}
 \parbox{\puw}{\usebox{\boxpu}} \times \overline{ \parbox{\paw}{\usebox{\boxpa}} }.
\end{equation}
Combining (\ref{eq:setup2a}) with (\ref{eq:setup2b}), we find that the reflected entropy can be evaluated by the following approximated correlator,
\newsavebox{\boxpv}
\sbox{\boxpv}{\includegraphics[width=190pt]{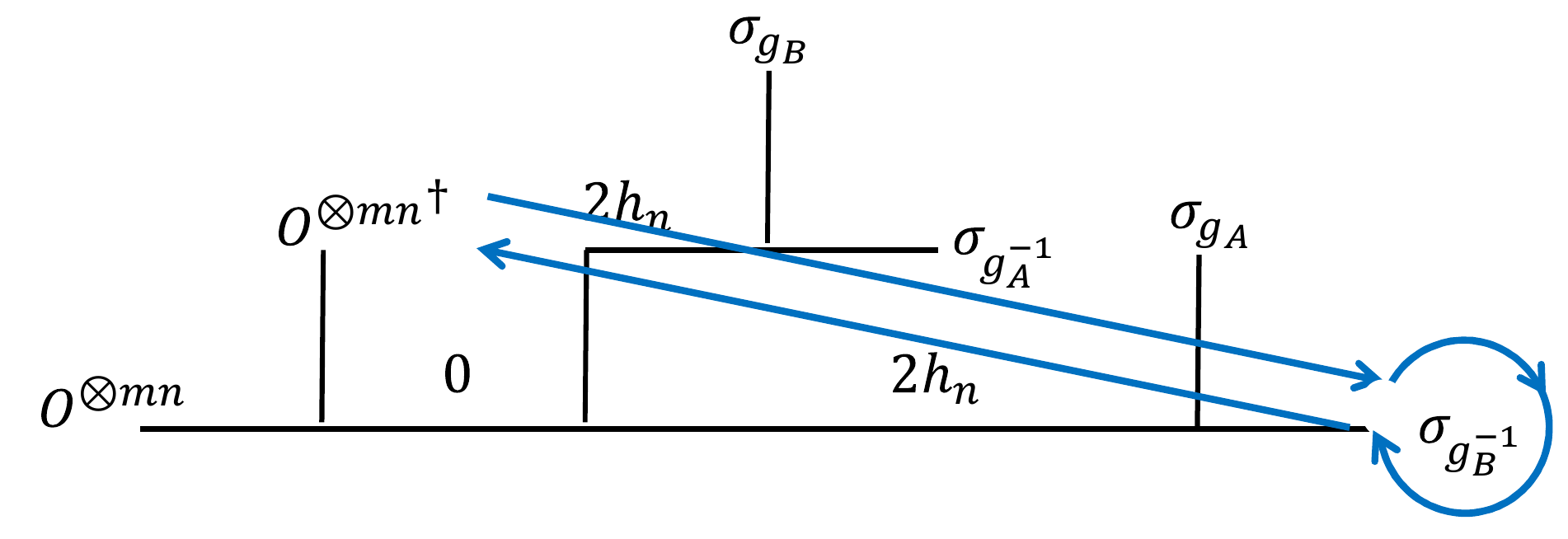}}
\newlength{\pvw}
\settowidth{\pvw}{\usebox{\boxpv}} 

\begin{equation}
{(C_{n,m})}^2 \parbox{\pvw}{\usebox{\boxpv}}   \times \overline{ \parbox{\pzgw}{\usebox{\boxpzg}}}    .
\end{equation}
Note that one of the effects of crossing branch cut cancels the monodromy around $z=u_2$ displayed in (\ref{eq:setup2a}).
We can apply the same technique as (\ref{eq:F}) to calculate each these left and light blocks and then we obtain
\begin{equation}
S_R (A:B)[O]=
\fr{c}{6}\log\fr{1+\s{\fr{(-u_1+v_1)(t-u_2)}{(-u_1+u_2)(-v_1+t)}    }}{1- \s{\fr{(-u_1+v_1)(t-u_2)}{(-u_1+u_2)(-v_1+t)} }  }
+
\fr{c}{6}\log\fr{1+\s{\fr{(-u_1+v_1)(v_2+t)}{(-u_1-t)(-v_1+v_2)}    }}{1- \s{\fr{(-u_1+v_1)(v_2+t)}{(-u_1-t)(-v_1+v_2)} }  },
\ \ \ \ 
\text{ if }
v_2<t<u_1.
\end{equation}
In a similar manner, we can also evaluate the reflected entropy for the other time region and thus we obtain
\begin{equation}
\begin{aligned}
S_R (A:B)[O]&=\left\{
    \begin{array}{ll}
    \fr{c}{3}\log\fr{1+\s{\fr{(-u_1+v_1)(v_2-u_2)}{(-u_1+u_2)(-v_1+v_2)}    }}{1- \s{\fr{(-u_1+v_1)(v_2-u_2)}{(-u_1+u_2)(-v_1+v_2)} }  } ,& \text{if } t<u_2 ,\\ \\
    \fr{c}{6}\log\fr{1+\s{\fr{(-u_1+v_1)(v_2-t)}{(-u_1+t)(-v_1+v_2)}    }}{1- \s{\fr{(-u_1+v_1)(v_2-t)}{(-u_1+t)(-v_1+v_2)} }  }
		+
		\fr{c}{6}\log\fr{1+\s{\fr{(-u_1+v_1)(v_2-u_2)}{(-u_1+u_2)(-v_1+v_2)}    }}{1- \s{\fr{(-u_1+v_1)(v_2-u_2)}{(-u_1+u_2)(-v_1+v_2)} }  }
		,  & \text{if } u_2<t<v_2 ,\\ \\
\fr{c}{6}\log\fr{1+\s{\fr{(-u_1+v_1)(t-u_2)}{(-u_1+u_2)(-v_1+t)}    }}{1- \s{\fr{(-u_1+v_1)(t-u_2)}{(-u_1+u_2)(-v_1+t)} }  }
+
\fr{c}{6}\log\fr{1+\s{\fr{(-u_1+v_1)(v_2+t)}{(-u_1-t)(-v_1+v_2)}    }}{1- \s{\fr{(-u_1+v_1)(v_2+t)}{(-u_1-t)(-v_1+v_2)} }  },
		  & \text{if } v_2<t<u_1 ,\\ \\
		\fr{c}{6}\log\fr{1+\s{\fr{(-u_1+t)(v_2-u_2)}{(-u_1+u_2)(-t+v_2)}    }}{1- \s{\fr{(-u_1+t)(v_2-u_2)}{(-u_1+u_2)(-t+v_2)} }  }
		+
		\fr{c}{6}\log\fr{1+\s{\fr{(-u_1+v_1)(v_2-u_2)}{(-u_1+u_2)(-v_1+v_2)}    }}{1- \s{\fr{(-u_1+v_1)(v_2-u_2)}{(-u_1+u_2)(-v_1+v_2)} }  }
		,  & \text{if } u_1<t<v_1 ,\\ \\
     \fr{c}{3}\log\fr{1+\s{\fr{(-u_1+v_1)(v_2-u_2)}{(-u_1+u_2)(-v_1+v_2)}    }}{1- \s{\fr{(-u_1+v_1)(v_2-u_2)}{(-u_1+u_2)(-v_1+v_2)} }  } ,& \text{if } v_1<t .\\ 
    \end{array}
  \right.\\
\end{aligned}
\end{equation}

However, this is not complete. In this setup, we have to take {\it disconnected channel} into account.
\footnote{
Here, we mean {\it disconnected} by the disconnected entanglement wedge, which generally leads to vanishing of mutual information.
 This is NOT the disconnected entanglement wedge cross section, like the right upper sketch of Figure \ref{fig:transition}.
}
The entanglement wedge for two sybsystems after a local quench was studied in \cite{Asplund2014}.
From this result, we can find that in the case $0<\e\ll u_2<v_2<u_1<v_1$, there is a possibility that the disconnected entanglement wedge is chosen as the minimal RT surface for two sybsystems.
If we set $\abs{u_1-v_1}=\abs{u_2-v_2}=l$ and $\abs{u_1-v_2}=d$ for simplicity, then the transition time between connected and disconnected entanglement wedge is given by
\begin{equation}
\begin{aligned}
     \bar{v}_2 \equiv  u_2+l+\fr{d^2(2l+d)}{d^2+dl-l^2}   , \ \ \ \ \ \ \ 
     \bar{u}_1 \equiv  u_2-\fr{l^3}{d^2+dl-l^2}    .
\end{aligned}
\end{equation}
Therefore, the correct reflected entropy is modified by
\begin{equation}\label{eq:disresult}
\begin{aligned}
S_R (A:B)[O]&=\left\{
    \begin{array}{ll}
    \fr{c}{3}\log\fr{1+\s{\fr{(-u_1+v_1)(v_2-u_2)}{(-u_1+u_2)(-v_1+v_2)}    }}{1- \s{\fr{(-u_1+v_1)(v_2-u_2)}{(-u_1+u_2)(-v_1+v_2)} }  } ,& \text{if } t<u_2 ,\\ \\
    \fr{c}{6}\log\fr{1+\s{\fr{(-u_1+v_1)(v_2-t)}{(-u_1+t)(-v_1+v_2)}    }}{1- \s{\fr{(-u_1+v_1)(v_2-t)}{(-u_1+t)(-v_1+v_2)} }  }
		+
		\fr{c}{6}\log\fr{1+\s{\fr{(-u_1+v_1)(v_2-u_2)}{(-u_1+u_2)(-v_1+v_2)}    }}{1- \s{\fr{(-u_1+v_1)(v_2-u_2)}{(-u_1+u_2)(-v_1+v_2)} }  }
		,  & \text{if } u_2<t<\bar{v}_2 ,\\ \\
0,
		  & \text{if } \bar{v}_2<t<\bar{u}_1 ,\\ \\
		\fr{c}{6}\log\fr{1+\s{\fr{(-u_1+t)(v_2-u_2)}{(-u_1+u_2)(-t+v_2)}    }}{1- \s{\fr{(-u_1+t)(v_2-u_2)}{(-u_1+u_2)(-t+v_2)} }  }
		+
		\fr{c}{6}\log\fr{1+\s{\fr{(-u_1+v_1)(v_2-u_2)}{(-u_1+u_2)(-v_1+v_2)}    }}{1- \s{\fr{(-u_1+v_1)(v_2-u_2)}{(-u_1+u_2)(-v_1+v_2)} }  }
		,  & \text{if } \bar{u}_1<t<v_1 ,\\ \\
     \fr{c}{3}\log\fr{1+\s{\fr{(-u_1+v_1)(v_2-u_2)}{(-u_1+u_2)(-v_1+v_2)}    }}{1- \s{\fr{(-u_1+v_1)(v_2-u_2)}{(-u_1+u_2)(-v_1+v_2)} }  } ,& \text{if } v_1<t .\\ 
    \end{array}
  \right.\\
\end{aligned}
\end{equation}

\subsection{Quench inside Region $A$ and $B$} 

In this subsection, we consider a local excitation inside the interval $A$.
We can accomplish the evaluation for this state by using the same 6-point correlator (\ref{eq:Renyi}) with $0<\e\ll v_1<u_2<v_2<-u_1$ .
The early time reflected entropy is again that for the vacuum due to the same reason as in the calculation of (\ref{eq:vacreduction}).
For $v_1<t<u_2$, the dominant channel is

\newsavebox{\boxpzc}
\sbox{\boxpzc}{\includegraphics[width=190pt]{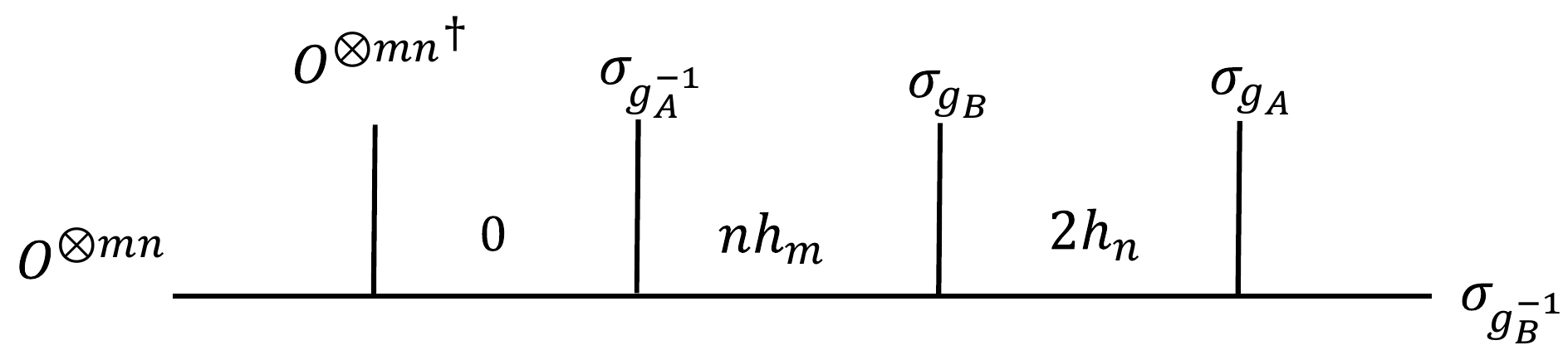}}
\newlength{\pzcw}
\settowidth{\pzcw}{\usebox{\boxpzc}} 

\newsavebox{\boxps}
\sbox{\boxps}{\includegraphics[width=190pt]{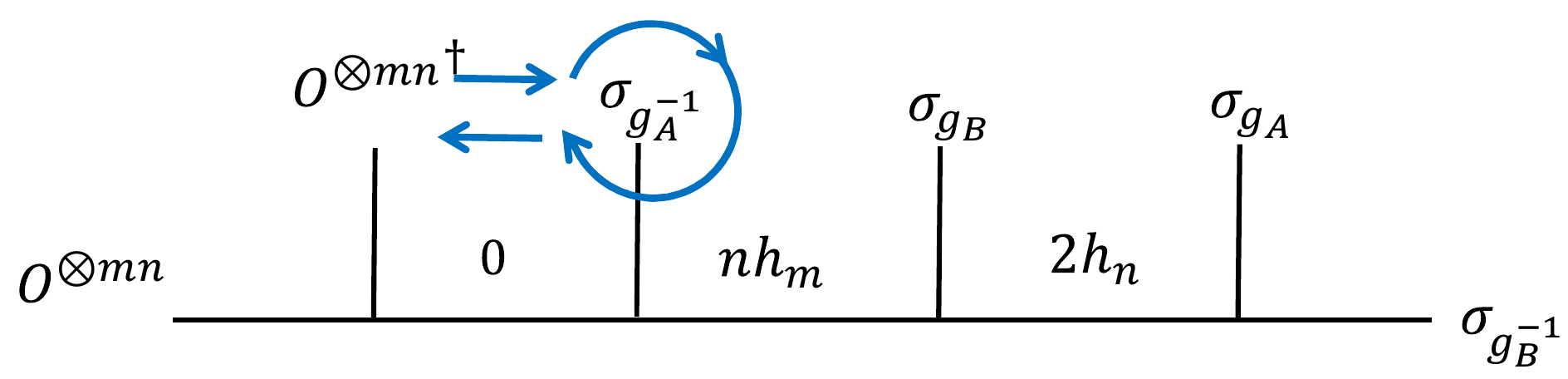}}
\newlength{\psw}
\settowidth{\psw}{\usebox{\boxps}} 

\newsavebox{\boxpzb}
\sbox{\boxpzb}{\includegraphics[width=190pt]{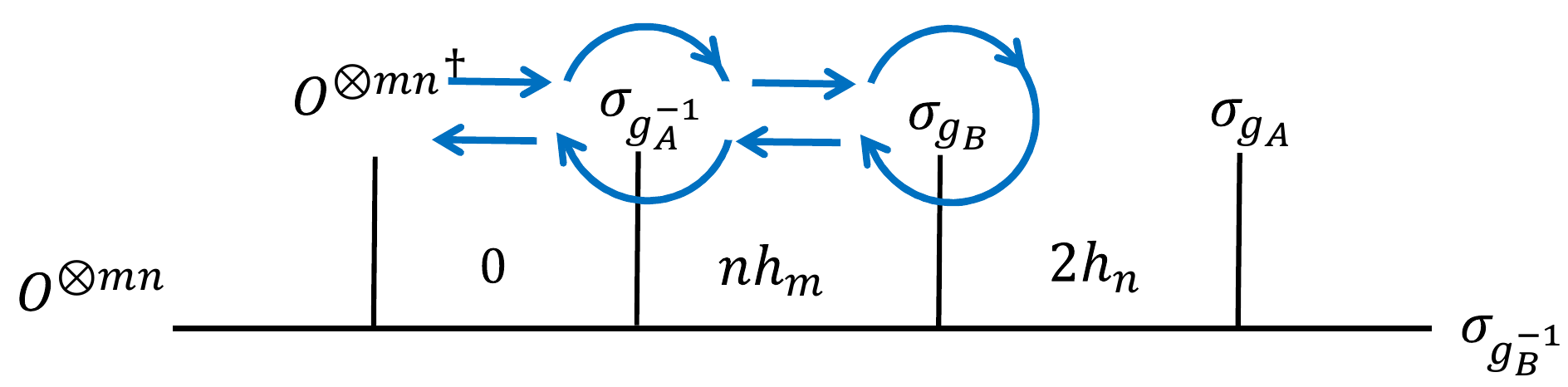}}
\newlength{\pzbw}
\settowidth{\pzbw}{\usebox{\boxpzb}}

\begin{equation}
{(C_{n,m})}^2 \parbox{\pzcw}{\usebox{\boxpzc}} \times (\text{anti-holomorphic part}),
\end{equation}
and the monodromy effect by crossing the branch cut is
\footnote{
Here, we choose the principle sheet on where the operators ${O^{\otimes mn}}$ and $\dg{{O^{\otimes mn}}}$ are inserted at $t=0$.
}

\begin{equation}
\parbox{\psw}{\usebox{\boxps}}    \times  \overline{   \parbox{\pzcw}{\usebox{\boxpzc}}     }.
\end{equation}
Therefore, the Regge limit of the 6-point function for $v_1<t<u_2$ is approximated by
\begin{equation}
{(C_{n,m})}^2 \parbox{\psw}{\usebox{\boxps}}    \times  \overline{   \parbox{\pzcw}{\usebox{\boxpzc}}     }.
\end{equation}
From this approximated 6-point function, we obtain
\begin{equation}
S_R (A:B)[O]=
    \fr{c}{6}\log\fr{1+\s{\fr{(u_1-t)(v_2-u_2)}{(-u_1+u_2)(t-v_2)}    }}{1- \s{\fr{(u_1-t)(v_2-u_2)}{(-u_1+u_2)(t-v_2)} }  } 
	+
    \fr{c}{6}\log\fr{1+\s{\fr{(u_1-v_1)(v_2-u_2)}{(-u_1+u_2)(v_1-v_2)}    }}{1- \s{\fr{(u_1-v_1)(v_2-u_2)}{(-u_1+u_2)(v_1-v_2)} }  }  ,
\ \ \  \ \ \text{if } v_1<t<u_2.
\end{equation}
This is a similar result to (\ref{eq:result1}), because both of them is based on almost the same monodromy trajectory.
On the other hand, the monodromy effect at $u_2<t<v_2$ is quite different from the case discussed in the subsection \ref{subsec:outside}.

For $u_2<t<v_2$, the dominant channel is again
\begin{equation}
{(C_{n,m})}^2 \parbox{\pzcw}{\usebox{\boxpzc}} \times (\text{anti-holomorphic part}),
\end{equation}
and the effect of crossing the branch cut is

\begin{equation}
 \parbox{\pzbw}{\usebox{\boxpzb}} \times  \parbox{\pgw}{\usebox{\boxpg}}  .
\end{equation}
Combining these monodromy, we obtain a similar result to (\ref{eq:resultA}),
\begin{equation}\label{eq:resultD}
\fr{c}{6} \log \BR{   \fr{4(t-u_1)(t-u_2)(t-v_1)(t-v_2)}{\e^2(u_2-v_1)(u_1-v_2)}   \pa{\fr{\sinh \pi \bar{\g}}{\bar{\g}}}^2   }
+\fr{c}{6}\log\fr{1+\s{\fr{(-u_1+v_1)(v_2-u_2)}{(-u_1+u_2)(-v_1+v_2)}    }}{1- \s{\fr{(-u_1+v_1)(v_2-u_2)}{(-u_1+u_2)(-v_1+v_2)} }  } ,  \ \ \ \ \    \text{if } u_2<t<v_2 .
\end{equation}
The  other choice (\ref{eq:conchannel}) can be neglected for the same reason.

We do not show the calculation of the reflected entropy for $t>v_2$ because what we need to do is just to repeat the above. The result is as follows,

\begin{equation}
\begin{aligned}
S_R (A:B)[O]&=\left\{
    \begin{array}{ll}
    \fr{c}{3}\log\fr{1+\s{\fr{(-u_1+v_1)(v_2-u_2)}{(-u_1+u_2)(-v_1+v_2)}    }}{1- \s{\fr{(-u_1+v_1)(v_2-u_2)}{(-u_1+u_2)(-v_1+v_2)} }  } ,
			& \text{if } t<v_1 ,\\ \\
    \fr{c}{6}\log\fr{1+\s{\fr{(-u_1+t)(v_2-u_2)}{(-u_1+u_2)(-t+v_2)}    }}{1- \s{\fr{(-u_1+t)(v_2-u_2)}{(-u_1+u_2)(-t+v_2)} }  }
		+
		\fr{c}{6}\log\fr{1+\s{\fr{(-u_1+v_1)(v_2-u_2)}{(-u_1+u_2)(-v_1+v_2)}    }}{1- \s{\fr{(-u_1+v_1)(v_2-u_2)}{(-u_1+u_2)(-v_1+v_2)} }  }
		,  & \text{if } v_1<t<u_2 ,\\ \\
      \fr{c}{6} \log \BR{   \fr{4(t-u_1)(t-u_2)(t-v_1)(t-v_2)}{\e^2(u_2-v_1)(u_1-v_2)}   \pa{\fr{\sinh \pi \bar{\g}}{\bar{\g}}}^2   }
+\fr{c}{6}\log\fr{1+\s{\fr{(-u_1+v_1)(v_2-u_2)}{(-u_1+u_2)(-v_1+v_2)}    }}{1- \s{\fr{(-u_1+v_1)(v_2-u_2)}{(-u_1+u_2)(-v_1+v_2)} }  } 
	,  & \text{if } u_2<t<v_2 ,\\ \\
		\fr{c}{6}\log\fr{1+\s{\fr{(-t+v_1)(v_2-u_2)}{(-t+u_2)(-v_1+v_2)}    }}{1- \s{\fr{(-t+v_1)(v_2-u_2)}{(-t+u_2)(-v_1+v_2)} }  }
		+
		\fr{c}{6}\log\fr{1+\s{\fr{(-u_1+v_1)(v_2-u_2)}{(-u_1+u_2)(-v_1+v_2)}    }}{1- \s{\fr{(-u_1+v_1)(v_2-u_2)}{(-u_1+u_2)(-v_1+v_2)} }  }
		,  & \text{if } v_2<t<\s{-u_1 v_2} ,\\ \\
		\fr{c}{6}\log\fr{1+\s{\fr{(t+v_1)(v_2-u_2)}{(t+u_2)(-v_1+v_2)}    }}{1- \s{\fr{(t+v_1)(v_2-u_2)}{(t+u_2)(-v_1+v_2)} }  }
		+
		\fr{c}{6}\log\fr{1+\s{\fr{(-u_1+v_1)(v_2-u_2)}{(-u_1+u_2)(-v_1+v_2)}    }}{1- \s{\fr{(-u_1+v_1)(v_2-u_2)}{(-u_1+u_2)(-v_1+v_2)} }  }
		,  & \text{if } \s{-u_1 v_2}<t<-u_1 ,\\ \\
     \fr{c}{3}\log\fr{1+\s{\fr{(-u_1+v_1)(v_2-u_2)}{(-u_1+u_2)(-v_1+v_2)}    }}{1- \s{\fr{(-u_1+v_1)(v_2-u_2)}{(-u_1+u_2)(-v_1+v_2)} }  } ,& \text{if } -u_1<t .\\ 
    \end{array}
  \right.\\
\end{aligned}
\end{equation}

\section{Entanglement Entropy as Pure State Limit}\label{sec:purelimit}

The reflected entropy measures correlations between $A$ and $B$. In particular, if we restrict ourselves to a pure state (e.g., $\rho=\ket{\Psi(t)} \bra{\Psi(t)}$, studied in Section \ref{sec:localCFT}) and set $B=\bar{A}$, then one can find that this measure reduces to entanglement entropy. In general, the reflected entropy has the following property,
\begin{equation}\label{eq:EE=RE}
S_R(A:B)=2S(A), \ \ \ \ \ \text{ if } \rho_{AB}  \text{ is a pure state.}
\end{equation}
We should check that our result is consistent with this property.

We consider a local excitation in an interval $A=[l_1,l_2]$ with $0<l_1<l_2$.
The entanglement entropy for this locally excited state had studied in \cite{Caputa2014a, Asplund2015} (non-perturbatively in \cite{Kusuki2018b,Kusuki2018c,Kusuki2019}) and the result is
\begin{equation}\label{eq:EE}
\begin{aligned}
S(A)[O]
&=\left\{
    \begin{array}{ll}
    \fr{c}{3} \log \pa{ \fr{  l_2-l_1}{\m}}   ,& \text{if }  t<l_1 ,\\ \\
    \fr{c}{6}  \log \pa{  \fr{ (l_2-t)(t-l_1)}{\e(l_2-l_1)} \fr{\sinh(\pi \bar{\g})}{\bar{\g}}} + \fr{c}{3} \log \pa{ \fr{  l_2-l_1}{\m}}   ,
	& \text{if } l_1<t<l_2  ,\\ \\
    \fr{c}{3} \log \pa{ \fr{  l_2-l_1}{\m}}   ,& \text{if }   t>l_2, \\
    \end{array}
  \right.\\
\end{aligned}
\end{equation}
where a positive constant $\m$ is a UV cutoff to regulate the twist operators.
To compare the entanglement entropy with the reflected entropy (\ref{eq:resultD}), we take the {\it pure state limit} by setting,
\begin{equation}
v_1=l_1-\m, \ \ \ \ \ u_2=l_1+\m, \ \ \ \ \ v_2=l_2-\m, \ \ \ \ \ u_1=l_2+\m,
\end{equation}
and then take the limit $\m \to 0$. As a result, we obtain
\begin{equation}
  S_R(A:B)[O]  \ar{  \text{pure state limit}     }\fr{c}{3}  \log \pa{  \fr{ (l_2-t)(t-l_1)(l_2-l_1)}{\e \m^2} \fr{\sinh(\pi \bar{\g})}{\bar{\g}}},  \ \ \ \ \  \text{ if }  l_1<t<l_2 ,
\end{equation}
which perfectly reproduces the entanglement entropy (\ref{eq:EE}).
Note that reflected entropy can be used as a natural regulator for entanglement entropy in QFT \cite{Dutta2019} and here one can find that the reflected entropy plays a role as a regulator of the entanglement entropy after a local quench.

\section{Dynamics of Correlations}\label{sec:dynamics}\label{sec:dynamics}

In this section, we would like to understand how dynamics of the correlation measures is characterized.
To this end, we will show various plots of the reflected entropy and read off important nature of its dynamics.
It is important to emphasize that there is another useful correlation measure, {\it mutual information}. Therefore, it is very interesting to discuss similarities and differences between dynamics of reflected entropy and mutual information.
To simplify the comparison with mutual information, we show the explicit form of the mutual information after a local quench in the following,
\begin{itemize}
\item  Quench outside intervals ( $0<\e\ll u_2<-v_1<-u_1<v_2$ and $O$ is acted on $x=0$ at $t=0$.)

\begin{equation}
\begin{aligned}
I (A:B)[O]&=\left\{
    \begin{array}{ll}
    S(u_2,v_2)+S(u_1,v_1)-S(v_1,u_2)-S(u_1,v_2)   
		,   & \text{if } t<u_2 ,\\ \\
    S(u_2,v_2,t)+S(u_1,v_1)-S(v_1,u_2,t)-S(u_1,v_2)   
		,   & \text{if } u_2<t<\s{-u_2 v_1} ,\\ \\
    S(u_2,v_2,t)+S(u_1,v_1)-S(v_1,u_2,-t)-S(u_1,v_2)   
		,   & \text{if } \s{-u_2 v_1}<t<-v_1 ,\\ \\
    S(u_2,v_2,t)+S(u_1,v_1,-t)-S(v_1,u_2)-S(u_1,v_2)   
		,   & \text{if } -v_1<t<-u_1 ,\\ \\
    S(u_2,v_2,t)+S(u_1,v_1)-S(v_1,u_2)-S(u_1,v_2,-t)   
		,   & \text{if } -u_1<t<\s{-u_1 v_2} ,\\ \\
    S(u_2,v_2,t)+S(u_1,v_1)-S(v_1,u_2)-S(u_1,v_2,t)   
		,   & \text{if } \s{-u_1 v_2}<t<v_2 ,\\ \\
    S(u_2,v_2)+S(u_1,v_1)-S(v_1,u_2)-S(u_1,v_2)   
		,   & \text{if } v_2<t ,
    \end{array}
  \right.\\
\end{aligned}
\end{equation}

\item Quench inside intervals ($0<\e\ll v_1<u_2<v_2<-u_1$ and $\s{-v_1 u_1}<v_2$ and $O$ is acted on $x=0$ at $t=0$.)

\begin{equation}
\begin{aligned}
I (A:B)[O]&=\left\{
    \begin{array}{ll}
    S(u_2,v_2)+S(u_1,v_1)-S(v_1,u_2)-S(u_1,v_2)   
		,   & \text{if } t<v_1 ,\\ \\
    S(u_2,v_2)+S(u_1,v_1,t)-S(v_1,u_2,t)-S(u_1,v_2)   
		,   & \text{if } v_1<t<u_2 ,\\ \\
    S(u_2,v_2,t)+S(u_1,v_1,t)-S(v_1,u_2)-S(u_1,v_2)   
		,   & \text{if } u_2<t<\s{-u_1 v_1} ,\\ \\
    S(u_2,v_2,t)+S(u_1,v_1,-t)-S(v_1,u_2)-S(u_1,v_2)   
		,   & \text{if } \s{-u_1 v_1}<t<v_2 ,\\ \\
    S(u_2,v_2)+S(u_1,v_1,-t)-S(v_1,u_2)-S(u_1,v_2,t)   
		,   & \text{if } v_2<t<\s{-u_1 v_2} ,\\ \\
    S(u_2,v_2)+S(u_1,v_1,-t)-S(v_1,u_2)-S(u_1,v_2,-t)   
		,   & \text{if } \s{-u_1 v_2}<t<-u_1 ,\\ \\
    S(u_2,v_2)+S(u_1,v_1)-S(v_1,u_2)-S(u_1,v_2)   
		,   & \text{if } -u_1<t ,
    \end{array}
  \right.\\
\end{aligned}
\end{equation}

\end{itemize}
where we define
\begin{equation}
\begin{aligned}
S(x,y)&=\fr{c}{3}\log \fr{y-x}{\m}, \\
S(x,y,t)&=\fr{c}{6} \log \fr{\abs{(y-x)(y-t)(t-x)}}{\e\m^2} \fr{\sinh \pi \g}{\g},
\end{aligned}
\end{equation}
and we assume $\g=\bar{\g}=\s{\fr{24}{c} h_O-1}$ for simplicity.
This was already calculated in \cite{Asplund2014} from the bulk side and we can perfectly reproduce this holographic result in the way introduced in Section \ref{sec:localCFT}.
\footnote{
A similar CFT calculation can be found in \cite{Caputa2015a}, but it might not be rigorous because their calculation of the 6-point Virasoro block is based on a wrong assumption, even in the Regge limit,
\begin{equation}
\sigma_n \times \bar{\sigma}_n = \bb{I}+ \cdots.
\end{equation}
As shown in \cite{Kusuki2019}, the Regge limit of this OPE is dominated by a NON-vacuum state.
Actually, in a special case, this assumption somehow gives a correct result and their final expression becomes consistent with the bulk computation.
However, in general, this assumption leads to a wrong estimate. On the other hand, our method introduced in Section  \ref{sec:localCFT} can be applied to any situations.
}
It means that the falling particle bulk interpretation \cite{Nozaki2013} of a local quench state can be applied not only to the single interval entanglement entropy but also to more refined correlation measures, mutual information and reflected entropy.

Note that in \cite{Asplund2014}, the holographic mutual information is compared to not the local operator quench state but the joining quench state. Therefore, they find the difference between the bulk result and the CFT result. Particularly, the remarkable difference is that the long range entanglement is found only in the CFT side (which can be also found for negativity \cite{Wen2015}).
However, our approach shows that the local operator quench state perfectly reproduces the holographic mutual information and then we cannot find such a long range entanglement. It means that the long range entanglement is a particular feature of the joining quench state. We expect that this long range effect can be completely understood by the recent development of the bulk interpretation of the joining quench state \cite{Shimaji2018, Caputa2019}.

\begin{figure}[t]
 \begin{center}
  \includegraphics[width=7.0cm,clip]{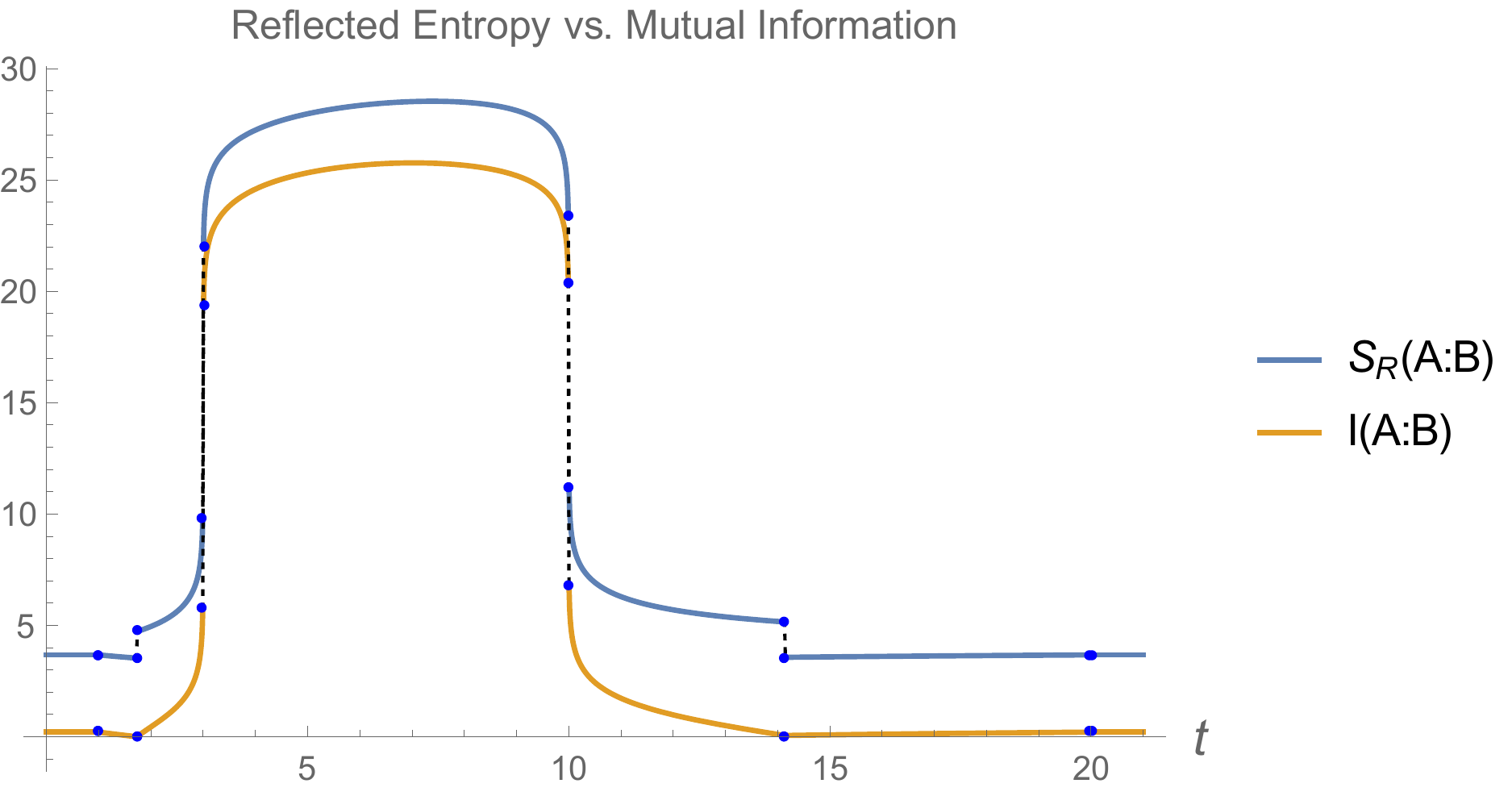}
  \includegraphics[width=7.0cm,clip]{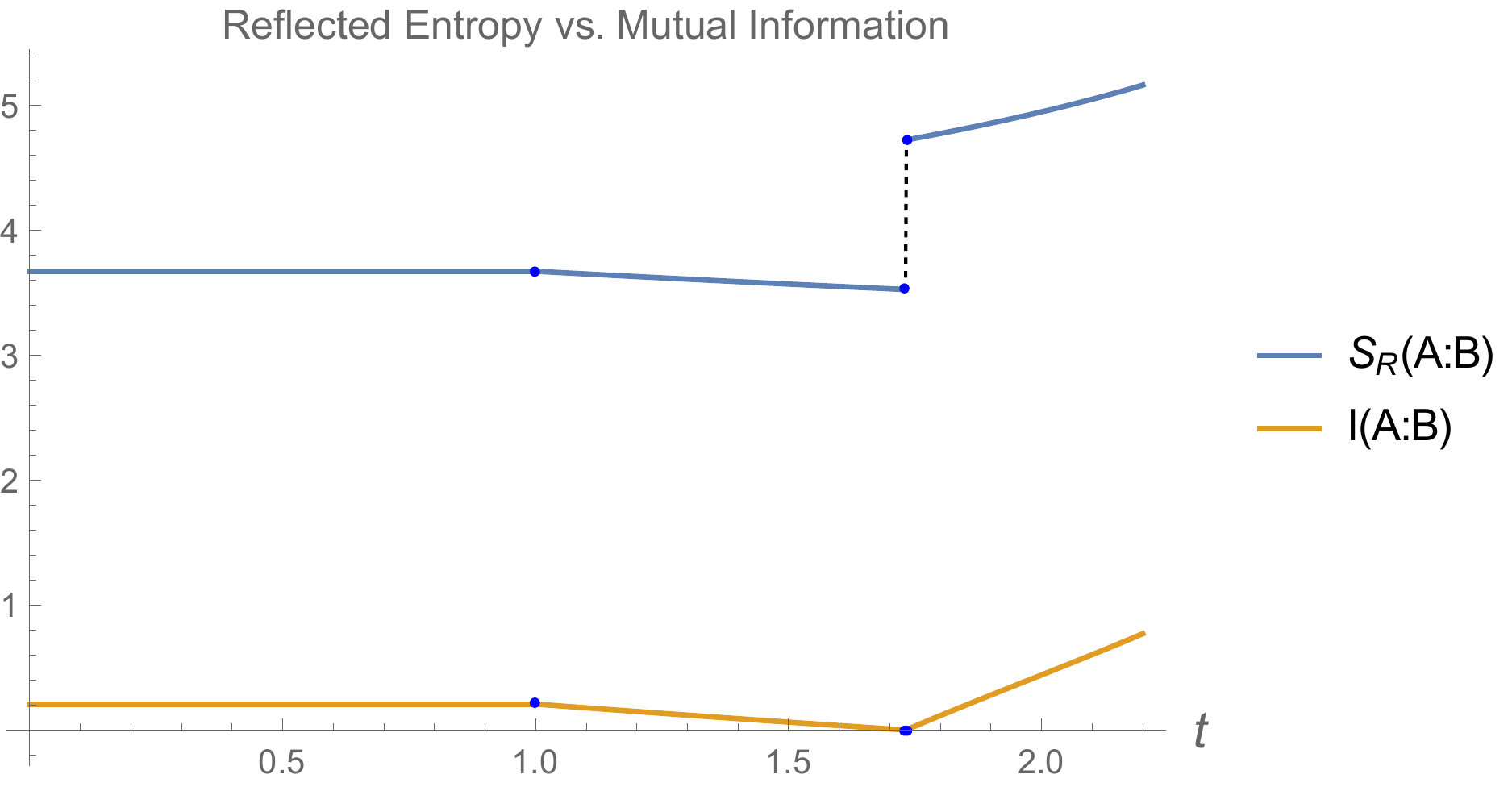}
  \includegraphics[width=7.0cm,clip]{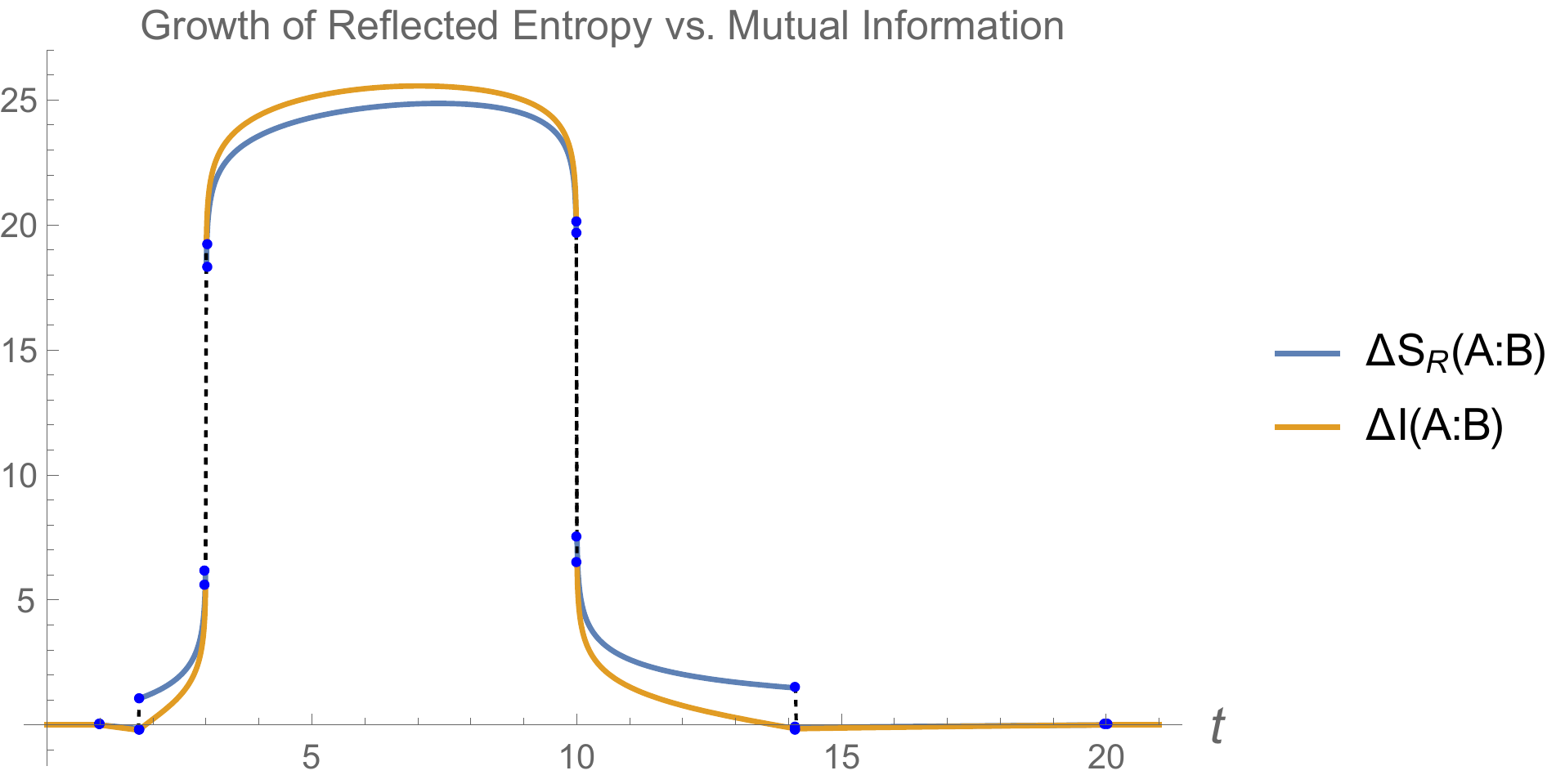}
  \includegraphics[width=7.0cm,clip]{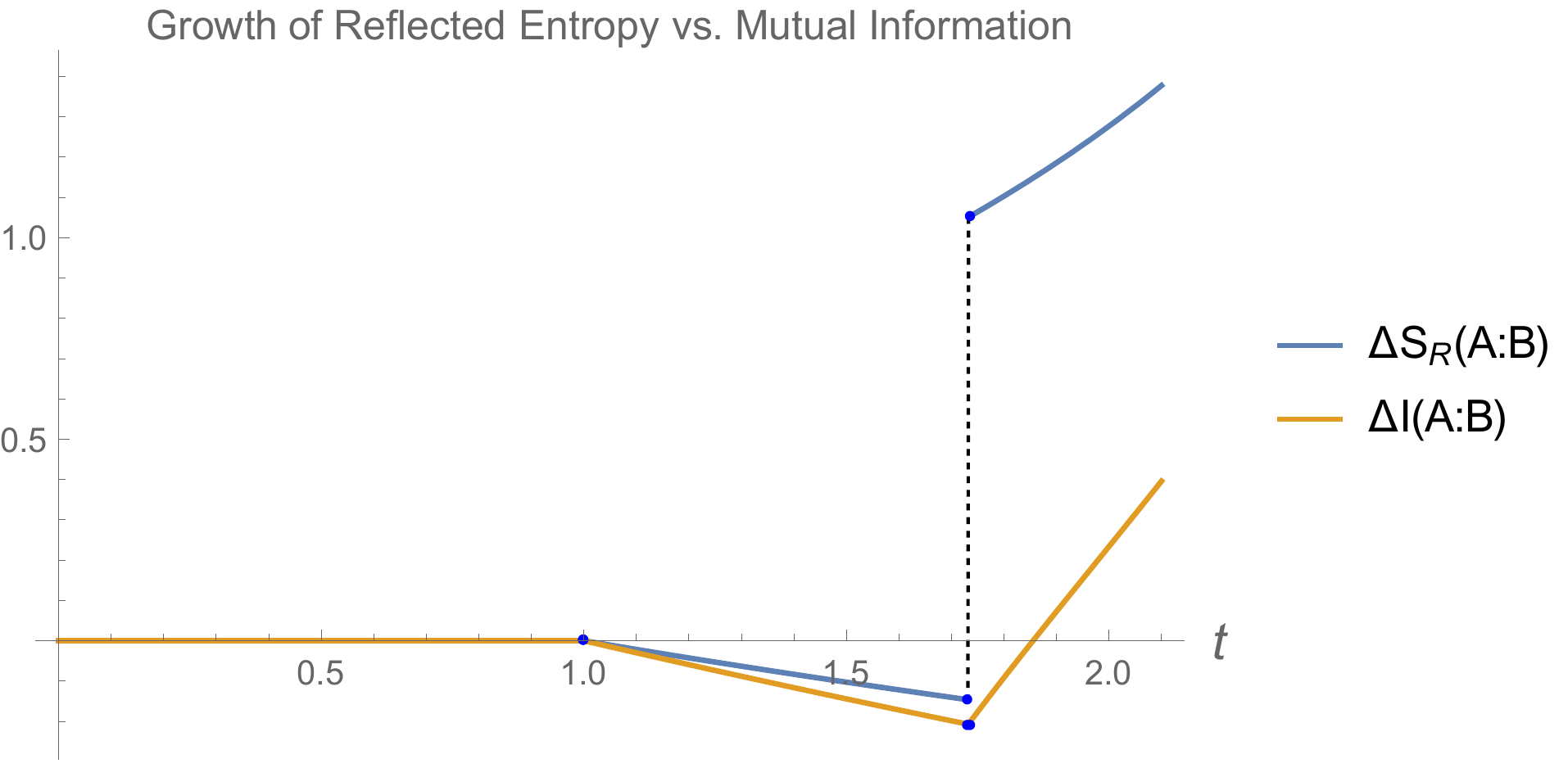}
 \end{center}
 \caption{Reflected entropy (blue) and mutual information (yellow) for a state locally quenched outside two intervals. Here we have set $(u_1,v_1,u_2,v_2)=(-10,-3,1,20)$, $\e=10^{-3}$, $\g=2$ and we remove the prefactor $\fr{c}{6}$.
We check that this parameter set satisfies the connected condition $0<\fr{(v_1-u_2)(u_1-v_2)}{(v_1-v_2)(u_1-u_2)}<\fr{1}{2}$.
Each blue dot shows a transition of itself or its first derivative.}
 \label{fig:plot1}
\end{figure}

In Figure \ref{fig:plot1}, we show the time-dependence of reflected entropy and mutual information in the setup ( $0<\e\ll u_2<-v_1<-u_1<v_2$). A first observation of this graph is that the reflected entropy is always larger than the mutual information.
In fact, as shown in \cite{Dutta2019}, the reflected entropy is bounded by the mutual information as
\begin{equation}
S_R(A:B)\geq I(A:B).
\end{equation}
That is, our result is perfectly consistent with this lower bound.
An important difference between mutual information and reflected entropy can be found at $t=\s{-u_2 v_1}, \s{-u_1 v_2}$,
the mutual information is continuous, on the other hand, the reflected entropy is discontinuous. To make it clear, we zoom into early time region in the right of the figure.
In the lower two plots, we show the difference between the local quench state and the vacuum state, 
\begin{equation}
\begin{aligned}
\D S_R(A:B)=S_R(A:B)[O] - S_R(A:B)[\bb{I}], \ \ \ \ \ \ \ \D I(A:B)=I(A:B)[O] - I(A:B)[\bb{I}],
\end{aligned}
\end{equation}
which measure a growth of correlations after a local quench.
In fact, they behave very similarly, but interestingly, we find the following inequalities for the mutual information and reflected entropy,
\footnote{
This does not contradict with $S_R(A:B)\geq I(A:B)$ because this is just a difference between the excited state and the vacuum state.  $S_R(A:B)\geq I(A:B)$  has already shown in Figure \ref{fig:plot1}.
} 
\begin{equation}\label{eq:DSR}
\begin{aligned}
\left\{
    \begin{array}{ll}
    \D S_R(A:B)  \geq \D I(A:B)  ,& \text{if } t\notin [-v_1,-u_1]  ,\\ \\
     \D S_R(A:B) \leq \D I(A:B)   ,& \text{if }  t\in [-v_1,-u_1] .\\
    \end{array}
  \right.\\
\end{aligned}
\end{equation}
It implies that the reflected entropy measure the dynamics of the correlations in a quite different way from the mutual information.
And this inequalities might be a key to understanding what correlations are measured by reflected entropy from the physical view point.
Possibly, it might be interpreted in the following.
The growth in $ t\in [-v_1,-u_1]$ is strongly caused by the quantum correlations, on the other hand, it would be expected that in $t\notin [-v_1,-u_1] $, the excitation changes both quantum correlations and classical correlations in a similar manner.
The point is that in the holographic CFT, the mutual information probes quantum correlations more purely than the reflected entropy.
\footnote{
This intuition comes from the inequality $S_R(A:B) \geq I(A:B)$. Moreover, the holographic mutual information satisfies the monogamy relation, while the holographic reflected entropy only satisfies the strong superadditivity, which is a weaker version of the monogamy relation. We do not have a further explanation for the reflected entropy, however, we can give a clearer explanation for the entanglement of purification by the fallowing inequality for any separable state,
\begin{equation}
E_P(A:B)\geq 2\fr{I(A:B)}{2}>\fr{I(A:B)}{2}.
\end{equation} 
}
Therefore, the quantum correlations in $ t\in [-v_1,-u_1]$ compared with the classical correlations result in the large growth of the mutual information, thus we obtain $ \D S_R(A:B) \leq \D I(A:B)$, while in $t\notin [-v_1,-u_1]$,  the change of the quantum correlations are not larger than the classical correlations enough to satisfy  $ \D S_R(A:B) \leq \D I(A:B)$.

Note that if we take two intervals $A=[-\infty, v_1]$ and $B=[u_2,\infty]$ and focus on the late time limit $t \gg \e$, then these two quantities approach
\begin{equation}\label{eq:decouple}
\D S_R(A:B) \sim \D I(A:B) \sim \D S(A)+ \D S(B),
\end{equation}
where $\D S(A)$ is the growth of the entanglement entropy for the interval $A$ after a local quench,
\begin{equation}
\D S(A)\ \sim \fr{c}{6} \log \fr{t}{\e}.
\end{equation}
This would be natural because in the late time limit, quasi particles do not interact with each other.

We have to comment that the reflected entropy is expected to be non-zero only in the time region $t\in[u_2,v_2]$ from the quasi particle picture \cite{Calabrese2005a, Calabrese2006a} and our result is perfectly consistent with this expectation. However, the behavior in the time-dependent region cannot be captured by the quasiparticle picture, which is one of the characteristics of the holographic CFT.
It would be worth mentioning that in the nontrivial time region $t\in[u_2,v_2]$, there are two phases as shown in the figure. The remarkable features in each phase is as follows:
\begin{itemize}

\item $ t\in [u_2, -v_1] \cup [-u_1, v_2]$

The reflected entropy is independent of the conformal dimension $h_O$ and does not include high energy scale (the UV cutoff parameter $\e$).

\item $ t \in [-v_1, -u_1] $

The reflected entropy depends on the conformal dimension $h_O$ and includes high energy scale.

\end{itemize}
It means that when the left or right moving excitation enters one interval, the excitation affects the reflected entropy but its effect is not so strong, on the other hand, if both left and right moving excitations enter two intervals, then the reflected entropy becomes much larger than that for the vacuum.
This strong effect comes from the entanglement between two intervals, which is created by the excitation.
However, we do not have any clear explanation of the small effect found in  $  t \notin [-v_1, -u_1] $.
Note that this small effect does not appear in RCFTs (see Section \ref{sec:other}).

\begin{figure}[t]
 \begin{center}
  \includegraphics[width=7.0cm,clip]{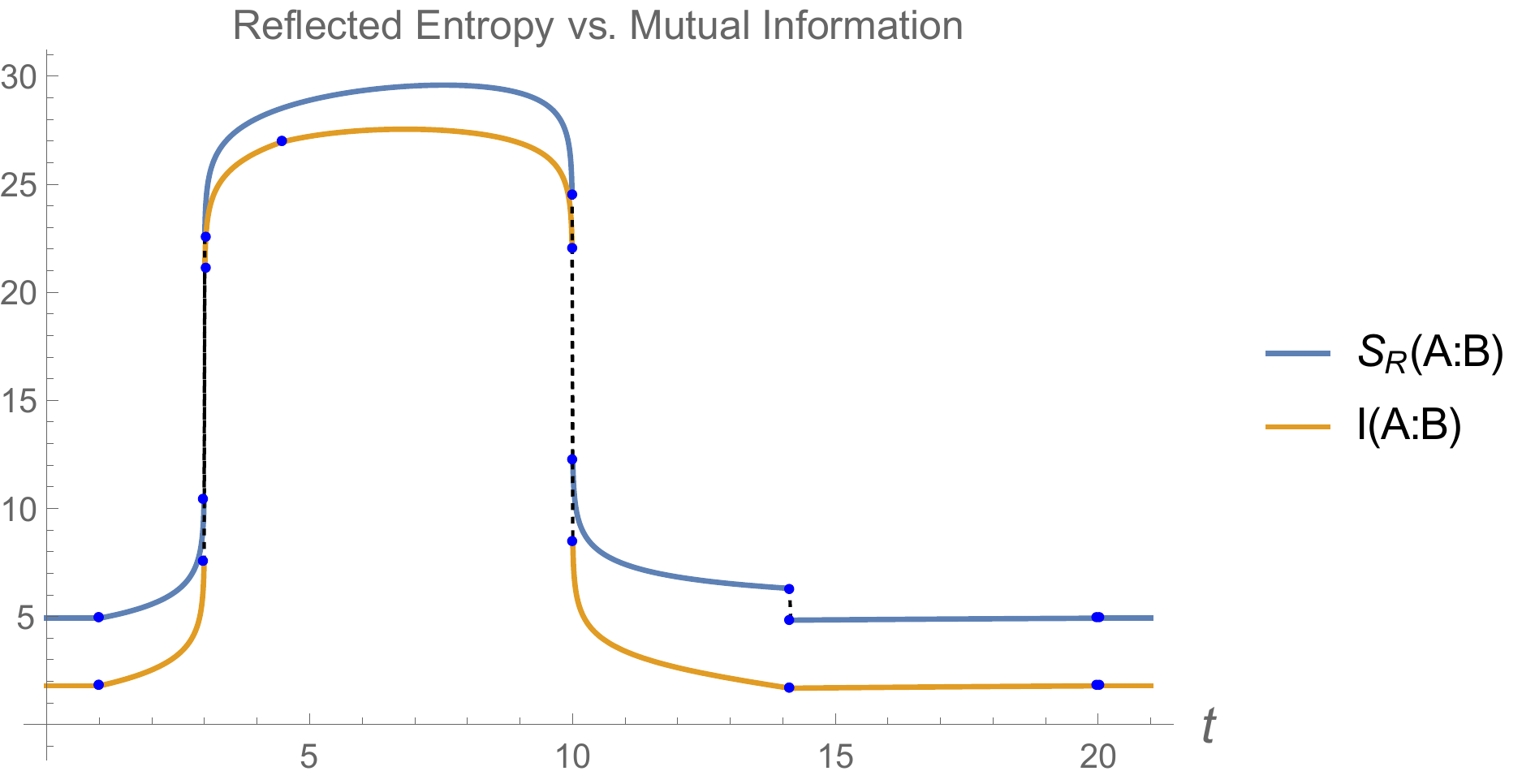}
  \includegraphics[width=7.0cm,clip]{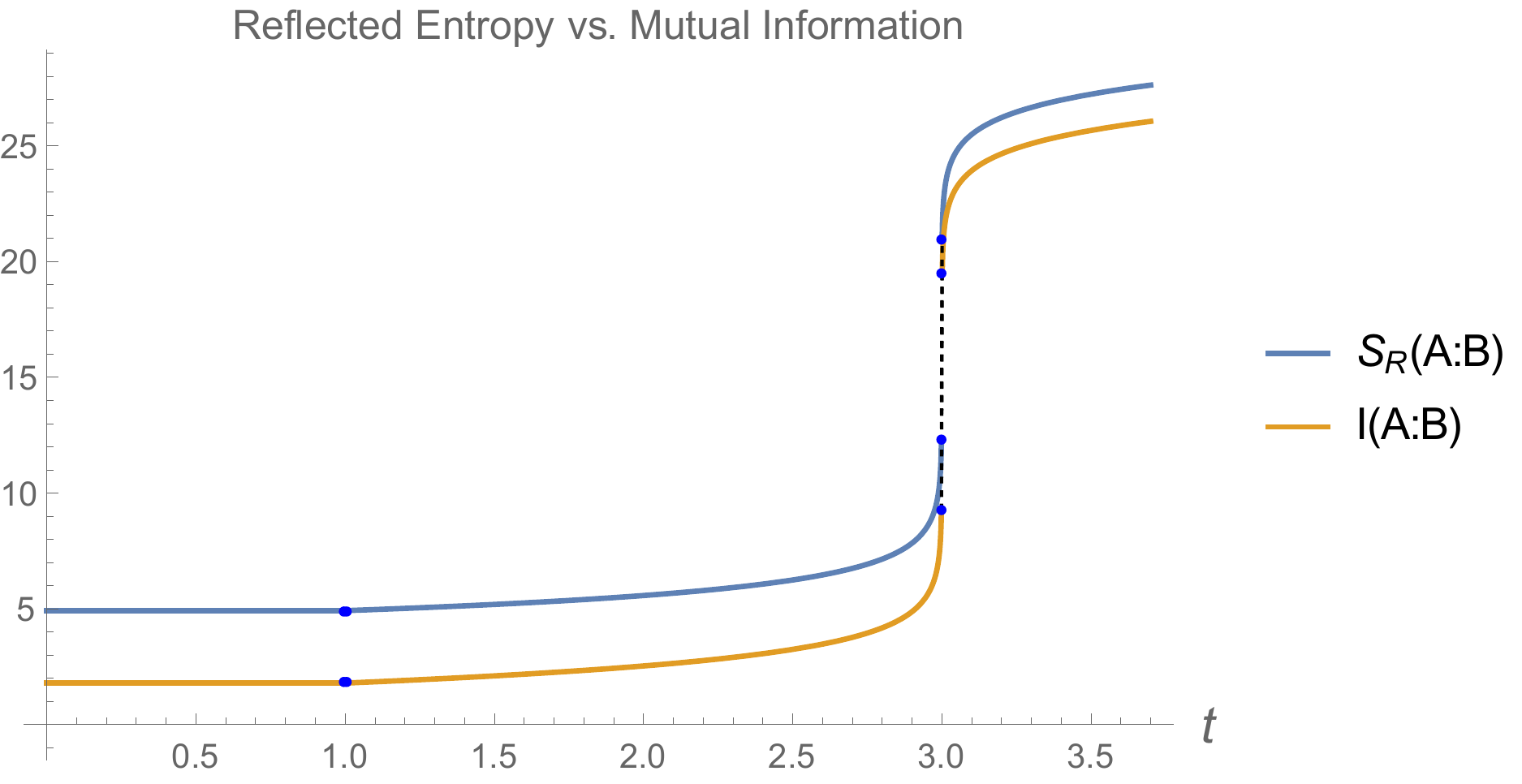}
  \includegraphics[width=7.0cm,clip]{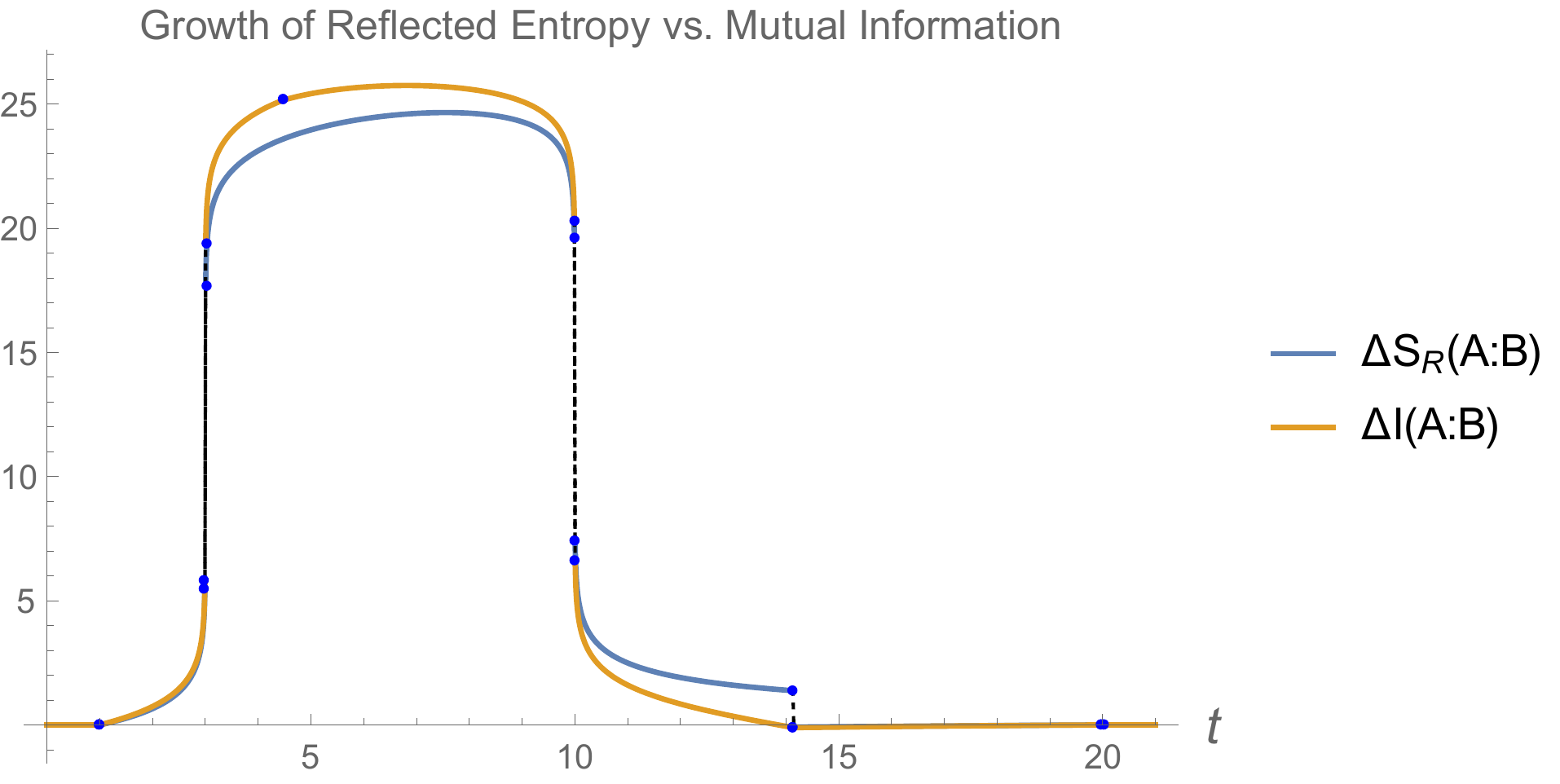}
  \includegraphics[width=7.0cm,clip]{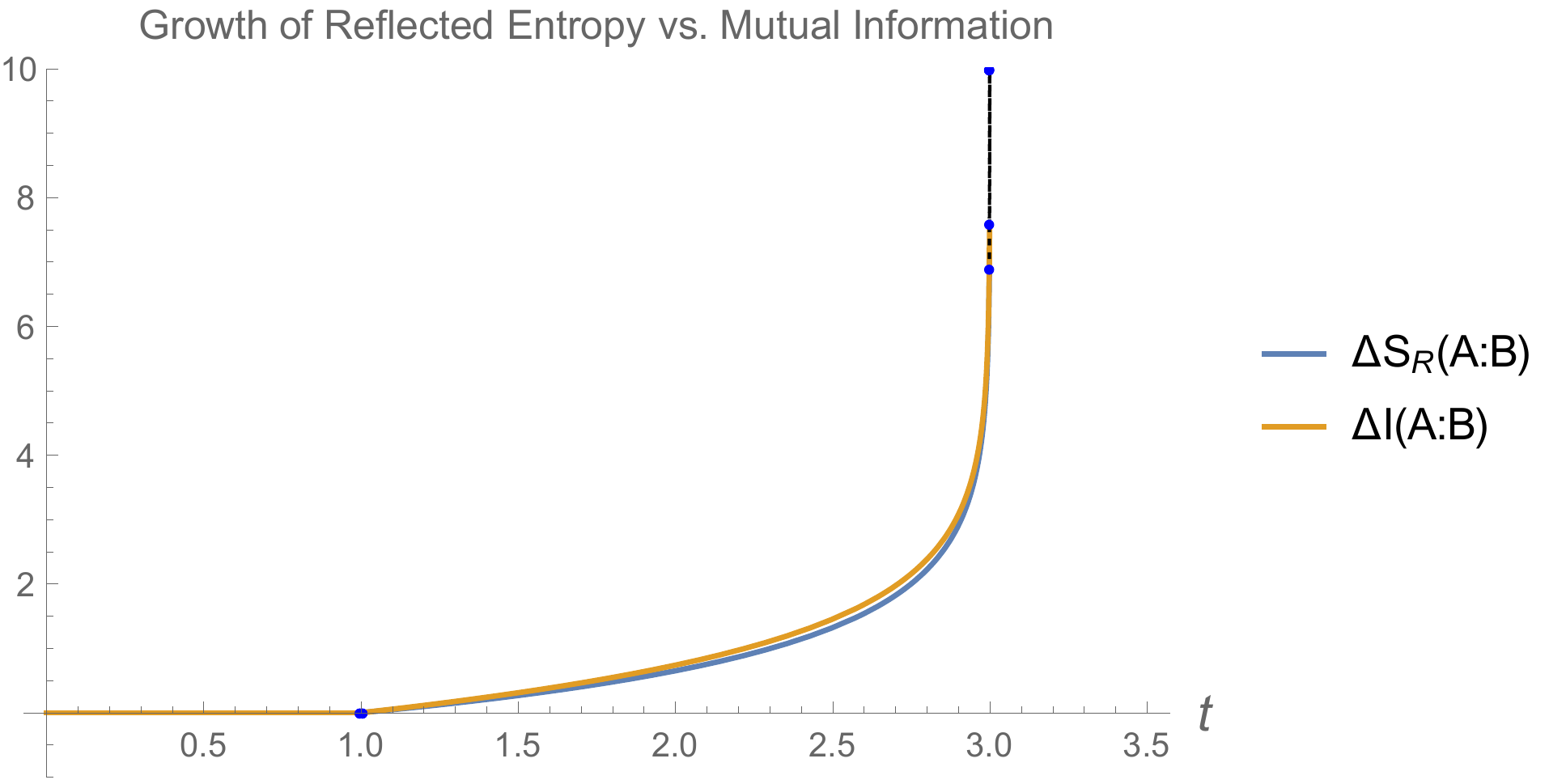}
 \end{center}
 \caption{Reflected entropy (blue) and mutual information (yellow) for a state locally quenched inside two intervals. Here we have set $(u_1,v_1,u_2,v_2)=(-20,1,3,10)$, $\e=10^{-3}$, $\g=2$ and we remove the prefactor $\fr{c}{6}$.
We check that this parameter set satisfies the connected condition $0<\fr{(v_1-u_2)(u_1-v_2)}{(v_1-v_2)(u_1-u_2)}<\fr{1}{2}$.
Each blue dot shows a transition of itself or its first derivative.}
 \label{fig:plot2}
\end{figure}

In Figure \ref{fig:plot2}, we show the reflected entropy and the mutual information in the different setup ($0<\e\ll v_1<u_2<v_2<-u_1$ and $\s{-v_1 u_1}<v_2$).
The main difference is that there is an additional transition for the mutual information, in that,  the first derivative of the mutual information is discontinuous at $t=\s{-u_1 v_1}$, which can not be observed for the reflected entropy.
Moreover, we can find the inequalities (\ref{eq:DSR}) and the transition of the reflected entropy at $t=\s{-u_1 v_2}$ as seen in Figure \ref{fig:plot2}. And also we find the agreement with the quasi particle picture in $t\notin[v_1,-u_1]$.

Finally, we would like to comment that our interpretation by comparing between the reflected entropy and the mutual information can be also applied to the entanglement of purification. This is because these two quantities reduces to the same entanglement wedge cross section in the holographic CFT.

\section{Reflected Entropy in Heavy State from CFT}\label{sec:Heavy}

We consider a CFT on a circle with length $L$. Then, the reflected entropy for a heavy state can be calculated from
\begin{equation}\label{eq:Renyi2}
\fr{1}{1-n} \log
	\fr{Z_{n,m}}
	{\pa{ Z_{1,m}}^n },
\end{equation}
where
\be\label{eq:Renyi2a}
Z_{n,m}=\Braket{  {O^{\otimes mn}}| \sigma_{g_A}(-u_1)\sigma_{g_A^{-1}}(-v_1)    \sigma_{g_B}(u_2) \sigma_{g_B^{-1}}(v_2)| {O^{\otimes mn}} }_{\text{CFT}^{\otimes mn}}.
\ee
Here, this correlator is defined on a cylinder. This can be mapped to the plane $(z,\bar{z})$ by
\begin{equation}
z=\ex{\fr{2 \pi i w}{L}}, \ \ \ \ \bar{z}=\ex{-  \fr{2\pi iw}{L}}.
\end{equation}
In this coordinates, we can also evaluate it by a single block approximation as in Section \ref{sec:localCFT}.

\begin{figure}[H]
 \begin{center}
  \includegraphics[width=14.0cm,clip]{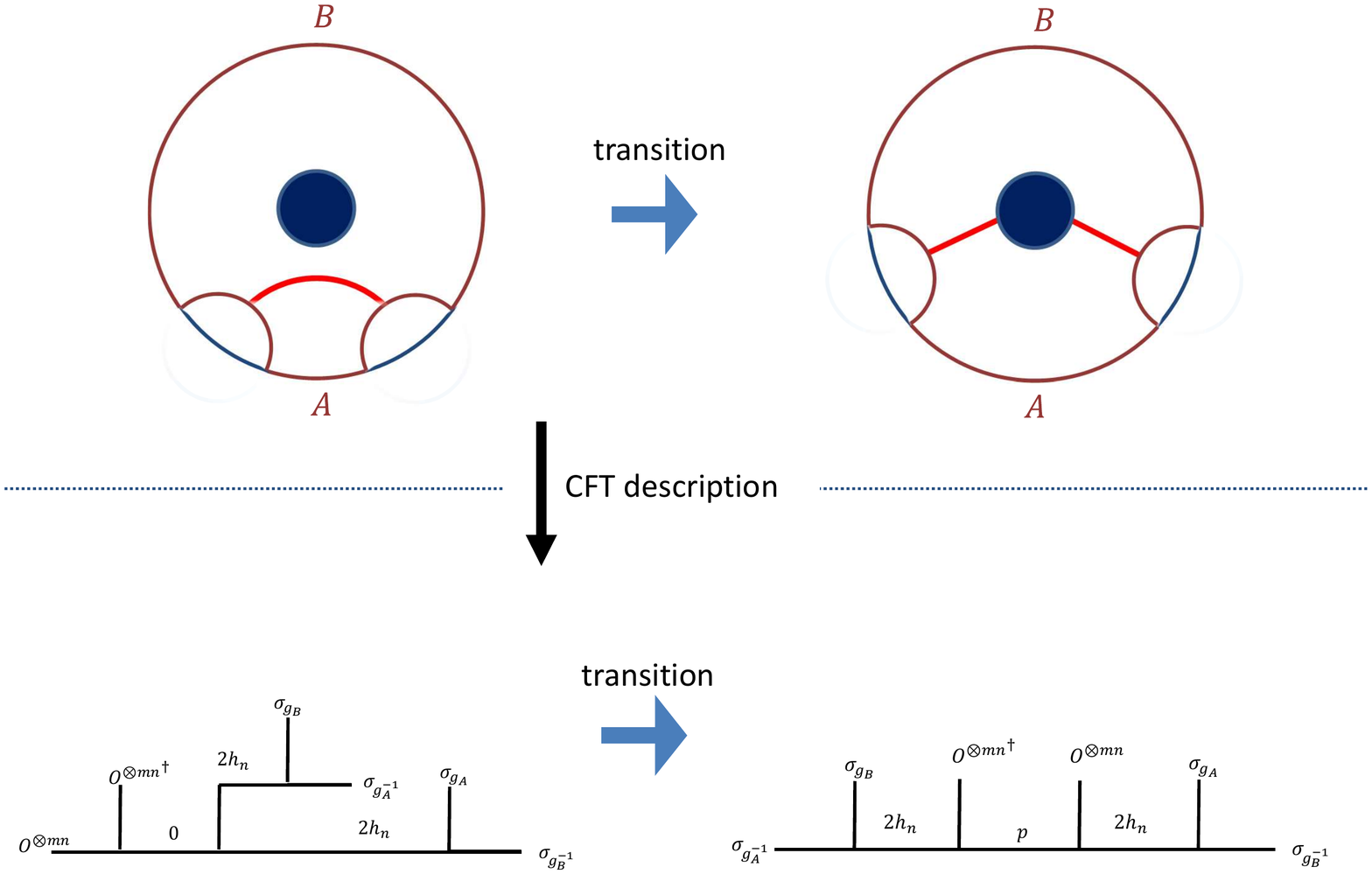}
 \end{center}
 \caption{The non-trivial entanglement wedge cross section in the BTZ background has two candidates. One is the connected codimension-2 surface and the other is the disconnected codimension-2 surfaces which have endpoints on the black hole horizon. The correct choice is the minimal one. If we could observe this transition in the CFT side, it should come from a change of the dominant channel in the large $c$ limit as shown in the lower of this figure.}
 \label{fig:transition}
\end{figure}

In this setup, we are very interested in a question, whether we can reproduce the transition of the entanglement wedge cross section or not.
It is known that the entanglement wedge cross section has a transition as shown in the upper of Figure \ref{fig:transition}. That is, 
 it is possible that the minimal cross section is given by the disconnected codimension-2 surfaces which have endpoints on the black hole horizon, instead of the connected surface. Actually, there is no reason to reproduce this transition from the reflected entropy (\ref{eq:Renyi2}) because our heavy state  (i.e., $\rho=\ket{O} \bra{O}$) is ``pure'' but the BTZ microstate is ``mixed''.
Nevertheless, it might be possible to find this transition from the CFT side, because the reduced density matrix could be approximated by that for a microstate of BTZ  in the large $c$ limit.
Naively, we can expect that the transition in the bulk side can be translated into a change of the dominant channel as shown in the lower of Figure \ref{fig:transition}. If this naive expectation is true, then the disconnected cross section should be reproduced from the following single block approximation,

\begin{equation}\label{eq:LLHHLL}
{(C_{n,m})}^2 {(C_{n,O})}^2   \parbox{\pzdw}{\usebox{\boxpzd}} \times \overline{ \parbox{\pzdw}{\usebox{\boxpzd}}  } .
\end{equation}
The intermediate state $p$ is dominated by $O^{\otimes 2}$ as explained below (\ref{eq:conchannel}).
The constant $C_{n,O}$ is the OPE coefficient between $O^{\otimes mn}$ and $\sigma_{g_B^{-1} g_A}$ and its asymptotics in the limits $c \to \infty$, $n \to 1$ is given by 
\begin{equation}\label{eq:CnO}
C_{n,O}\to \g^{h_n} \bar{\g}^{h_n},
\end{equation}
with $\g=\s{\fr{24}{c}h_O-1}$ and $\bar{\g}=\s{\fr{24}{c}\bar{h}_O-1}$.
This is justified in the holographic CFT \cite{Kraus2016}, which is explained in Appedix \ref{app:CnO}. Notice that we have no exponential suppression from the OPE coefficients. Moreover, the degeneracy of the primary fields should be also $1$ becasue we are taking the OPE including twist operators.

The limit $m \to 1$ of the denominator in (\ref{eq:Renyi2}) is 
\begin{equation}
\braket{O^{\otimes 2}|O^{\otimes 2 }}^n=1
\end{equation}
In the semiclassical limit, the LLHHLL block (\ref{eq:LLHHLL}) is simplified because only the contribution to the intermediate state is the primary exchange,
\footnote{We can show this fact by using the Virasoro algebra as in Appendix E of \cite{Fitzpatrick2014} and this is also justified by the monodromy method \cite{Banerjee2016}.
} 
which means that the LLHHLL block is decomposed into two HHLL blocks as

\newsavebox{\boxpd}
\sbox{\boxpd}{\includegraphics[width=190pt]{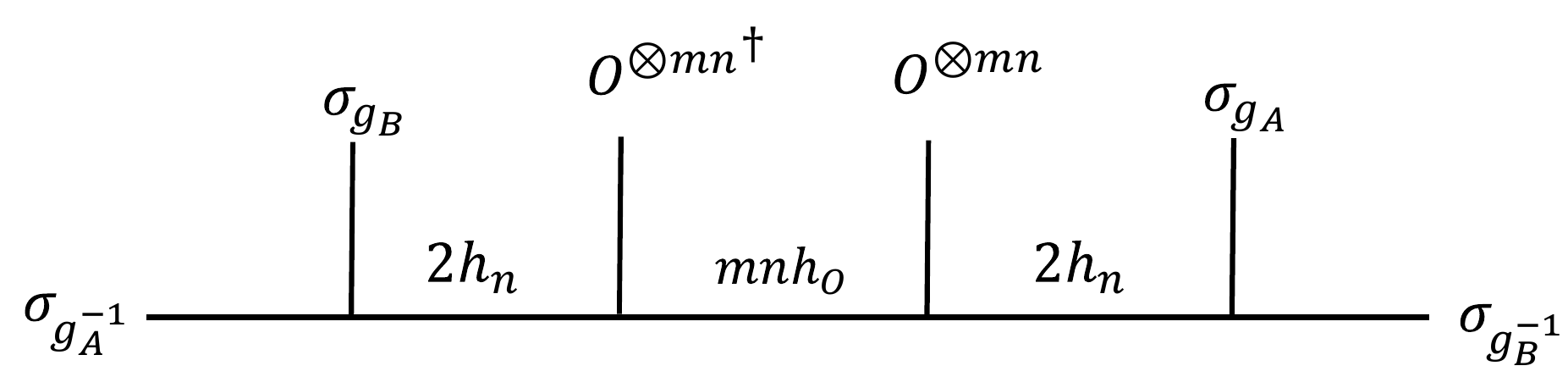}}
\newlength{\pdw}
\settowidth{\pdw}{\usebox{\boxpd}} 

\newsavebox{\boxpe}
\sbox{\boxpe}{\includegraphics[width=110pt]{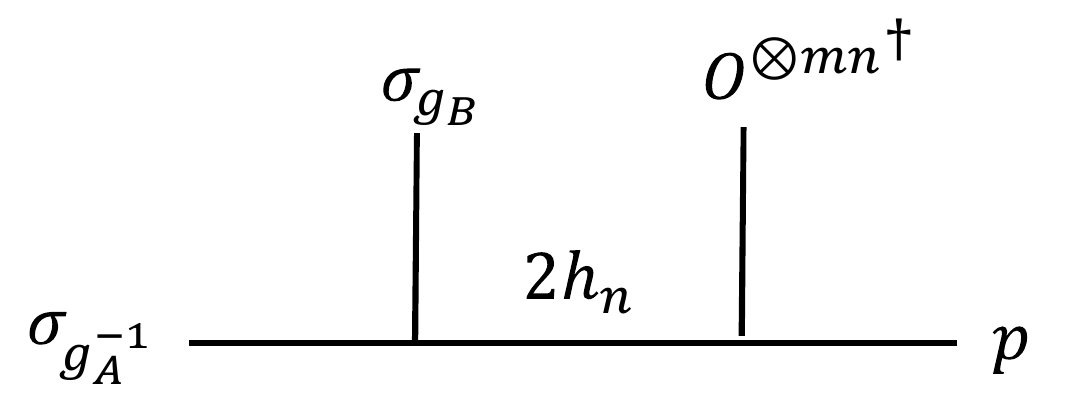}}
\newlength{\pew}
\settowidth{\pew}{\usebox{\boxpe}}

\newsavebox{\boxpf}
\sbox{\boxpf}{\includegraphics[width=110pt]{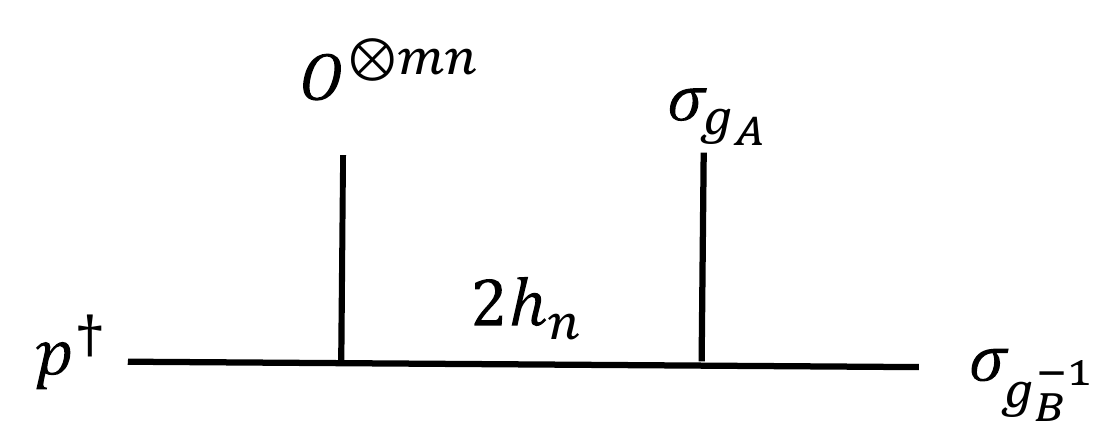}}
\newlength{\pfw}
\settowidth{\pfw}{\usebox{\boxpf}} 

\begin{equation}\label{eq:CorrinH}
\parbox{\pzdw}{\usebox{\boxpzd}} \ar{\text{HHLL}}  \parbox{\pew}{\usebox{\boxpe}}  \times  \parbox{\pfw}{\usebox{\boxpf}} .
\end{equation}
Note that this expression is precise only under the $m,n\rightarrow1$ limit. 
Thus, the reflected entropy is
\begin{equation}\label{eq:REinH}
S_R(A:B)=
\fr{c}{6} \log \pa{\coth\fr{ \pi \g   (v_1+u_2)}{2L}} 
+ \fr{c}{6} \log \pa{\coth\fr{ \pi \bar{\g}   (v_2+u_1)}{2L}}.
\end{equation}
The detailed calculation is shown in Appendix \ref{app:squareblock}.

This result perfectly matches the entanglement wedge cross section in the BTZ metric \cite{Takayanagi2018a}. 
It means that the thermalization in the large $c$ limit \cite{Fitzpatrick2015, Lashkari2018, Hikida2018, Romero-Bermudez2018, Brehm2018} can also be found in the reflected entropy.
Our result also answers the interesting physics question, what is the bulk dual of our quench state. We show that the surface ends at the horizon of the black hole. This can be explained by considering the horizon as an end of the world brane \cite{Takayanagi2011, Hartman2013, Almheiri2018, Cooper2018}. In this case, the surface can end at the horizon even if we consider a pure state black hole. We have to mention that this idea should be also applied to the entanglement entropy in a heavy state because the reflecte entropy (\ref{eq:REinH}) should reproduce the entanglement entropy by the relation (\ref{eq:EE=RE}) in the pure state limit. Note that the entanglement entropy from the pure state limit of the reflected entropy does not match the result in \cite{Asplund2015}. This is because their derivation assumes that the change of the dominant channel (i.e., the transition shown in Figure \ref{fig:transition}) does not happen. This was because we expected the OPE coefficients in another channel is suppressed exponentially under the large $c$ limit. However, as we have seen here, this is actually not the case at least under the $n\rightarrow1$ limit (namely, the entanglement entropy or the reflected entropy). 

This brane gives rise to another phase of the holographic entanglement entropy when size of the subsystem becomes larger than the energy scale of the heavy states. In particular, as increasing the energy (enlarging the horizon radius), the phase transition like figure \ref{fig:transition} should happen very quickly. Notice that the transition point is obviously less than half of the total subsystem size.  Therefore, the breaking of the eigenstate thermalization hypothesis (ETH) for the entanglement entropy must happen faster than we expected so far\cite{Garrison2018, Dymarsky2018}. We would like to make further comments on this point in near future. 

\section{Quantum Correction to Reflected Entropy} \label{sec:quantum}

In the calculation of entanglement entropy in the holographic CFT, the large $c$ limit commutes with the von-Neumann limit in usual setups (vacuum state, local quench state, etc.). However, we have to calculate the reflected entropy by taking first the limit $c \to \infty$ even if we consider the vacuum state. We discuss this problem in this section. 

To calculate  reflected entropy or entanglement entropy for two intervals $A$ and $B$ in the vacuum state, we start with the semiclassical block (\ref{eq:HHLLblock}) (and its anti-holomorphic block),
\begin{equation}
\ca{F}^{LL}_{HH}(h_p|z)   =   (1-z)^{h_L(\d-1)} \pa{ \fr{1-(1-z)^\d}{\d}}^{h_p-2h_L} \pa{ \fr{1+(1-z)^{\fr{\d}{2}}}{2}}^{-2h_p},
\end{equation}
with $\d=\s{1-\fr{24}{c}h_H}$ and then we obtain the entanglement entropy  \cite{Hartman2013a} by setting $h_p$ to be zero and $h_H=h_L=\fr{c}{24} \pa{n-\fr{1}{n}}$,
\begin{equation}\label{eq:SAB}
\begin{aligned}
S(A:B)=\fr{c}{3} \log \fr{z}{\m},
\end{aligned}
\end{equation}
and the reflected entropy  (\ref{eq:REvacuum}) by setting $h_H, h_L=0$ and $h_p=\fr{c}{12} \pa{n-\fr{1}{n}}$,
\begin{equation}\label{eq:SRAB}
\begin{aligned}
S_R(A:B)=\fr{c}{3} \log \fr{1+\s{1-z}}{1-\s{1-z}}.
\end{aligned}
\end{equation}
Here, we focus on the nontrivial case where the entanglement wedge is disconnected. The cross ratio is related to the coordinated as
\begin{equation}
z=\fr{(v_1-u_2)(u_1-v_2)}{(v_1-v_2)(u_1-u_2)},
\end{equation}
and the connected condition can be expressed in terms of the cross ratio as  $0<z<\fr{1}{2}$.

If one wants to take first the von-Neumann limit, one cannot use the semiclassical block because this block is defined in the limit $c \to \infty$ with $\fr{h_p}{c}, \fr{h_L}{c}, \fr{h_H}{c}$ fixed.
Actually, the von-Neumann limit also simplifies evaluating the block, for example, the entanglement entropy calculated by the following simplification at any $c>1$,
\begin{equation}
\ca{F}^{LL}_{LL}(0|z) \ar{h_L \to 0} 1-2h_L \log z+O(h_L^2),
\end{equation}
which perfectly reproduces (\ref{eq:SAB}). On the other hand, the reflected entropy is calculated by
\begin{equation}
\ca{F}^{LL}_{LL}(h_p|z) \ar{h_p \to 0} 1+h_p\log\fr{z}{\s{1-z}}+O(h_p^2),
\end{equation}
where we first take the limit $h_L \to 0$. The result is
\begin{equation}\label{eq:nonRE}
\ti{S}_R(A:B)=\fr{c}{3} \log \fr{4\s{1-z}}{z},
\end{equation}
which is quite different from (\ref{eq:SRAB}).
\footnote{
More precisely, in the calculations of entanglement entropy and reflected entropy, we assume that our CFT has $c>1$ and no extra currents besides the Virasoro current.
These global block reductions in the limits $h_L, h_p \to 0$ can be shown by the Virasoro algebra.
We also checked this global reduction formula of the Virasoro block by using the recursion relation up to order 6.
}
We have to mention that we take the large $c$ limit after the von-Neumann limit to approximate the correlator by a single block.
The motivation to reverse there two limits, $c \to \infty$ and the von-Neumann limit, is to understand non-perturbative effects to the reflected entropy.

\begin{figure}[H]
 \begin{center}
  \includegraphics[width=7.0cm,clip]{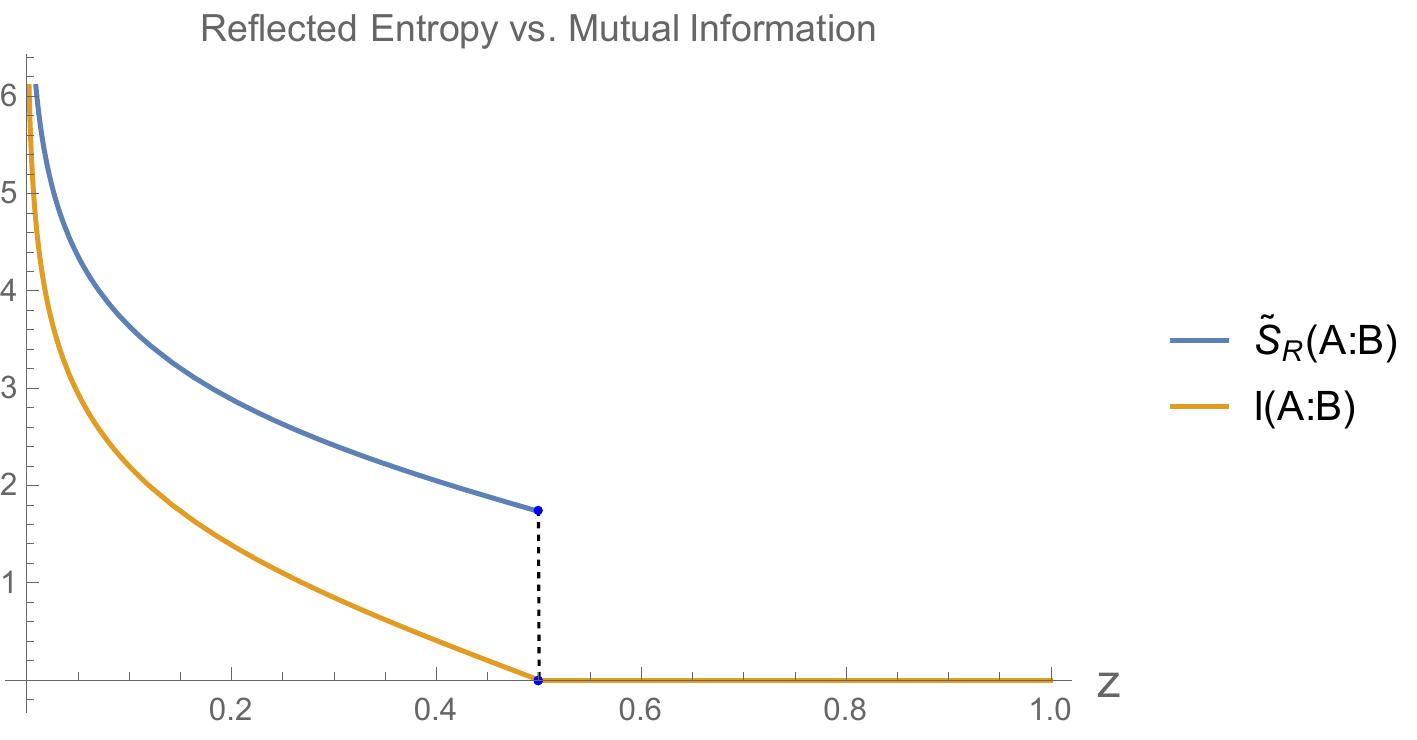} \ \  \ \ \ \ \ \ \ 
  \includegraphics[width=6.0cm,clip]{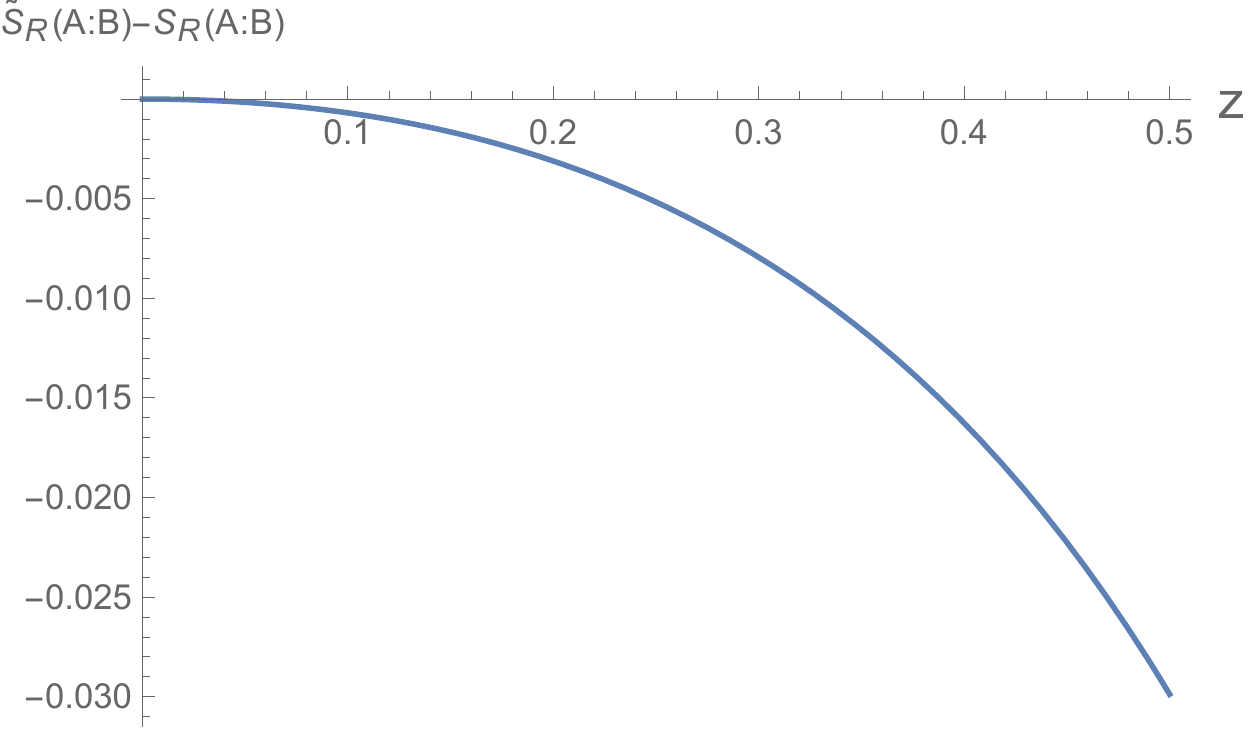}
 \end{center}
 \caption{(Left) This shows the $z$-dependence of the non-perturbative reflected entropy and the mutual information. We can check the inequality $\ti{S}_R(A:B)\geq I(A:B)$.
(Right) The difference between $\ti{S}_R(A:B)$ and $S_R(A:B)$. We can find that the quantum correction is always negative.
It might be natural because the quantum corrections should smooth the transition at $z=\fr{1}{2}$ in the left figure, in other words, the corrections should decrease the classical reflected entropy $S_R(A:B)$.
}
 \label{fig:SRvsMI}
\end{figure}

\begin{figure}[h]
 \begin{center}
  \includegraphics[width=8.0cm,clip]{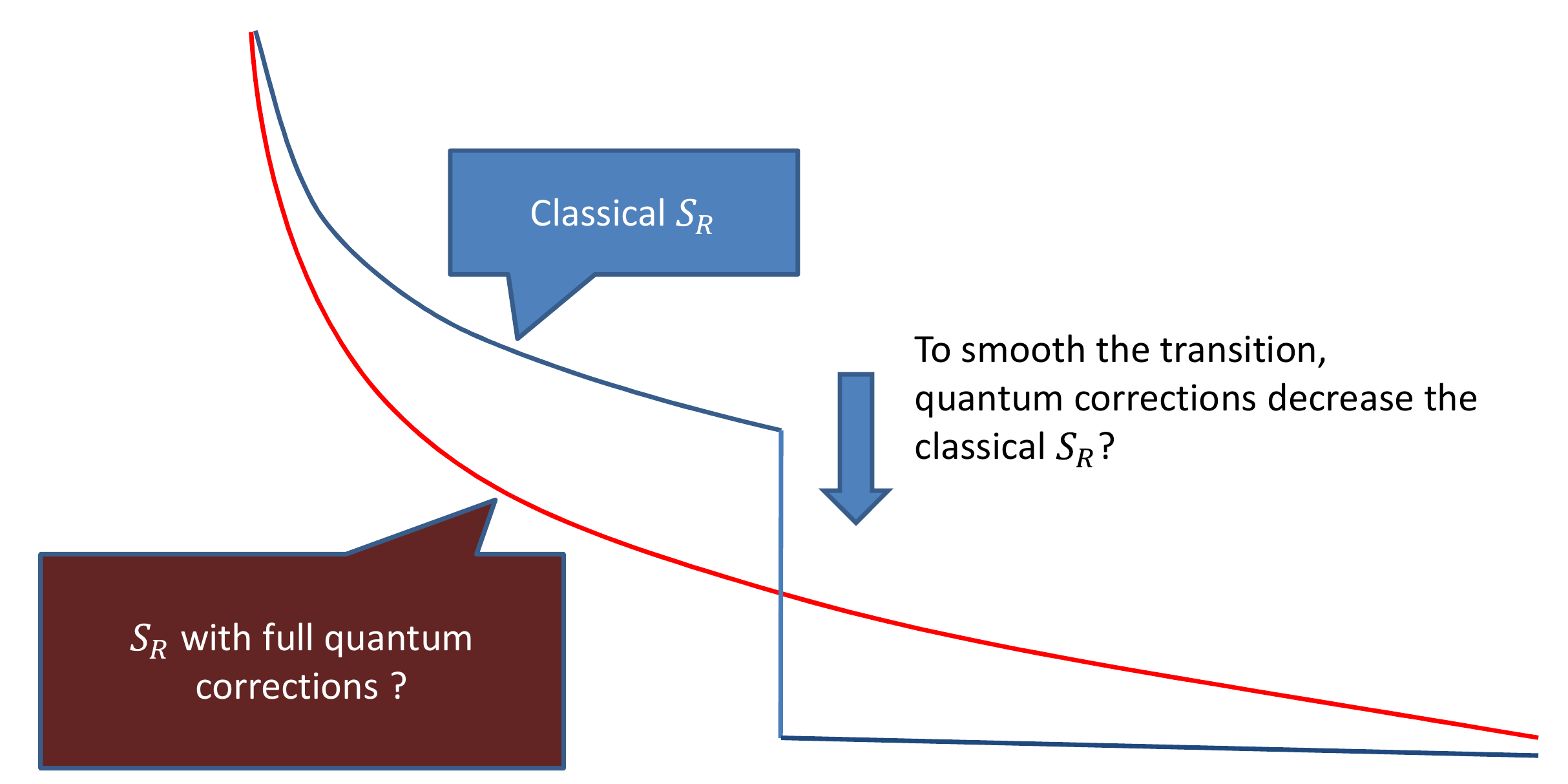}
 \end{center}
 \caption{Sketch of the effect of quantum corrections. It is naturally expected for the quantum corrections to decrease the classical reflected entropy to smooth the transition.}
 \label{fig:correction}
\end{figure}

The discrepancy between (\ref{eq:nonRE}) and (\ref{eq:SRAB}) means that the two limits $c \to \infty$ and $m,n \to 1$ do not commute with each other. In other words, there are non-perturbative effects in the reflected entropy, which cannot be found in the entanglement entropy because $c \to \infty$ and $n \to 1$ commute with each other in its calculation.
We can interpret $\ti{S}_R(A:B)$ as the reflected entropy including {\it quantum corrections}.
We can immediately find that the inequality $\ti{S}_R(A:B)\geq I(A:B)$ is satisfied from the left of Figure \ref{fig:SRvsMI} and also show the two monotonicity inequalities of the holographic reflected entropy,
\begin{equation}
\begin{aligned}
S_R(A:BC) &\geq I(A:B)+I(A:C), \\
S_R(A:BC) &\geq S_R(A:B). \\
\end{aligned}
\end{equation}
We plot the difference between  $\ti{S}_R(A:B)$ and $S_R(A:B)$ in the right of Figure \ref{fig:SRvsMI}.
From this, we can find that the quantum correction is always negative.
This is natural because the quantum correction should smooth the transition of the reflected entropy at $z=\fr{1}{2}$, therefore, the quantum correction should decrease the classical reflected entropy $S_R(A:B)$ in order to connect two disconnected lines at $z=\fr{1}{2}$ as sketched in Figure \ref{fig:correction}.
Note that other quantum corrections come from sub-leading conformal blocks. This effect can also be calculated in the same way and it expected to be negative. This is one of interesting directions for future research.

\begin{figure}[h]
 \begin{center}
  \includegraphics[width=7.0cm,clip]{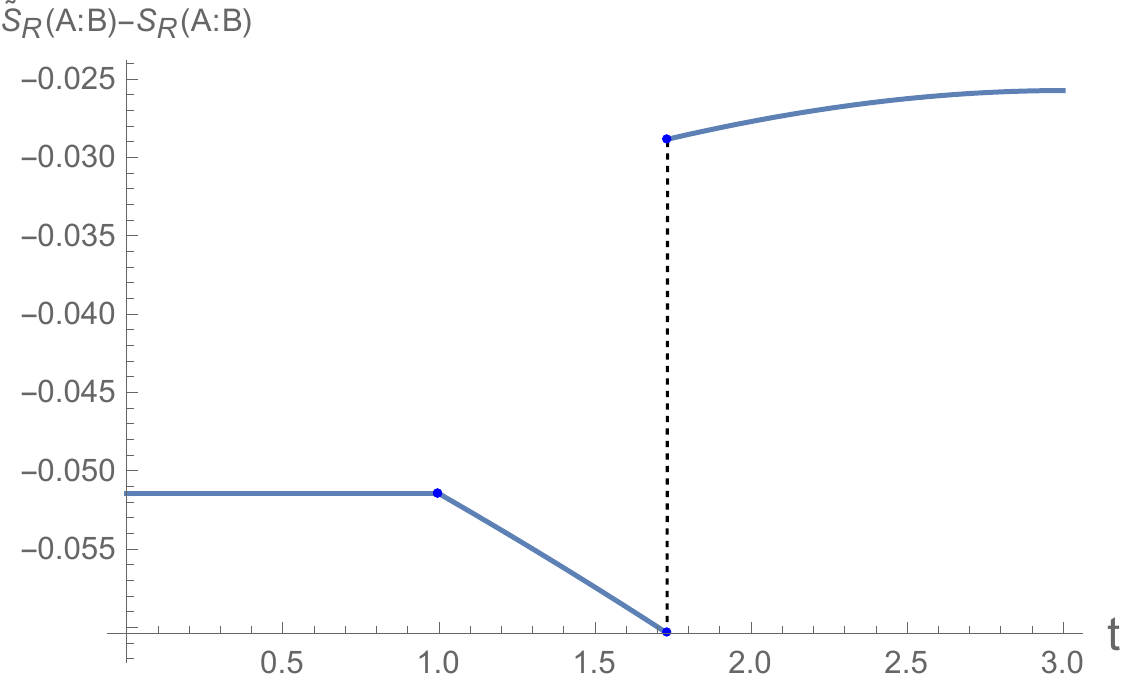}
 \end{center}
 \caption{The difference between $\ti{S}_R(A:B)$ and $S_R(A:B)$ for a local quench state. Here the parameters are set to be $(u_1,v_1,u_2,v_2)=(-10,-3,1,20)$, $\e=10^{-3}$, $\g=\bar{\g}=2$ and the prefactor $\fr{c}{6}$ is removed.}
 \label{fig:SRn}
\end{figure}

The non-perturbative effect for a local quench state can be also evaluated in the same way. 
In Figure \ref{fig:SRn}, we show the time-dependence of the non-perturbative effect in the same setup as in Figure \ref{fig:plot1}.
One can find that the non-perturbative effect after the transition at $t=\s{-u_2 v_1}$ becomes very small.
It is natural because this transition at $t=\s{-u_2 v_1} $ is attributed not by the large $c$ limit but by the $\e \to 0$ limit, hence, this discontinuity should not be resolved by the quantum correlations.
It would be very interesting to examine some inequalities for holographic reflected entropy in \cite{Dutta2019} for nontrivial states (e.g., local quench studied above) in the same non-perturbative way.
This trial could answer a question, which inequalities of the holographic reflected entropy break down by the quantum corrections.
We hope to return this issue in future work.

\section{Reflected Entropy in Integrable System} \label{sec:other}

It is very interesting to compare our result to the dynamics of the reflected entropy in other CFTs, in particular, integrable CFTs.
There are many works to study entanglement entropy after a local quench in various setups  \cite{He2014,Numasawa2016, Nozaki2013, Asplund2015, Caputa2014a, David2016, Caputa2017, He2017, Guo2018, Shimaji2018, Apolo2018}. Their motivation is to characterize CFT classes by the dynamics of entanglement. And from those results, this quantity is expected to capture the chaotic natures of CFTs.
On this background, it is naturally expected that by using a refined tool, reflected entropy, we can obtain more information to classify CFTs. In this section, we will briefly discuss  how the reflected entropy grows after a local quench in RCFTs and investigate whether the RCFT reflected entropy has a different growth from the holographic reflected entropy or not.

An important difference between the holographic CFT and RCFTs is that in the former, the OPE in the Regge limit does not contain the vacuum state, whereas in the later, the vacuum state can propagate even in the Regge limit. As a result, the time-dependence cannot be found in RCFTs. We will briefly explain this mechanism of the vanishing time-dependence by considering an analogy of (\ref{eq:F}) (see also (\ref{eq:AppF})) in RCFTs.

In our CFT, the Regge limit of this block is obtained by the monodormy matrix as
\footnote{
In the analytic continuation $m \to 1$, the exponent is replaced by $2mnh_O \to 4nh_O $ by the squaring rule (\ref{eq:squaring}).
}

\newsavebox{\boxpo}
\sbox{\boxpo}{\includegraphics[width=190pt]{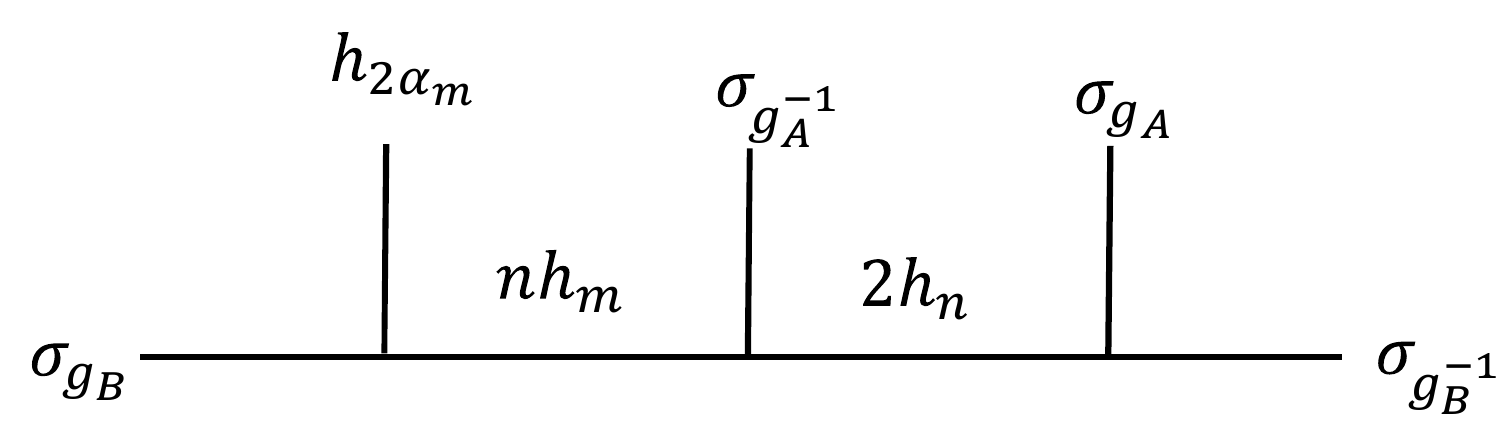}}
\newlength{\pow}
\settowidth{\pow}{\usebox{\boxpo}} 

\begin{equation}\label{eq:holomono}
\begin{aligned}
 &\parbox{\piw}{\usebox{\boxpi}}  \\
&\ar{\e \to0}
 \ti{\ca{M}}
_{0, 2\a_m}^{(-)}
   \left[
    \begin{array}{cc}
    \a_m   & \a_m   \\
     \a_{O}   &   \a_{O} \\
    \end{array}
  \right]
	\times (2i \e)^{h_{2\a_m}-2nmh_O}
 \parbox{\pow}{\usebox{\boxpo}},
\end{aligned}
\end{equation}
where $h_a=\a(Q-\a)$. We would like to mention that the time-dependence is encapsulated in the position of the external operator $h_{2\a_m}$.
On the other hand, if we consider the Regge limit in RCFTs,

\newsavebox{\boxpRa}
\sbox{\boxpRa}{\includegraphics[width=190pt]{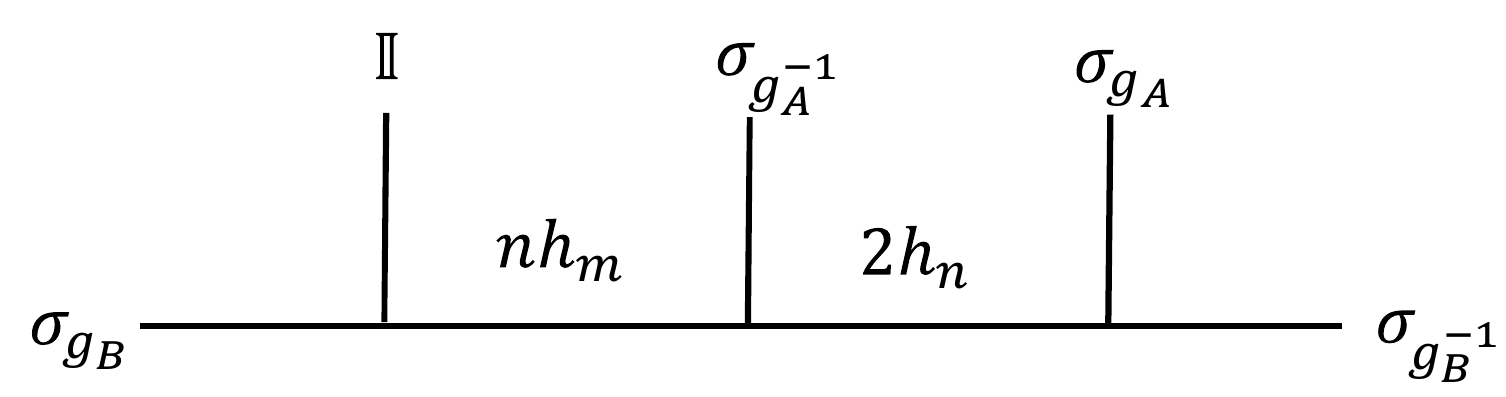}}
\newlength{\pRaw}
\settowidth{\pRaw}{\usebox{\boxpRa}} 

\begin{figure}[t]
 \begin{center}
  \includegraphics[width=10.0cm,clip]{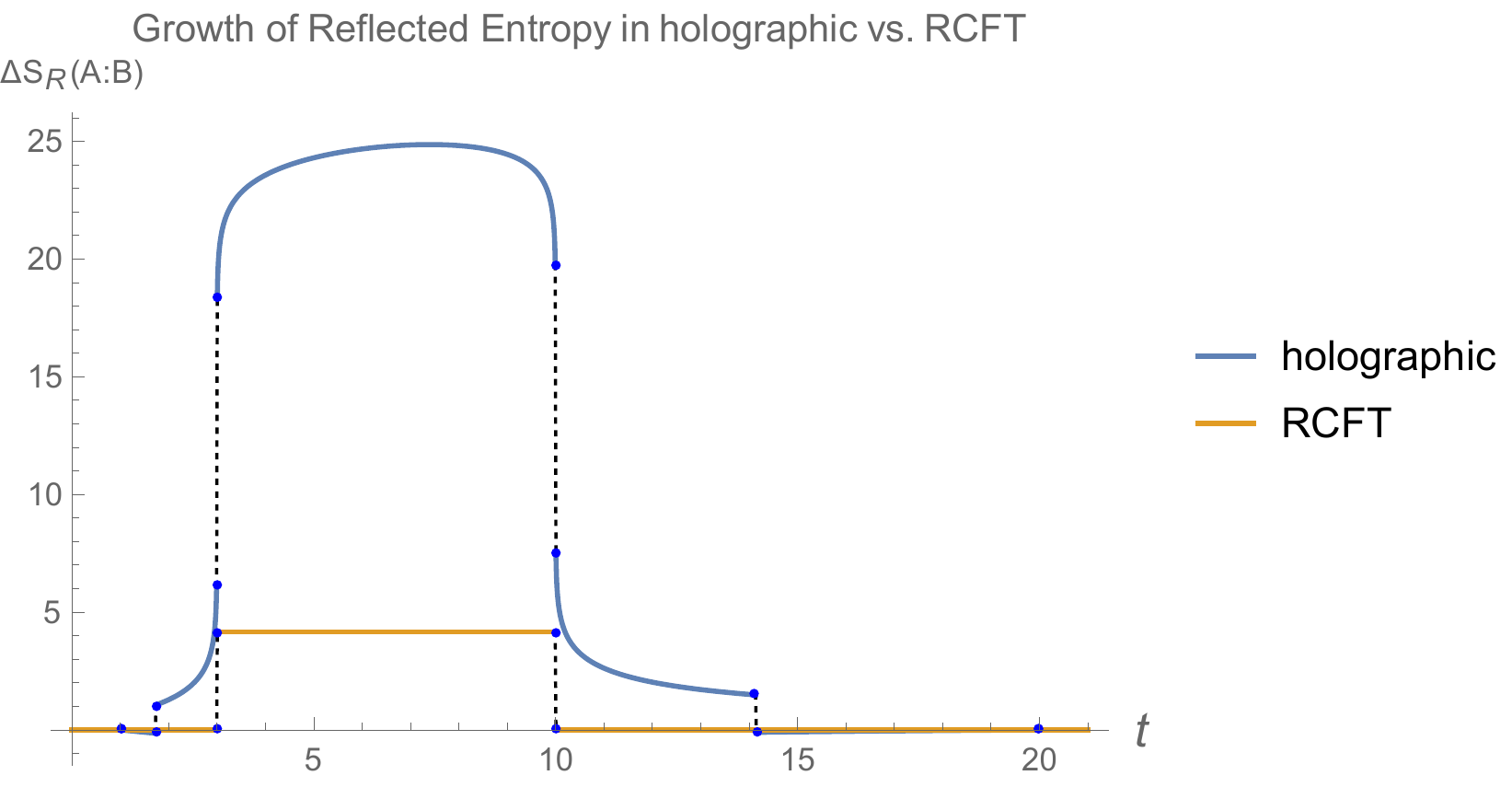}
 \end{center}
 \caption{ The growth of reflected entropy in holographic CFT (blue) and Ising model (yellow). 
$\D S_R$ means the difference between the excited state and the vacuum state.
Here $(u_1,v_1,u_2,v_2)=(-20,1,3,10)$, $\e=10^{-3}$
and we divide them by $\fr{c}{6}$. We choose $\g=2$ in holographic CFT and $O=\sigma$ in Ising model.
Each blue dot shows a transition of itself or its first derivative.
}
 \label{fig:RCFTplot}
\end{figure}

\newsavebox{\boxpRb}
\sbox{\boxpRb}{\includegraphics[width=160pt]{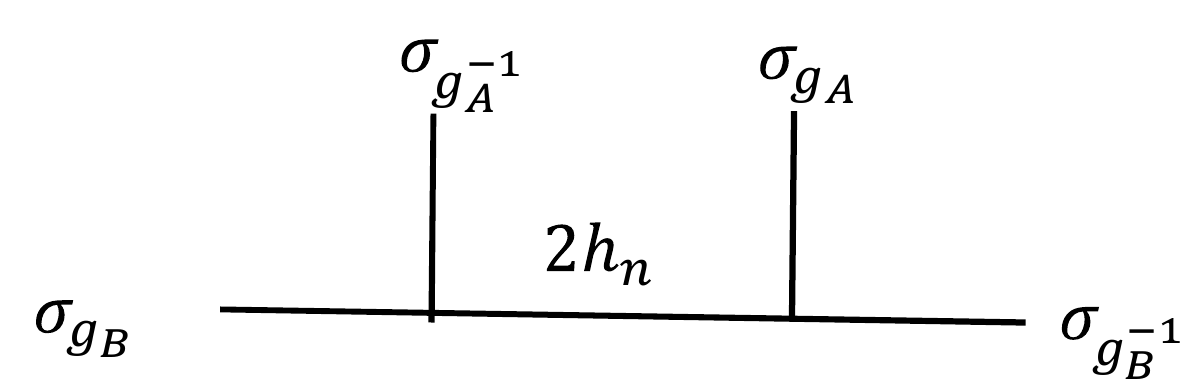}}
\newlength{\pRbw}
\settowidth{\pRbw}{\usebox{\boxpRb}} 

\begin{equation}\label{eq:RCFTmono}
\begin{aligned}
 &\parbox{\piw}{\usebox{\boxpi}} \\
&\ar{\e \to0}
 \ca{M}_{0, 0}^{(-)}
   \left[
    \begin{array}{cc}
    \a_m   & \a_m   \\
     \a_{O}   &   \a_{O} \\
    \end{array}
  \right]
	\times (2i \e)^{-2nmh_O}
\parbox{\pRaw}{\usebox{\boxpRa}}.
\end{aligned}
\end{equation}
The key point is that the operator $h_{2\a_m}$ is replaced by the identity, therefore, this 5-point block reduces a 4-point block,
\begin{equation}
\parbox{\pRaw}{\usebox{\boxpRa}}=\parbox{\pRbw}{\usebox{\boxpRb}}.
\end{equation}
This means that the time-dependence disappears in this single block approximation.
The way to calculate the reflected entropy in RCFTs is just repeating the calculation in Section \ref{sec:localCFT} replacing (\ref{eq:holomono}) by (\ref{eq:RCFTmono}). As a result, if we consider the setup ( $0<\e\ll u_2<-v_1<-u_1<v_2$ and $O$ is acted on $x=0$ at $t=0$.) for example, we obtain
\begin{equation}\label{eq:DSRR}
\begin{aligned}
\D S_R(A:B)[O]&=\left\{
    \begin{array}{ll}
  0 
		,   & \text{if } t<-v_1 ,\\ \\
  2\log d_O 
		,   & \text{if } -v_1<t<-u_1,\\ \\
  0
		,   & \text{if } -u_1<t ,
    \end{array}
  \right.\\
\end{aligned}
\end{equation}
where $d_O$ is a constant, so-called {\it quantum dimension}, which is re-expressed in terms of the modular S matrix as \cite{Numasawa2016,He2014}
\begin{equation}
d_O=\fr{S_{0O}}{S_{00}}.
\end{equation}
We would like to comment that this result is consistent with the relation (\ref{eq:EE=RE}). Namely, in the pure state limit ($\bar{A}=B$), the reflected entropy reduces to the entanglement entropy, which implies
\begin{equation}
\D S_R(A:B)[O]=2\D S(A)[O]=2 \log d_O.
\end{equation} 
This is consistent with the previous result $\D S(A)[O]=\log d_O$ in \cite{He2014}.
We will show the detailed calculation in a future paper about reflected entropy in finite $c$ CFTs.
Consequently, we can conclude that the reflected entropy in RCFTs cannot grow as in the holographic CFT and in fact, the dynamics can be fully captured by the quasi-particle picture. More concretely, if the quasi-particle enters a interval, then entanglement is created between the interval and its complement. In terms of the reflected entropy, this phenomena can be observed as a non-zero constant characterized by the quantum dimension, like entanglement entropy \cite{Numasawa2016,He2014}. In other words, the RCFT reflected entropy after a local quench is characterized by a step function, which is quite different from the holographic case.
We show the comparison of the reflected entropy between holographic CFT and Ising model in Figure \ref{fig:RCFTplot}.
One can find two significant differences from this Figure,
\begin{itemize}
\item The small effect in  $ t\in [u_2, -v_1] \cup [-u_1, v_2]$ does not appear in RCFTs, unlike the holographic CFT (see also Section \ref{sec:dynamics}).

\item The holographic CFT shows the logarithmic growth in $ t\in [-v_1, -u_1]$, on the other hand, the growth of RCFT approaches a finite constant.

\end{itemize}
This difference between RCFT and holographic CFT means that the reflected entropy might be also related to a nature of chaos in a given CFT, therefore, we expect that by making use of the reflected entropy, we can also study the information scrambling \cite{Asplund2015a, Kusuki2018c}, which might be a interesting direction for future work.
It would be interesting to note that this growth pattern (\ref{eq:DSRR}) is exactly the same as that of the mutual information.
\footnote{
Here, we mean not $S_R(A:B)=I(A:B)$ but $\D S_R(A:B)=\D I(A:B)$. 
}
This is a stronger version of the decoupling relation (\ref{eq:decouple}), which is quite natural because in RCFTs, the quasi particle picture can be applied in any time region.

These properties are quite different from the holographic case as show in (\ref{eq:DSR}), therefore, we could classify CFTs by studying whether the growth of reflected entropy and mutual information are different or not. Further studies in this direction shed light on what correlations are measured by reflected entropy.
We would also like to mention that the quantum  dimension can be interpreted as an effective degrees of freedom included in the operator $O$ and our result suggests that the reflected entropy captures this degrees of freedom, like entanglement entropy.

\section{Odd Entanglement Entropy} \label{sec:odd}

As mentioned in the introduction, the odd entanglement entropy in holographic CFTs also matches the reflected entropy (the entanglement wedge cross section) in our dynamical setup. These agreement can be understood from a similarity of the methods to calculate the odd entanglement entropy and the reflected entropy especially in the holographic CFTs. (Interestingly, this agreement is also the case for RCFTs. ) In this section, we will sketch the proof of this coincidence. An interesting point is that this quantity is not based on the purification, therefore, it is nontrivial in this sense that this quantity also reproduces the entanglement wedge cross section, like the reflected entropy. 

Following the definition \eqref{eq:OEE}, the odd entanglement entropy in our setup can be obtained from the following correlation function,
\be\label{eq:OEEq}
	\tr \pa{\rho_{AB}^{T_B}}^{n}=\fr{\Braket{\sigma_{n}(u_1)\bar{\sigma}_{n}(v_1)  {O^{\otimes n}}(w_1,\bar{w}_1)  \dg{{O^{\otimes n}}}(w_2,\bar{w}_2)   \bar{\sigma}_{n}(u_2) \sigma_{n}(v_2) }_{\text{CFT}^{\otimes n}}}
	{   \Braket{O(w_1,\bar{w}_1)  \dg{{O}} (w_2,\bar{w}_2)} ^{n} },
\ee
where $\sigma_n$ and $\bar{\sigma}_n$ correspond to the usual twist operators with twist number $\pm1$ and $n$ is the analytic continuation of an {\it odd} integer. If one assumes an even integer analytic continuation, the \eqref{eq:OEEq} is nothing but the one for the negativity.  
Note that for the odd entanglement entropy the complications from the decoupling effect (as like reflected entropy and negativity) do not appear. This is just because we take here the analytic continuation of an odd integer, thus no decoupling of the replica sheet happens\cite{Calabrese2012,Calabrese2013a}. Therefore, we can safely use the Virasoro conformal blocks for the calculation of the odd entanglement entropy as like the entanglement entropy in holographic CFTs. 

If one evaluates the \eqref{eq:OEEq} in the holographic CFTs, one can again approximate it as a single semiclassical conformal block.
The semiclassical conformal block (more precisely, the linearized semicalssical block \cite{Alkalaev2016c}) has the following form,
\begin{equation}
\log \ca{F}(z_i) \sim h f_0 (z_i) +h_p f_p (z_i)+O(h,h_p),
\end{equation}
where external dimensions $h$ and internal dimensions $h_p$ are given by the form,
\begin{equation}
h \sim h_p \sim \sigma c \ \ \ \ \ \ \ \ \ \ \text{with } \sigma \ll 1,
\end{equation}
and the functions $f_0$ and $f_p$ are of order one.
The Landau symbol $O(x,y)$ stands for various quantities vanishing as $x^n y^m \to 0$ with $n+m\geq2$. Here the entanglement entropy is obtained by $f_0(z_i)$ because the corresponding correlator is dominated by the vacuum block (i.e., $h_p=0$) \cite{Hartman2013a}. 

Let us recall the case of the reflected entropy. After all the reflected entropy came from this $f_p(z_i)$ because we take the  limit  $h \to 0$ of the external operators (i.e., $m \to 1$ limit).
In other words, the numerator of the 6-point function (\ref{eq:Renyi}) can be re-expressed by a series expansion in its internal dimension $h_n$ as
\begin{equation}
\log \ca{F}(z_i)= <\text{denominator in } (\ref{eq:Renyi})> + 2h_n f_p(z_i)+O\pa{(1-n)^2}.
\end{equation}
Here the first term is compensated by the denominator and the second term $2f_p$ corresponds to the value of reflected entropy. Remind that the factor $2$ of $2f_p$ comes from ``doubling of Virasoro block'' due to the doubling of the Hilbert space (namely, the even integer analytic continuation). 

Thus, we can immediately show that
\begin{equation}
S_O(A:B)[O]-S(A:B)[O] = \fr{c}{12} f_p (z_i) = \fr{1}{2} S_R(A:B)[O],
\end{equation}
where we used the fact that the conformal block related to the odd entanglement entropy has the intermediate dimension $h_p=h_n$ as shown in  \cite{Tamaoka2019}.
It means that the calculation of the odd entanglement entropy is just a repetition of that in section \ref{sec:localCFT}. 

\begin{figure}[t]
 \begin{center}
  \includegraphics[width=12.0cm,clip]{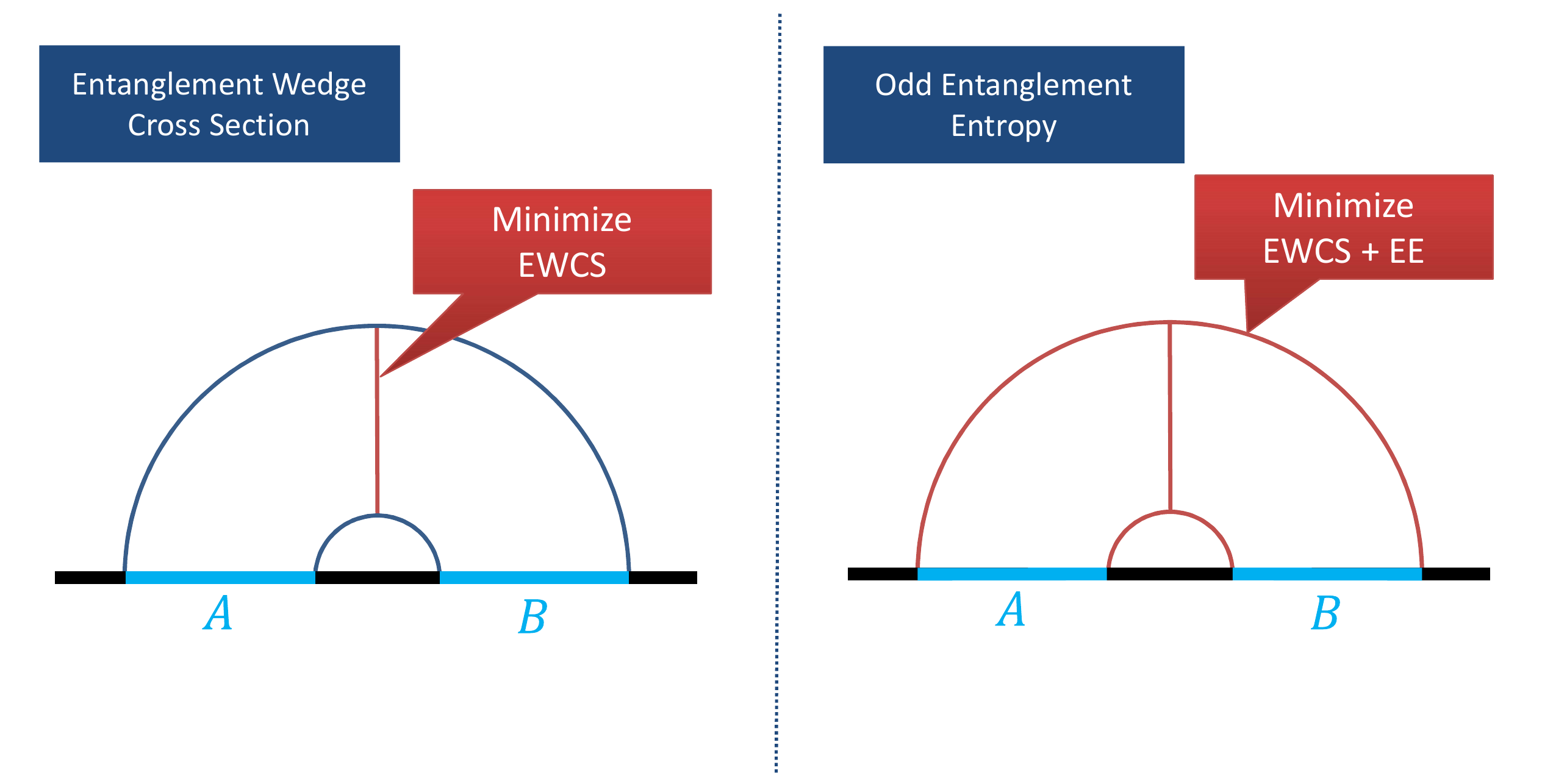}
 \end{center}
 \caption{The reflected entropy is given by minimizing the red line in the left, on the other hand, the odd entanglement entropy is given by minimizing the red lines in the right, which is the sum of two RT surfaces and the entanglement wedge cross section. }
 \label{fig:GEE}
\end{figure}

Strictly speaking, it might happen to find the disagreement between reflected entropy and odd entanglement entropy, because reflected entropy is based on the minimal of the entanglement wedge cross section, on the other hand, the odd entanglement entropy computes the minimal of the {\it sum} of two RT surfaces and the ``entanglement wedge cross section'' (see Figure \ref{fig:GEE})
\footnote{
We abused the word ``entanglement wedge cross section'' (precisely, the minimal surface which ends at two RT surfaces). It is not necessary that this corresponds to the (minimal) entanglement wedge cross section.}.
This could cause a change of the dominant channel of the single block approximations. 
However, we can easily check the agreement between the reflected entropy and the odd entanglement entropy (up to prefactor $2$) by assuming $\mu\ll\e\ll1$. We expect that these two minimizing problems provide the same result\footnote{In the regime $\mu\sim\e$, the area of the ``entanglement wedge cross section'' could be comparable to area of the two RT surfaces. In such regimes, we potentially have this deviation. Clarifying such possibilities in more general dynamical setup might be an interesting future direction.}.

Since the reflected entropy for RCFT in section \ref{sec:other} relies on the single conformal block approximation due to the Regge limit, we can also show
\be
\D S_O(A:B)[O]-\D S(A:B)[O]=  \dfrac{1}{2}\D S_R(A:B)[O] \;\;\;(\textrm{for RCFT}).
\ee
Therefore, we can use $S_O(A:B)$ as a signature of the chaos as like the reflected entropy. However, we suspect that the ``bare values'', $S_O(A:B)[O]-S(A:B)[O]$ and $S_R(A:B)[O]$ for RCFT, should behave quite differently. 
\section{Entanglement Wedge Cross Section for Falling Particle Geometry}\label{sec:bulk}

In this section, we consider the entanglement wedge cross section in the Poincare AdS${}_3$ geometry,
\be
\dd s^2=\dfrac{\dd z^2-\dd t^2+\dd x^2}{z^2},
\ee
with a falling particle whose trajectory is given by
\be
z^2-t^2=\epsilon^2, x=0. \label{eq:ifp}
\ee
Here $\epsilon$ corresponds not to the cutoff for radial direction (UV cutoff in CFT side) but to the size of the particle. We will define the cutoff for radial direction by $\mu$. We also set AdS radius $\ell_{\textrm{AdS}}\equiv1$ for simplicity. This geometry is expected to be dual to the local operator quench at $(x,t)=(0,0)$ in the holographic CFT\cite{Nozaki2013}. 

Since the falling particle gets boosted under the time evolution, we must take into account the back-reaction due to the boosted particle. By using the global coordinates, one can put the falling particle always on the center and represent the back-reacted geometry outside of the particle\cite{Nozaki2013, Horowitz1999} as
\be
\dd s^2=-(r^2+1-M)\dd t^2+\dfrac{\dd r^2}{r^2+1-M}+r^2\dd \theta^2,
\ee
where $M$ characterizes the mass of the particle. 
For $M<1$, this metric describes the geometry with a conical deficit located at $r=0$. For $M\geq1$, it gives rise to the static BTZ geometry with mass $M-1$. In particular, we are interested in the latter BTZ setup. To this end, one can analytically continue the former results to the latter ones $\sqrt{1-M}\rightarrow i\sqrt{M-1}\equiv i\gamma$. Note that one can identify the present $\gamma=\sqrt{M-1}$ with the same one introduced in CFT analysis $\gamma=\s{\fr{24}{c}h_O-1}$. The static BTZ corresponds to $\gamma=\bar{\gamma}$. In section \ref{subsec:rot}, we will briefly discuss the $\gamma\neq\bar{\gamma}$ case, dual to the rotating BTZ blackhole.

Since the above geometries are locally AdS${}_3$, it is very useful to write them by using the embedding coordinates in $\mathbb{R}^{2,2}$:
\be
\dd s^2=\eta_{AB}\dd X^A \dd X^B=-\dd X^2_0-\dd X^2_1+\dd X^2_2+\dd X^3_3,
\ee
with
\be
X^2=-1,
\ee
where we defined
\be
X\cdot Y\equiv \eta_{AB}X^AY^B.
\ee
Then the geometry \eqref{eq:ifp} is given by
\begin{subequations}\label{subeq:ifpcoord}
\begin{align}
X_0&=\dfrac{t}{z},\\
X_1&=\dfrac{\epsilon+\epsilon^{-1}(z^2+x^2-t^2)}{2z},\\
X_2&=\dfrac{x}{z},\\
X_3&=\dfrac{-\epsilon+\epsilon^{-1}(z^2+x^2-t^2)}{2z}.
\end{align}
\end{subequations}

On the other hand, one can describe the back-reacted geometry in global coordinates as the following coordinates:
\begin{subequations}\label{subeq:btzcoord}
\begin{align}
X_0&=\sqrt{\dfrac{r^2+1-M}{1-M}}\sin\left( \sqrt{1-M}\tau\right),\\
X_1&=\sqrt{\dfrac{r^2+1-M}{1-M}}\cos\left( \sqrt{1-M}\tau\right),\\
X_2&=\dfrac{r}{\sqrt{1-M}}\sin\left( \sqrt{1-M}\theta\right),\\
X_3&=\dfrac{r}{\sqrt{1-M}}\cos\left( \sqrt{1-M}\theta\right),
\end{align}
\end{subequations}
where we chose $\tau\in[0,\pi]$ ($\tau\in[-\pi,0]$) for $t\geq0$ ($t\leq0$) and $\theta\in[0,\pi]$ ($\theta\in[-\pi,0]$) for $x\geq0$ ($x\leq0$). 
Note that we also imposed identification along the angular direction $\theta\sim\theta+2\pi$ which will become important for later analysis. Having this identification in mind, we can easily relate these two geometries by using the above embedding coordinates. 
\subsection{Geodesics between two minimal surfaces}
First, we derive the geodesic distance between two geodesics anchored on the boundary points. This will be very useful to obtain the entanglement wedge cross section of our interests. In the embedding coordinates, the length of the geodesics ending on the bulk points $X_i$ and $X_j$ is given by
\begin{subequations}\label{subeq:geodist}
\begin{align}
\sigma(X_i,X_j)&=\log(\xi^{-1}_{ij}+\sqrt{\xi^{-1}_{ij}-1}\sqrt{\xi^{-1}_{ij}+1})\\
\xi^{-1}_{ij}&=-X_i\cdot X_j.
\end{align}
\end{subequations}
On the other hand, the spacelike geodesics $\gamma_{ij}$ anchored on two bulk points $X_i$ and $X_j$ is given by
\be
X_{ij}^A(\la)=m^A\ex{-\la}+n^A\ex{\la},
\ee
where
\be
m^2=n^2=0,\; 2m\cdot n=-1.
\ee
If we have
\be
X(\la_i)=X_i,\; X(\la_j)=X_j,
\ee
as a boundary condition and if both $X_i$ and $X_j$ are sufficiently close to the boundary, we can write
\be
X_{ij}^A(\la)=\dfrac{X^A_i\ex{-\la}+X^A_j\ex{\la}}{\sqrt{-2X_i\cdot X_j}},
\ee 
where
\be
\ex{-\la_i}=\ex{\la_j}=\sqrt{-2X_i\cdot X_j}(\equiv\sqrt{2\xi^{-1}_{ij}}).
\ee
We would like to find the pair of parameters $(\la,\la^\prime)=(\la_\ast,\la_\ast^\prime)$ which minimizes (extremizes) the length of geodesics $\sigma(\la,\la^\prime)\equiv\sigma(X_{14}(\la),X_{23}(\la^\prime))$. \if (To this end, one can show that
\be
\dfrac{\p}{\p \la}\sigma(\la,\la^\prime)\Bigg|_{\la=\la_\ast,\la^\prime=\la^\prime_\ast}=\dfrac{\p}{\p \la^\prime}\sigma(\la,\la^\prime)\Bigg|_{\la=\la_\ast,\la^\prime=\la^\prime_\ast}=0,
\ee
is equivalent to
\be
\dfrac{\p}{\p \la}\xi^{-1}(\la,\la^\prime)\Bigg|_{\la=\la_\ast,\la^\prime=\la^\prime_\ast}=\dfrac{\p}{\p \la^\prime}\xi^{-1}(\la,\la^\prime)\Bigg|_{\la=\la_\ast,\la^\prime=\la^\prime_\ast}=0.
\ee
)\fi
As a result, we find
\be
\la_\ast=\dfrac{1}{4}\log\left[\dfrac{(\xi^{-1}_{12})(\xi^{-1}_{14})}{(\xi^{-1}_{24})(\xi^{-1}_{34})}\right],\;\;\; \la^\prime_\ast=\dfrac{1}{4}\log\left[\dfrac{(\xi^{-1}_{12})(\xi^{-1}_{24})}{(\xi^{-1}_{14})(\xi^{-1}_{34})} \right],
\ee
and
\be
\xi^{-1}_{ij}(\la_\ast,\la^\prime_\ast)=\dfrac{1}{\sqrt{v}}(1+\sqrt{u}).
\ee
Here $u$ and $v$ are given by,
\begin{align}
u=\dfrac{\xi^{-1}_{12}\xi^{-1}_{34}}{\xi^{-1}_{13}\xi^{-1}_{24}},\; v=\dfrac{\xi^{-1}_{14}\xi^{-1}_{23}}{\xi^{-1}_{13}\xi^{-1}_{24}},
\end{align}
and reduce to the standard cross ratio in the CFT side. 
Therefore, we have obtained
\be
E_W=\dfrac{1}{4G}\sigma(\la_\ast,\la^\prime_\ast)=\dfrac{1}{4G}\log\left(\dfrac{1+\sqrt{u}+\sqrt{(1+\sqrt{u})^2-v}}{\sqrt{v}}\right), \label{eq:ewcs_ifp}
\ee
which has effectively the same form as  the AdS${}_3$ one in the embedding coordinates\cite{Hirai2018}. 
Here we introduced the Newton constant by $G$, which is related to the central charge as $c=\fr{3}{2G}$ by the AdS/CFT dictionary \cite{Brown1986}.
We will apply the above formula \eqref{eq:ewcs_ifp} in order to obtain the entanglement wedge cross section in the falling particle geometry. Notice that, however, we had the identification $\theta\sim\theta+2\pi$ along the angular direction. Therefore, we have multiple solutions, most of which correspond to the solutions with non-trivial winding around the deficit angle (or the blackhole). What we need to pick up is the one which reproduces the correct minimal surfaces (namely, the correct entanglement wedge) and gives the minimal cross section of the entanglement wedge.

\subsection{An example: Quench outside Region A and B}
Here we illustrate an example of the holographic local quench. In section \ref{subsubsec:zero}, we will see the perfect agreement with the CFT analysis. Quite similar analysis show the agreement even in other setups. To avoid redundancy, we will not present other examples here. In section \ref{subsubsec:non-zero}, we also comment on the non-zero size case.

\subsubsection{Dominant phase for small particle limit}\label{subsubsec:zero}
\begin{figure}[t]
\begin{center}
\resizebox{120mm}{!}{\includegraphics{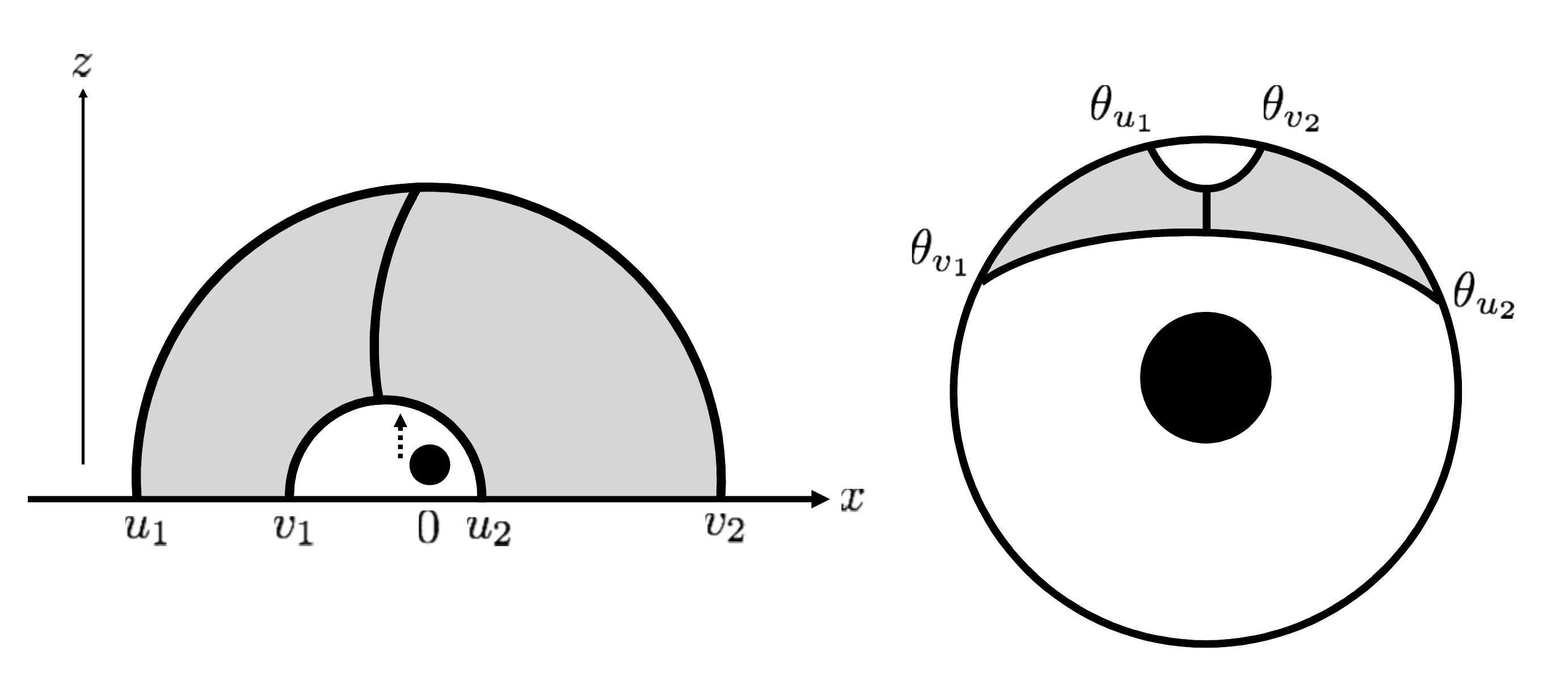}}
\caption{Left: our setup in the Poincare coordinates. Black curves ending on the boundary  are minimal surfaces and the shaded region corresponds to the (time slice of) entanglement wedge. Another solid curve anchored on the minimal surfaces represents the minimal cross section of the entanglement wedge. Right: The back-reacted geometry in the global coordinates. To be precise, each ``boundary'' points map to the different time and radial slices, thus the right panel is quite schematic. For each figure, the black-colored circle represents the black hole.}\label{fig:xfp_i1}
\end{center}
\end{figure}

\begin{figure}[t]
\begin{center}
\resizebox{140mm}{!}{\includegraphics{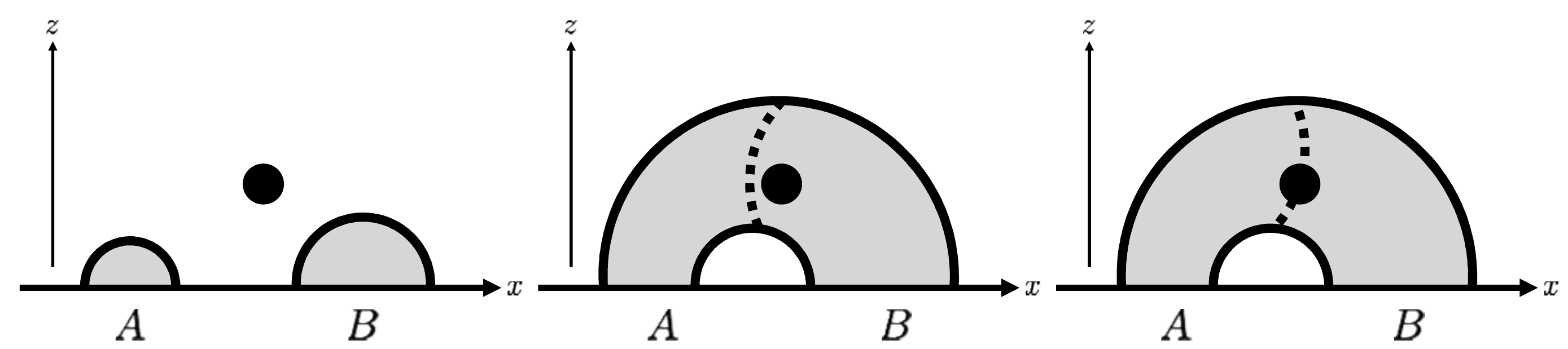}}
\caption{Three possibilities for entanglement wedge (shaded regions) and its cross section (dotted lines): disconnected (left), connected (center) and splitting cross sections (right). In the small particle limit $\epsilon\rightarrow0$, we can fix our phase either disconnected (left) or connected (center) for every time regions.}\label{fig:xfp_i0}
\end{center}
\end{figure}

Let us consider the bulk dual of a local (heavy) operator quench outside region between $A$ and $B$. Namely, we assume $A=[u_1,v_1]$ and $B=[u_2,v_2]$ where $0<u_2<-v_1<-u_1<v_2$ (see Figure \ref{fig:xfp_i1}). To make life simpler and for comparison with the CFT results, we focus on the small particle limit $\epsilon\rightarrow0$. Without this assumption, we will observe many transitions between the three phases (see Figure \ref{fig:xfp_i0}). We comment on these transitions briefly in the upcoming subsection. 
At the first time, $0<\epsilon\ll t<u_2$, the falling particle is outside of the entanglement wedge and does not affect any back-reaction to its inside. Indeed we can compute geodesics in global coordinates and then back to the original metric by using the following relation:
\begin{align}
(\tau_{u_1}, \theta_{u_1}, r_{u_1})&=\left(\frac{2t\epsilon}{u_1^2-t^2}, \pi-\frac{2u_1\epsilon}{u_1^2-t^2}, \frac{|u_1^2-t^2|}{2\mu\epsilon}\right),\\
(\tau_{v_1}, \theta_{v_1}, r_{v_1})&=\left(\frac{2t\epsilon}{v_1^2-t^2}, \pi-\frac{2v_1\epsilon}{v_1^2-t^2}, \frac{|v_1^2-t^2|}{2\mu\epsilon}\right), \\
(\tau_{u_2}, \theta_{u_2}, r_{u_2})&=\left(\frac{2t\epsilon}{u_2^2-t^2}, \pi-\frac{2u_2\epsilon}{u_2^2-t^2}, \frac{|u_2^2-t^2|}{2\mu\epsilon}\right),\\
(\tau_{v_2}, \theta_{v_2}, r_{v_2})&=\left(\frac{2t\epsilon}{v_2^2-t^2}, \pi-\frac{2v_2\epsilon}{v_2^2-t^2}, \frac{|v_2^2-t^2|}{2\mu\epsilon}\right).
\end{align}
Thus, we obtain
\begin{align}
E_W&=\frac{c}{6}\log\dfrac{1+\sqrt{\frac{(v_1 - u_1) (v_2 - u_2)}{(u_2 - u_1) (v_2 - v_1)}}}{1-\sqrt{\frac{(v_1 - u_1) (v_2 - u_2)}{(u_2 - u_1) (v_2 - v_1)}}},\;\,(\textrm{if}\;\,0<t<u_2),
\end{align}
at the leading order of $\epsilon$ expansion. Notice that this is just the same cross section as one for Poincare AdS${}_3$. 

\begin{figure}[t]
\begin{center}
\resizebox{120mm}{!}{\includegraphics{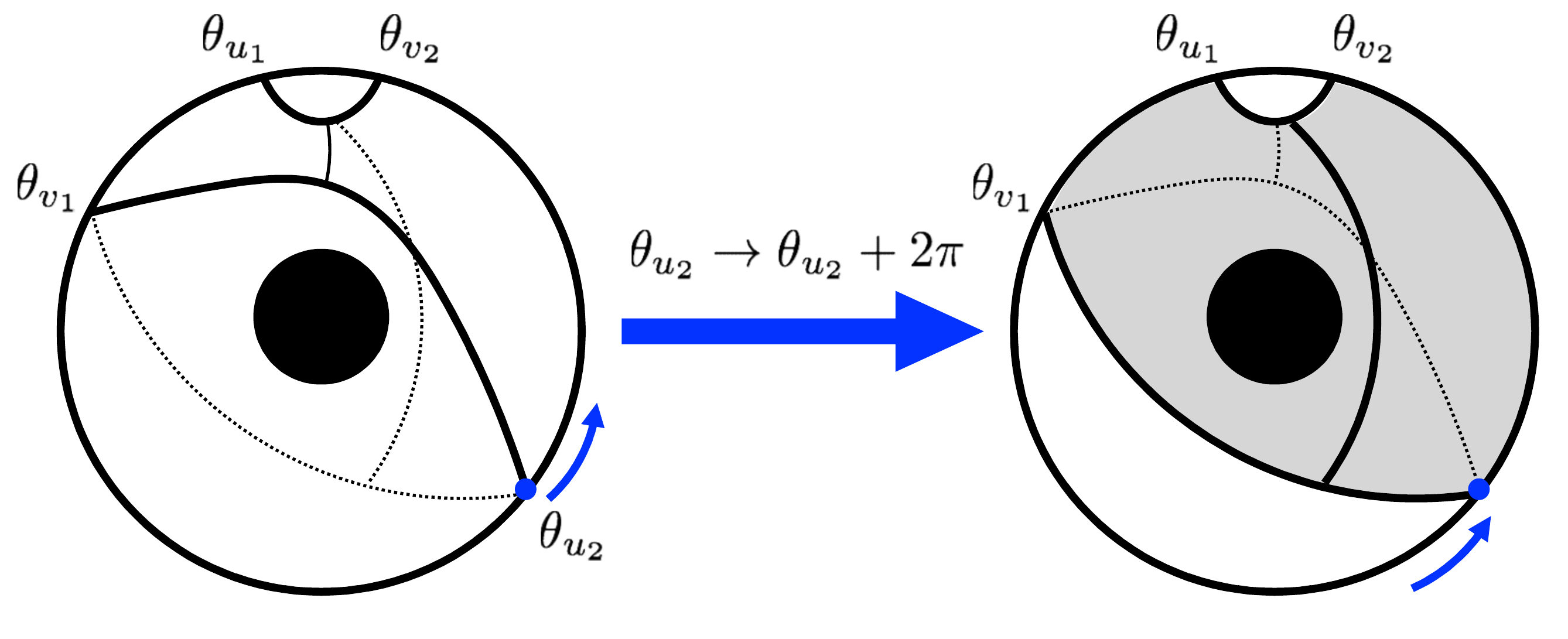}}
\caption{The manipulation in order to obtain the correct entanglement wedge. In the right panel, dotted lines describe the non-minimal surfaces and ``cross section'' obtained naively from \eqref{eq:ewcs_ifp}. After shifting $\theta_{u_2}\rightarrow\theta_{u_2}+2\pi$, we achieve the left panel which describes the correct entanglement wedge and its cross section.}\label{fig:xfp_i2}
\end{center}
\end{figure}

In the regime $u_2<t<\sqrt{-u_2v_1}$, the falling particle is getting closer to the entanglement wedge, but still outside of the entanglement wedge. Since the coordinates across the singularity on $u_2$, the relation between two coordinates changes slightly,
\begin{align}
(\tau_{u_2}, \theta_{u_2}, r_{u_2})&=\left(\frac{2t\epsilon}{u_2^2-t^2}, \pi-\frac{2u_2\epsilon}{u_2^2-t^2}, \frac{|u_2^2-t^2|}{2\mu\epsilon}\right)\rightarrow\left(\pi+\frac{2t\epsilon}{u_2^2-t^2}, -\frac{2u_2\epsilon}{u_2^2-t^2}, \frac{|u_2^2-t^2|}{2\mu\epsilon}\right),
\end{align}
whereas that for other coordinates ($u_1,v_1,v_2$) does not change. From the CFT viewpoint, this effect can be seen as the monodromy transformation in \eqref{eq:effect1} although here we have no distinction between the left and right moving.
We will take the same replacement for each coordinate ($u_1,v_1,v_2$) when the time $t$ exceeds each (absolute) value.
In this regime, the back-reaction to the minimal surfaces becomes visible, so the entanglement wedge cross section does,
\begin{align}
E_W&=\frac{c}{12}\log\dfrac{1+\sqrt{\frac{(v_2-t) (v_1-u_1)}{(t-u_1) (v_2-v_1)}}}{1-\sqrt{\frac{(v_2-t) (v_1-u_1)}{(t-u_1) (v_2-v_1)}}}+\frac{c}{12}\log\dfrac{1+\sqrt{\frac{(v_1 - u_1) (v_2 - u_2)}{(u_2 - u_1) (v_2 - v_1)}}}{1-\sqrt{\frac{(v_1 - u_1) (v_2 - u_2)}{(u_2 - u_1) (v_2 - v_1)}}},\;\,(\textrm{if}\;\,u_2<t<\sqrt{-u_2v_1}).
\end{align}
When the particle enters the entanglement wedge ($\sqrt{-u_2v_1}<t<-v_1$), we cannot use the formula naively. This is because the original one captures the non-minimal surfaces (see left panel of Figure \ref{fig:xfp_i2}). Thus, we should utilize the identification so that we have correct entanglement wedge. This can be achieved by shifting the $\theta_{u_2}\rightarrow\theta_{u_2}+2\pi$, which is the same manipulation when one computes the holographic entanglement entropy (see right panel of Figure \ref{fig:xfp_i2}).

\begin{figure}[t]
\begin{center}
\resizebox{120mm}{!}{\includegraphics{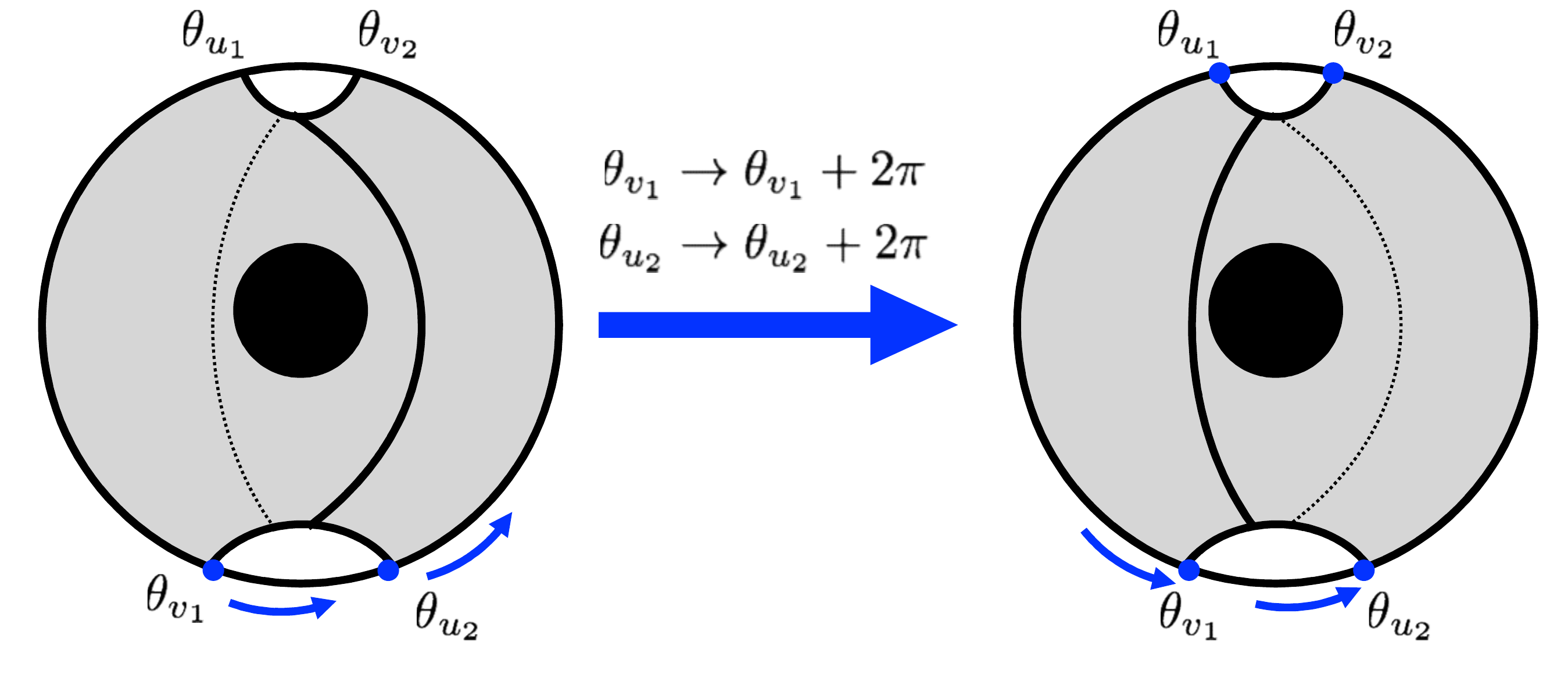}}
\caption{The manipulation in order to obtain the minimal cross section of the entanglement wedge. }\label{fig:xfp_i3}
\end{center}
\end{figure}

\noindent
Then we get
\begin{align}
E_W&=\frac{c}{12}\log\dfrac{1+\sqrt{\frac{(t+v_2) (v_1-u_1)}{(-t-u_1) (v_2-u_2)}}}{1-\sqrt{\frac{(t+v_2) (v_1-u_1)}{(-t-u_1) (v_2-u_2)}}}+\frac{c}{12}\log\dfrac{1+\sqrt{\frac{(v_1 - u_1) (v_2 - u_2)}{(u_2 - u_1) (v_2 - v_1)}}}{1-\sqrt{\frac{(v_1 - u_1) (v_2 - u_2)}{(u_2 - u_1) (v_2 - v_1)}}},\;\,(\textrm{if}\;\,\sqrt{-u_2v_1}<t<-v_1).
\end{align}
Note that the aforementioned manipulation corresponds to choosing the unusual conformal block in \eqref{eq:ch2} which is compensated by the monodromy in the holomorphic part. 

When the particle is falling near the center of the entanglement wedge ($-v_1<t<-u_1$), the corresponding minimal cross section acquires the significant effects on the back-reaction. The minimal one can be obtained from the \eqref{eq:ewcs_ifp} by shifting $\theta_{v_1}\rightarrow\theta_{v_1}+2\pi$ and $\theta_{u_2}\rightarrow\theta_{u_2}+2\pi$ (see Figure \ref{fig:xfp_i3}), 
\begin{align}
E_W&=\frac{c}{6}\log\frac{2\sinh \gamma\pi}{\gamma\epsilon}\sqrt{\frac{(t+u_1)(t+u_2)(t+v_1)(t+v_2)}{(u_2-v_1)(u_1-v_2)}}\nonumber\\
&\hspace{2cm}+\frac{c}{12}\log\dfrac{1+\sqrt{\frac{(v_1 - u_1) (v_2 - u_2)}{(u_2 - u_1) (v_2 - v_1)}}}{1-\sqrt{\frac{(v_1 - u_1) (v_2 - u_2)}{(u_2 - u_1) (v_2 - v_1)}}},\;\,(\textrm{if}\;\,-v_1<t<-u_1).
\end{align}
For $-u_1<t$, we can repeat the similar analysis. In summary, we have obtained,
\begin{align}
E_W&=\left\{\begin{array}{ll}
\frac{c}{6}\log\frac{1+\sqrt{\frac{(v_1 - u_1) (v_2 - u_2)}{(u_2 - u_1) (v_2 - v_1)}}}{1-\sqrt{\frac{(v_1 - u_1) (v_2 - u_2)}{(u_2 - u_1) (v_2 - v_1)}}}&(\textrm{if}\;\,0<t<u_2)\vspace{1mm}\\
\frac{c}{12}\log\frac{1+\sqrt{\frac{(v_2-t) (v_1-u_1)}{(t-u_1) (v_2-v_1)}}}{1-\sqrt{\frac{(v_2-t) (v_1-u_1)}{(t-u_1) (v_2-v_1)}}}+\frac{c}{12}\log\frac{1+\sqrt{\frac{(v_1 - u_1) (v_2 - u_2)}{(u_2 - u_1) (v_2 - v_1)}}}{1-\sqrt{\frac{(v_1 - u_1) (v_2 - u_2)}{(u_2 - u_1) (v_2 - v_1)}}} & (\textrm{if}\;\,u_2<t<\sqrt{-u_2v_1})\vspace{1mm}\\
\frac{c}{12}\log\frac{1+\sqrt{\frac{(t+v_2) (v_1-u_1)}{(-t-u_1) (v_2-u_2)}}}{1-\sqrt{\frac{(t+v_2) (v_1-u_1)}{(-t-u_1) (v_2-u_2)}}}+\frac{c}{12}\log\frac{1+\sqrt{\frac{(v_1 - u_1) (v_2 - u_2)}{(u_2 - u_1) (v_2 - v_1)}}}{1-\sqrt{\frac{(v_1 - u_1) (v_2 - u_2)}{(u_2 - u_1) (v_2 - v_1)}}} & (\textrm{if}\;\,\sqrt{-u_2v_1}<t<-v_1)\vspace{1mm}\\
\frac{c}{6}\log\frac{2\sinh \gamma\pi}{\gamma\epsilon}\sqrt{\frac{(t+u_1)(t+u_2)(t+v_1)(t+v_2)}{(u_2-v_1)(u_1-v_2)}}+\frac{c}{12}\log\frac{1+\sqrt{\frac{(v_1 - u_1) (v_2 - u_2)}{(u_2 - u_1) (v_2 - v_1)}}}{1-\sqrt{\frac{(v_1 - u_1) (v_2 - u_2)}{(u_2 - u_1) (v_2 - v_1)}}} & (\textrm{if}\;\,-v_1<t<-u_1) \vspace{1mm}\\
\frac{c}{12}\log\frac{1+\sqrt{\frac{(t+u_2) (u_1-v_1)}{(t+v_1) (u_1-u_2)}}}{1-\sqrt{\frac{(t+u_2) (u_1-v_1)}{(t+v_1) (u_1-u_2)}}}+\frac{c}{12}\log\frac{1+\sqrt{\frac{(v_1 - u_1) (v_2 - u_2)}{(u_2 - u_1) (v_2 - v_1)}}}{1-\sqrt{\frac{(v_1 - u_1) (v_2 - u_2)}{(u_2 - u_1) (v_2 - v_1)}}} & (\textrm{if}\;\,-u_1<t<\sqrt{-u_1v_2}) \vspace{1mm}\\ 
\frac{c}{12}\log\frac{1+\sqrt{\frac{(t-u_2) (u_1-v_1)}{(t-v_1)(u_1-u_2)}}}{1-\sqrt{\frac{(t-u_2) (u_1-v_1)}{(t-v_1)(u_1-u_2)}}}+\frac{c}{12}\log\frac{1+\sqrt{\frac{(v_1 - u_1) (v_2 - u_2)}{(u_2 - u_1) (v_2 - v_1)}}}{1-\sqrt{\frac{(v_1 - u_1) (v_2 - u_2)}{(u_2 - u_1) (v_2 - v_1)}}} & (\textrm{if}\;\,\sqrt{-u_1v_2}<t<v_2)\vspace{1mm}\\
\frac{c}{6}\log\frac{1+\sqrt{\frac{(v_1 - u_1) (v_2 - u_2)}{(u_2 - u_1) (v_2 - v_1)}}}{1-\sqrt{\frac{(v_1 - u_1) (v_2 - u_2)}{(u_2 - u_1) (v_2 - v_1)}}} & (\textrm{if}\;\,v_2<t)
\end{array}
\right.
\end{align}
These results perfectly agree with the CFT results (\ref{eq:main}).

\subsubsection{Another phase without small particle limit}\label{subsubsec:non-zero}
\begin{figure}[t]
\begin{center}
\resizebox{100mm}{!}{\includegraphics{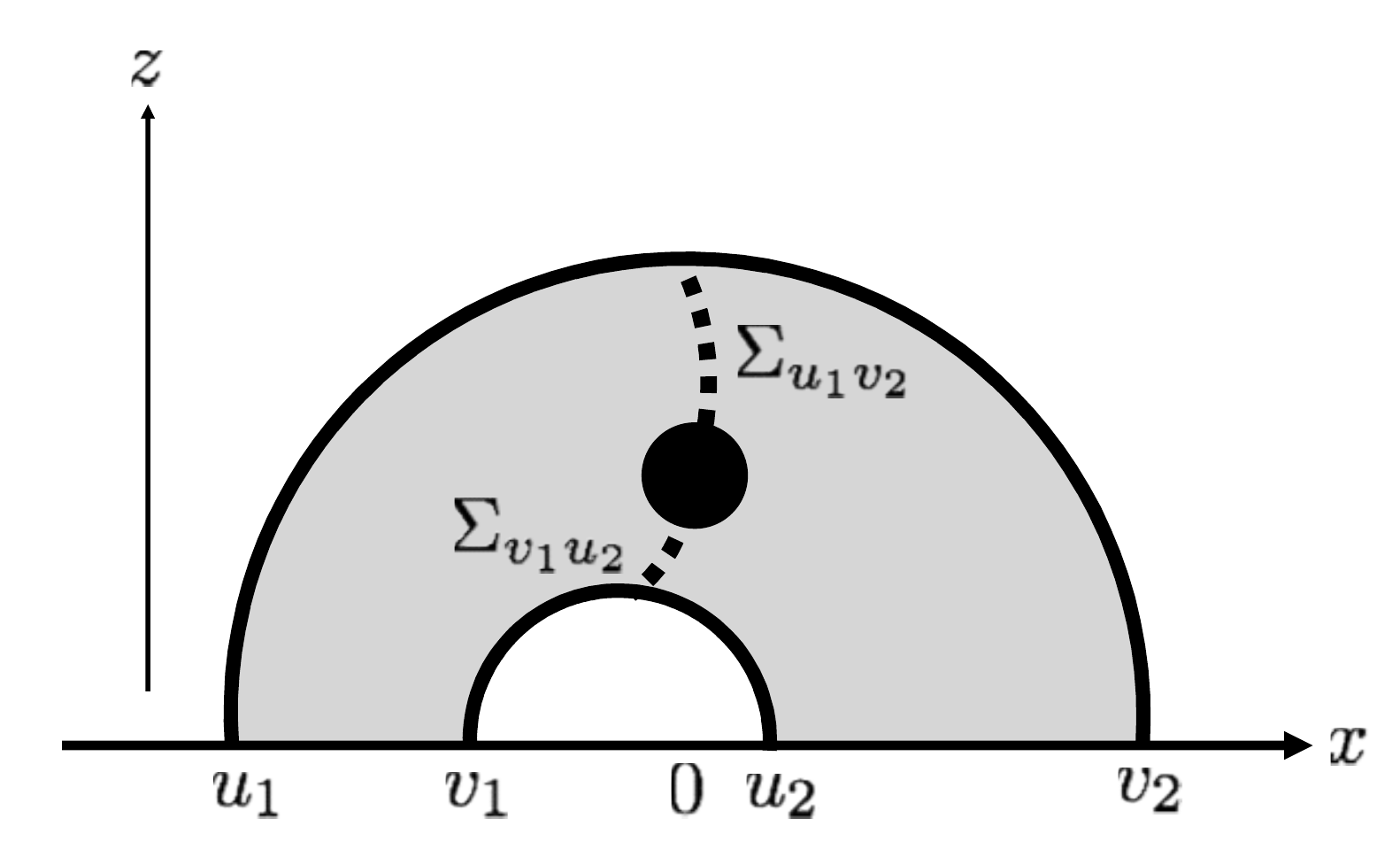}}
\caption{Two dotted lines ($\Sigma_{u_1v_2}$ and $\Sigma_{v_1u_2}$) show another possibility for the entanglement wedge cross section which ends on the horizon. This happens only when the particle is falling inside the entanglement wedge.}\label{fig:fp_ic}
\end{center}
\end{figure}

\begin{figure}[t]
\begin{center}
\resizebox{110mm}{!}{\includegraphics{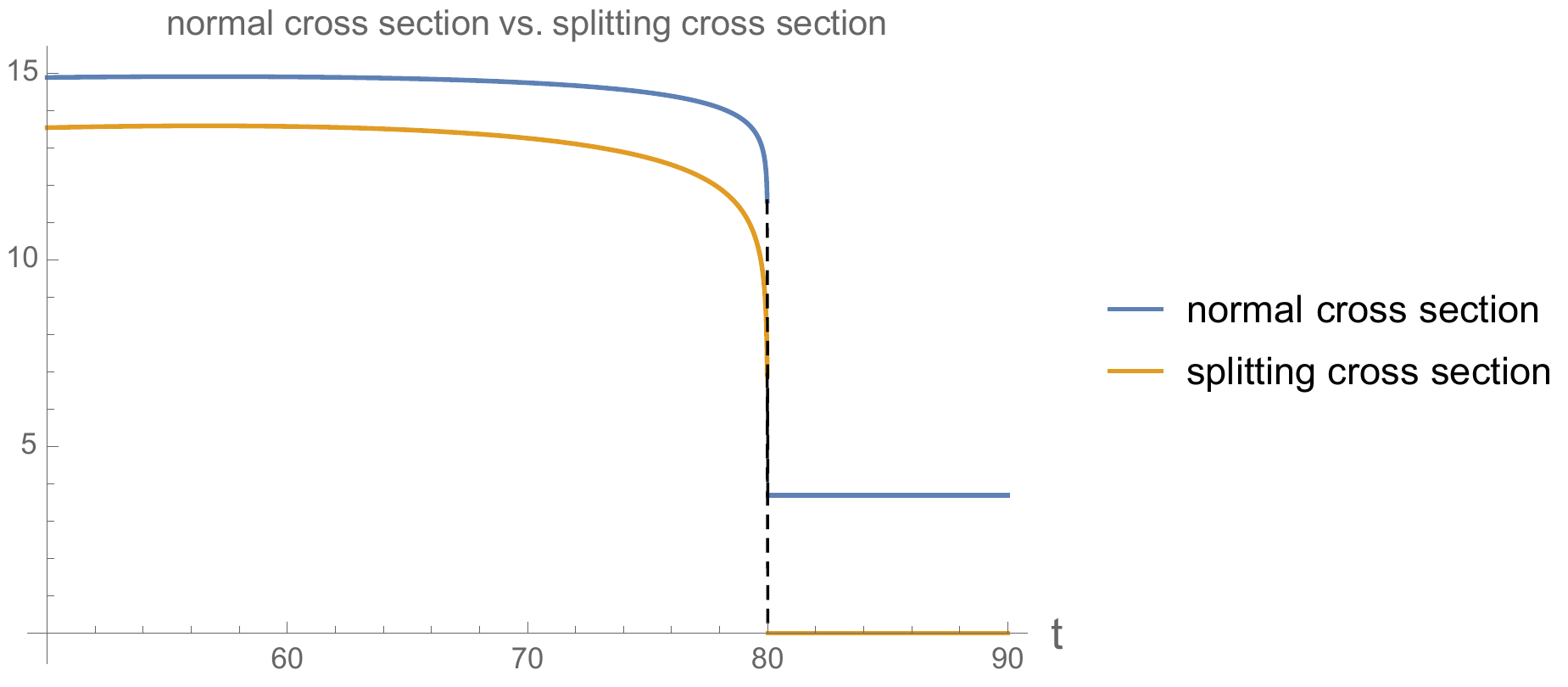}}
\caption{This plot shows the time dependence of the splitting cross section (Figure \ref{fig:fp_ic}) and the normal cross section discussed in the previous section (middle panel of Figure \ref{fig:xfp_i0}).  Here we set $-u_1=v_2=80, -v_1=u_2=2, \gamma=2,$ and $\epsilon=0.1$. In this setup, the splitting cross section becomes a minimum one. }\label{fig:plot_disc_ewcs}
\end{center}
\end{figure}

There is another interesting possibility --- the entanglement wedge cross section splits into two pieces and each of them ends on the falling particle (see Figure \ref{fig:fp_ic}). Although it will never become dominant contribution for the $\epsilon\rightarrow0$ limit, this phase can become the dominant one when the size of the particle $\epsilon$ has a comparable length scale with each interval ($A$, $B$ and distance between them). 
Even in this case, one can use \eqref{subeq:geodist} for each segment ($\Sigma_{u_1v_2}$ and $\Sigma_{v_1u_2}$) in the Figure \ref{fig:fp_ic} and find the ``minimal'' ones\footnote{As discussed in the below, we must minimize not the segments of cross sections but the minimal surfaces, otherwise what we compute is no longer the minimal cross section of the entanglement wedge. }. After all, we obtain 
\begin{align}
E_W&=\dfrac{1}{4G}(\sigma(\Sigma_{u_1v_2})+\sigma(\Sigma_{v_1u_2})),\\
\sigma(\Sigma)&=\log\dfrac{r_\ast+\sqrt{r_\ast^2-(M-1)}}{\sqrt{M-1}},
\end{align}
where $r_\ast$ corresponds to the ``turning point'' in the geodesics anchored on the boundary points. One can see the $r_\ast$ in the literature\cite{Nozaki2013} (see also appendix B of \cite{Asplund2014}):
\be
r_\ast=\dfrac{\sqrt{1-A^2-B^2(1-M)+\sqrt{(1-A^2-B^2(1-M))^2+4B^2(1-M)}}}{\sqrt{2}B}.
\ee
Here we defined
\begin{align}
A&=\Bigg|\dfrac{\sin(\sqrt{1-M}\D\tau_\infty)}{\sin(\sqrt{1-M}\D\theta_\infty)}\Bigg|,\\
B&=\Bigg|\dfrac{\cos(\sqrt{1-M}\Delta\tau_\infty)-\cos(\sqrt{1-M}\Delta\theta_\infty)}{\sqrt{1-M}\sin(\sqrt{1-M}\D\theta_\infty)}\Bigg|,
\end{align}
where $(\D\tau_\infty,\D\theta_\infty)=(\tau_{v_1}-\tau_{u_2},\theta_{v_1}-\theta_{u_2})$ or $(\tau_{u_1}-\tau_{v_2},\theta_{u_1}-\theta_{v_2})$. We also have the possibilities $(\D\tau_\infty,\D\theta_\infty)\rightarrow(\D\tau_\infty,2\pi-\D\theta_\infty)$.
Note that the minimum value of the $\sigma(\Sigma)$ does not always correspond to the correct entanglement wedge. We must carefully choose the one which {\it minimizes the area of the minimal surfaces}. For example, in the small size limit $\epsilon\rightarrow0$, the analytic expression in $-v_1<t<-u_1$ is given by
\be
E_W=\dfrac{c}{6}\log\frac{\sqrt{(t^2-u_2^2)(t^2-v_1^2)(u_1^2-t^2)(v_2^2-t^2)}}{\gamma^2\epsilon^2 (u_2-v_1)(u_1-v_2)},\;\,(\textrm{if}\;\,-v_1<t<-u_1).
\ee
Obviously, this possibility is excluded from the $\epsilon$ dependence. However, this is what we have seen in \eqref{eq:resultAano} as a (non-dominant) conformal block. Rather interestingly, we can confirm large $c$ conformal blocks nicely tell us the each possibility for each phase. Moreover, this splitting cross section can be a dominant one if $\epsilon$ becomes non-zero (see Figure \ref{fig:plot_disc_ewcs} as an example). 

\subsection{Rotating case}\label{subsec:rot}
We can easily extend the previous calculations to the rotating BTZ black hole with angular momentum $J$. In the CFT side, we let the local operator have the scaling dimension $h_O\neq \bar{h}_O$. The embedding coordinates in the rotating case are given by
\begin{subequations}\label{subeq:btzcoord_rot}
\begin{align}
X_0&=\sqrt{\dfrac{r^2-r_+^2}{r_+^2-r_-^2}}\sin\left(r_+\tau-r_-\theta\right),\\
X_1&=\sqrt{\dfrac{r^2-r_+^2}{r_+^2-r_-^2}}\cos\left(r_+\tau-r_-\theta\right),\\
X_2&=\sqrt{\dfrac{r^2-r_-^2}{r_+^2-r_-^2}}\sin\left(r_+\theta-r_-\tau\right),\\
X_3&=\sqrt{\dfrac{r^2-r_-^2}{r_+^2-r_-^2}}\cos\left(r_+\theta-r_-\tau\right),
\end{align}
\end{subequations}
where $r_+ (r_-)$ correspond to the radius of the outer (inner) horizon,
\begin{align}
r_+-r_-&=\sqrt{M-1-J}\equiv\gamma,\\
r_++r_-&=\sqrt{M-1+J}\equiv\bar{\gamma}.
\end{align}
Note that the above coordinates cover only the region $r>r_+$. Here $\gamma$ and $\bar{\gamma}$ are the same one in the CFT.
In the previous subsections, we assumed $J=0$, hence $\gamma=\bar{\gamma}$. By using the above coordinates, one can check that the local heavy operator with $h_O\neq \bar{h}_O$ consistently reproduces the rotating BTZ results. 

\section{Discussion} 

We will propose some remaining questions and interesting future works at the end of this
paper:

\begin{itemize}
\item information spreading

One of our basic questions is how question information spreads in a strongly coupled system.
A useful tool to probe how information spreads is {\it mutual information} as studied in \cite{Asplund2015a}.
It is natural to expect that reflected entropy provides new information about this problem. What we need to calculate reflected entropy in their setup is the light cone singularities of the 6-point conformal blocks. Nevertheless there are currently no explicit forms of the light-cone singularity of Virasoro blocks, they are recently investigated by numerically \cite{Kusuki2018,Kusuki2018b} and analytically \cite{Kusuki2018a} in large $c$ and \cite{Kusuki2018c, Collier2018} in general $c>1$. Now that we have all tools to accomplish this task, it would be very interesting to study information spreading by making use of reflected entropy.
(In the bulk side,  a first step in this direction has already been taken in \cite{Wang2019})

\item monotonicity

The entanglement of purification has some useful properties and the holographic reflected entropy satisfies all these inequalities.
However, if we leave the holographic CFT, some of them break down.
It would be interesting to clarify how the quantum corrections break down them.
In particular, there is little knowledge about the monotonicity for reflected entropy,
\begin{equation}
S_R(A:BC)\geq S_R(A:B).
\end{equation}
Our approach developed in this paper can be applied {\it non-perturbatively} to {\it non-trivial} state, therefore, we believe that our approach makes it clear before long.

\item relation to negativity

There is an interesting proposal for the relation between entanglement wedge cross section and {\it negativity} in \cite{Kudler-Flam2019a, Kusuki2019b}.
The negativity can also be calculated in CFT by the replica trick.
It would be very interesting to compare the reflected entropy and the negativity for a local quench state.
This trial should reveal differences and similarities of them.
We believe that our approach developed in this paper is useful to calculate the negativity after a local quench and it will bring about a deep understanding of this relation.

\item Renyi reflected entropy

As shown in this paper, the reflected entropy in the holographic CFT is approximated by the thermal reflected entropy.
However, it is not trivial for the Renyi reflected entropy to also show this thermalization.

Another motivation to study Renyi reflected entropy is to compare the gravity side.
As mentioned in the main text, the Renyi reflected entropy has an obvious replica transition as the replica number $n$ is varied. (Similar transitions can be found in \cite{Metlitski2009,Belin2013,Belin2015,Belin2017,Dong2018, Kusuki2018c}.)
This might be related to the instability and we could find a transition accompanied by this instability in the bulk side.
Further understanding of this transition is one of interesting future directions.
Note that a sturdy of the Renyi reflected entropy is already started in \cite{Jeong2019}, however, the result is only perturbative, which does not enable us to observe the transition.

\item joining quench, global quench, splitting quench, double quench

In this paper, we only focus on the local operator quench introduced in \cite{Nozaki2014}. Aside from this system, there are many variable ways to excite the vacuum state (e.g., joining quench \cite{Calabrese2007}, global quench \cite{Calabrese2006a,Calabrese2005},  splitting quench \cite{Shimaji2018} and double quench  \cite{Guo2018, Caputa2019, Kusuki2019c})
It would be interesting to study dynamics of reflected entropy in these setups and identify similarities and differences.

\item finite $c$

An important future work is to understand how the dynamics of reflected entropy behaves in other CFTs.
This is motivated by the fact that the dynamics of entanglement entropy captures the chaotic nature of a given CFT.
That is, its time-dependence in the holographic CFT \cite{Nozaki2013,Caputa2014a, Asplund2015}, in RCFTs \cite{Numasawa2016,He2014}, and in another irrational CFT \cite{Caputa2017} are very different from each other.
It is naturally expected for reflected entropy to be more useful to characterize CFTs, in particular, to identify the holographic CFT.
In fact, our method allows us to calculate the reflected entropy even in finite $c$ CFTs and we have shown a part of results in this paper. We hope to give complete results in a future paper.

There is another motivation to study the reflected entropy in finite $c$ systems.
The reflected entropy is very recently invented in \cite{Dutta2019}, therefore, we have very limited knowledge about its properties (e.g., the monotonicity is satisfied or not). Against this backdrop, this challenge gives a key to understanding the reflected entropy.

\item odd entanglement entropy

Our natural expectation is that the odd entanglement entropy also contains information about correlations between two intervals and capture the chaotic nature in some sense.
However, we have little knowledge about the odd entanglement entropy itself.
An immediate future work is to investigate its properties in various setups and find out universality.
It is particularly interesting for us to find a property which only holds in the holographic CFT. We expect that this quantity could be a good tool to identify the holographic CFT.

Our result strongly suggests that the odd entanglement entropy in the holographic CFT perfectly captures the entanglement wedge cross section even in more general systems. We hope to prove this statement in a rigorous and general way in future.

\end{itemize}

\section*{Acknowledgments}
We thank Souvik Dutta, Jonah Kudler-Flam, Thomas Hartman, Masamichi Miyaji, Masahiro Nozaki, Tokiro Numasawa, Tadashi Takayanagi and Koji Umemoto for fruitful discussions and comments.
YK is supported by the JSPS fellowship. KT is supported by JSPS Grant-in-Aid for Scientific Research (A) No.16H02182 and Simons Foundation through the ``It from Qubit'' collaboration. We are grateful to the conference “Quantum Information and String Theory 2019” in YITP and ``Strings 2019''.

\clearpage

\appendix

\if(
\section{Virasoro Block Approximation} \label{app:Vir}
\if(
In general, we cannot approximate the blocks of the orbifold theory (``orbifold blocks'') by the Virasoro blocks. 
However, in the limits $n,m \to 1$ and $c \to \infty$, this approximation is eventually justified. 
In this section, we will explain why this approximation works in this case. 
The distinction between the above two blocks is actually crucial when we consider the RCFT, for example. (We also discuss this point later.)
In what follows, we basically discuss the particular cases which are relevant to the calculation of the reflected entropy. In the end of the present section, we also comment on the case of the odd entanglement entropy. 

Since we analytically continue an even integer $m$ to the real number, replica sheets labelled by $m=0,\dots,\fr{m}{2}-1$ and ones for $\fr{m}{2},\dots,m-1$ should decouple. A similar decoupling of the replica sheets is well-known in the context of the logarithmic negativity because in this case we consider the analytic continuation of an even integer \cite{Calabrese2012,Calabrese2013a} (see also \cite{Kusuki2019b}).
To make it clear, we introduce the following notations:

\begin{table}[H]
  \begin{tabular}{|c|l|} 
\hline
     $O_{(k,l)}$
				&
		Operator on $(k,l)$-sheet. ($k=0,\dots, m-1$ and $l=0,\dots, n-1$.)
\\  \hline
     $O^{\otimes n}_{(k)}$
				& $\bigotimes^{n-1}_{l=0}O_{(k,l)} $ 
\\  \hline

     $\sigma_n^{(0)}$
				&
		$O_{(k,l)}(\ex{2\pi i}z)\sigma_n^{(0)}(0) = O_{(k,l+1)}(z)\sigma_n^{(0)}(0) $, \ \ \ \ \ \ (if $k=0$),  \\
		&$O_{(k,l)}(\ex{2\pi i}z)\sigma_n^{(0)}(0) = O_{(k,l)}(z)\sigma_n^{(0)}(0) $, \ \ \ \ \ \ \ \ \  (otherwise). 
\\  \hline
     $\sigma_n^{(m/2)}$
				&
		$O_{(k,l)}(\ex{2\pi i}z)\sigma_n^{(m/2)}(0) = O_{(k,l+1)}(z)\sigma_n^{(m/2)}(0) $, \ \ \ \ \ \   (if $k=\fr{m}{2} $),  \\
		&$O_{(k,l)}(\ex{2\pi i}z)\sigma_n^{(m/2)}(0) = O_{(k,l)}(z)\sigma_n^{(m/2)}(0) $, \ \ \ \ \ \ \ \ \ (otherwise). 
\\  \hline
     $\sigma_m^{\otimes n}$
				&
		$O_{(k,l)}(\ex{2\pi i}z)\sigma_m^{\otimes n}(0) = O_{(k+1,l)}(z)\sigma_m^{\otimes n}(0) $.
\\  \hline
     $\bar{\sigma}_m^{\otimes n}$
				&
		$O_{(k,l)}(\ex{2\pi i}z)\bar{\sigma}_m^{\otimes n}(0) = O_{(k-1,l)}(z)\bar{\sigma}_m^{\otimes n}(0) $.
\\  \hline
     ${\sigma'}_m^{\otimes n}$
				&
		$O_{(k,l)}(\ex{2\pi i}z){\sigma'}_m^{\otimes n}(0) = O_{(k+1,l+1)}(z){\sigma'}_m^{\otimes n}(0) $,  \ \ \ \ \ (if $k=0,\fr{m}{2}$),   \\
		&$O_{(k,l)}(\ex{2\pi i}z){\sigma'}_m^{\otimes n}(0) = O_{(k+1,l)}(z){\sigma'}_m^{\otimes n}(0) $, \ \ \ \ \ \ \ \ (otherwise) .
\\  \hline
     $\bar{\sigma'}_m^{\otimes n}$
				&
		$O_{(k,l)}(\ex{2\pi i}z)\bar{\sigma'}_m^{\otimes n}(0) = O_{(k-1,l-1)}(z)\bar{\sigma'}_m^{\otimes n}(0) $,  \ \ \ \ \ (if $k=0,\fr{m}{2}$),   \\
		&$O_{(k,l)}(\ex{2\pi i}z)\bar{\sigma'}_m^{\otimes n}(0) = O_{(k-1,l)}(z)\bar{\sigma'}_m^{\otimes n}(0) $, \ \ \ \ \ \ \ \ (otherwise) .
\\  \hline
  \end{tabular}
\end{table}
\ \\

Then, the operator $O^{\otimes mn}$ in \eqref{eq:Renyi2a} can be written as
\be
O^{\otimes mn}=O^{\otimes n}_{(0)} \otimes \cdots \otimes O^{\otimes n}_{(m/2)} \otimes \cdots.
\ee
Throughout this paper, we suppress the transposition acting on the operators on second half sheets concerning to $m$. 
We have to emphasize that in the analytic continuation of even $m$, the operator $O^{\otimes mn}$ does NOT reduce to $O$ but
the ``square'' of $O$ as
\footnote{
One can understand this squaring by considering a simple case.
Let us prepare our state by a pure state $\rho_{A \bar{A}}$, then the canonical putrefied state is given by the ``square'' of $\rho_A$ as 
$\rho_A \otimes \rho^T_A$.
}

\begin{equation}
\lim_{m \in \text{even} \to 1} O^{\otimes mn} \to  O^{\otimes n}_{(0)} \otimes O^{\otimes n}_{(1/2)}.
\end{equation}
One can also find the same decoupling in the original paper \cite{Dutta2019}, where the analytic continuation leads to
\begin{equation}
\lim_{m \in \text{even} \to 1} \sigma_{g_A^{-1} g_B}\to \sigma_n^{(0)} \otimes \bar{\sigma}_n^{(1/2)}.
\end{equation}
It means that this tricky analytic continuation provides two decoupled sheets labeled by $0$ and $1/2$.

The relation between the above notations and the twist operators in the main text is given by
\begin{equation}
\begin{aligned}
\sigma_{g_B}={\sigma}_m^{\otimes n}, \ \ \ \ \ 
\sigma_{g_B^{-1}}=\bar{\sigma}_m^{\otimes n}, \ \ \ \ \ 
\sigma_{g_A}={\sigma'}_m^{\otimes n}, \ \ \ \ \ 
\sigma_{g_A^{-1}}=\bar{\sigma'}_m^{\otimes n}, \ \ \ \ \ 
\sigma_{g_A^{-1} g_B}=\sigma_n^{(0)} \otimes \bar{\sigma}_n^{(m/2)}, \ \ \ \ \ 
\end{aligned}
\end{equation}
and the conformal block can be re-expressed by
\newsavebox{\boxpCa}
\sbox{\boxpCa}{\includegraphics[width=250pt]{blockCa.pdf}}
\newlength{\pCaw}
\settowidth{\pCaw}{\usebox{\boxpCa}} 

\begin{equation}
 \parbox{\pCaw}{\usebox{\boxpCa}},
\end{equation}
where $\cdots$ means the rest of $O^{\otimes mn}$, that is, $\bigotimes_{l=\hat{0},1,2,\dots, \hat{\fr{m}{2}}, \dots, n-1}  O^{\otimes n}_{(l)}$, which is not important because it disappears in the limit $m \to 1$.

The point is that $\{ O^{\otimes n}_{(0)}, \sigma_n^{(0)} \}$ do not interact with $\{ O^{\otimes n}_{(m/2)}, \sigma_n^{(m/2)} \}$, therefore, the component of the conformal block (i.e., three point block) is decoupled into two parts, for example,
\begin{equation}\label{eq:square2}
\braket{  \sigma_n^{(0)}  \otimes  \bar{\sigma}_n^{(m/2)}  |  O^{\otimes n}_{(0)} \otimes   O^{\otimes n}_{(m/2)}   | \sigma_n^{(0)}  \otimes \bar{\sigma}_n^{(m/2)}  } = 
\braket{  \sigma_n^{(0)}   |  O^{\otimes n}_{(0)} |   \sigma_n^{(0)}  } 
\braket{  \sigma_n^{(m/2)}   |  O^{\otimes n}_{(m/2)} |   \sigma_n^{(m/2)}  }.
\end{equation}
Let us highlight this decoupling by
\newsavebox{\boxpCb}
\sbox{\boxpCb}{\includegraphics[width=250pt]{blockCb.pdf}}
\newlength{\pCbw}
\settowidth{\pCbw}{\usebox{\boxpCb}} 

\begin{equation}\label{eq:decoupleblock}
 \parbox{\pCbw}{\usebox{\boxpCb}}.
\end{equation}
)\fi
Then, the precise Regge limit considered in (\ref{eq:monoB}) is given by 
\begin{equation}
 {\bold M}_{0, 2\a_n}^{(-)}
   \left[
    \begin{array}{cc}
    \a_n   & \a_n   \\
     \a_{O}   &   \a_{O} \\
    \end{array}
  \right]
{\red
 {\bold M}_{0, 2\a_n}^{(-)}
   \left[
    \begin{array}{cc}
    \a_n   & \a_n   \\
     \a_{O}   &   \a_{O} \\
    \end{array}
  \right]
}
	\times (2i \e)^{2h_{2\a_n}-4nh_O} 
	\times (\text{remaining block}),
\end{equation}
where $\fr{c}{24}\pa{n-\fr{1}{n}}=\a_n(Q-\a_n)$ and we extract the leading part in the limits $n,m \to 1$. Notice that each $\sigma_n^{(i)}$ has the Liouville momentum $\a_n$ as like standard twist operators in $\mathbb{Z}_n$ orbifold theories. 
By using (\ref{eq:M1}), we can show that 
\begin{equation}\label{eq:monomono}
 {\bold M}_{0, 2\a_n}^{(-)}
   \left[
    \begin{array}{cc}
    \a_n   & \a_n   \\
     \a_{O}   &   \a_{O} \\
    \end{array}
  \right]
{\red
 {\bold M}_{0, 2\a_n}^{(-)}
   \left[
    \begin{array}{cc}
    \a_n   & \a_n   \\
     \a_{O}   &   \a_{O} \\
    \end{array}
  \right]
}
=
 {\bold M}_{0, 2\b_n}^{(-)}
   \left[
    \begin{array}{cc}
    \b_n   & \b_n   \\
     \a_{O}   &   \a_{O} \\
    \end{array}
  \right].
\end{equation}
This is the reason why the constant part can be calculated correctly and then reproduce the bulk result.
The time dependent part is calculated by the global block, which is the approximation of the Virasoro block in a special limit. This global block approximation can be also applied to the block with more symmetries.
Therefore, we can also correctly calculate the time dependent part and reproduce the gravity result.

We have to emphasize that this Virasoro block approximation is justified only when we restrict ourselves to the von-Neumann limit and the holographic CFT. Hence, if one wants to calculate the Renyi reflected entropy or the reflected entropy in other CFTs, one has to take care of the fact that the block associated to the reflected entropy is NOT the Virasoro block.
For example, we cannot show the reflected entropy growth in RCFTs (\ref{eq:DSRR}) by a naive approximation. The prefactor $2$ in (\ref{eq:DSRR}) is just because of the decoupling into 2 parts (\ref{eq:decoupleblock}), which cannot be explained by the naive approximation of the ``orbifold block''. 
)\fi
\section{Semiclassical Fusion and Monodromy Matrix} \label{app:FM}

In this appendix, we show the detailed derivation of the semiclassical monodromy matrix.
We have the closed expression for the fusion and monodromy matrix, therefore, it is possible to evaluate their semiclassical limits by using them as in \cite{Collier2018}. However, the simplest way to calculate them is to make use of the closed form of the HHLL Virasoro block. We have to emphasize that what we need here is not the usual HHLL block introduced in \cite{Fitzpatrick2015},
\begin{equation}\label{eq:HHLLblock2}
\ca{F}^{LL}_{HH}(h_p|z)   = (1-z)^{h_L(\d-1)} \pa{ \fr{1-(1-z)^\d}{\d}}^{h_p-2h_L} {}_2 F_1(h_p,h_p,2h_p;1-(1-z)^\d),
\end{equation}
but the semiclassical block derived by the monodromy method \cite{Hijano2015a},
\footnote{
The semiclassical conformal block with the $z_i$-dependences, which are not fixed by the global conformal transformation, is shown in \cite{Brehm2018, Anous2019}.
}
\begin{equation}\label{eq:HHLLblock}
\ca{F}^{LL}_{HH}(h_p|z)   =   (1-z)^{h_L(\d-1)} \pa{ \fr{1-(1-z)^\d}{\d}}^{h_p-2h_L} \pa{ \fr{1+(1-z)^{\fr{\d}{2}}}{2}}^{-2h_p},
\end{equation}
where $\d=\s{1-\fr{24}{c}h_H}$.
The former is derived in the large $c$ limit with $\fr{h_H}{c}, h_L,h_p$ fixed, on the other hand, the later is calculated in a different regime of parameter space, in the large $c$ limit with $\fr{h_H}{c}, \fr{h_L}{c}, \fr{h_p}{c}$ fixed and set $h_H \gg h_L, h_p$ (which is discussed in \cite{Alkalaev2016c, Brehm2018}) . Therefore, these two HHLL blocks are different from each other.
For convenience, we call the former {\it HHLL limit} and the later {\it semiclassical limit}.
We have to choose the later in our calculation because we take first the large $c$ limit of the block with the twist operators, whose conformal dimensions are proportional to $c$.
Note that the HHLL block and the semiclassical block can be related through
\begin{equation}\label{eq:appHHLL}
 {}_2 F_1(h_p,h_p,2h_p;z) \ar{h_p \to \infty} \pa{\fr{1+\s{1-z}}{2}}^{-2h_p},
\end{equation}
which is shown by using the following identity,
\begin{equation}
 {}_2 F_1(h_p,h_p-\fr{1}{2} ,2h_p;z) =\pa{\fr{1+\s{1-z}}{2}}^{1-2h_p}.
\end{equation}

The fusion transformation leads to the relation,
\begin{equation}
\ca{F}^{LL}_{HH}(h_p|z) \ar{z \to 1} 
 \ca{F}_{\a_p, \a_H} 
   \left[
    \begin{array}{cc}
    \a_L   & \a_L   \\
     \a_H   &   \a_H \\
    \end{array}
  \right] 
(1-z)^{h_L(1-\d)},
\end{equation}
where we introduce the Liouville momentum as
\begin{equation}
\a_L\pa{Q-\a_L}=h_L, \ \ \ \ \  \a_H\pa{Q-\a_H}=h_H, \ \ \ \ \  \a_p\pa{Q-\a_p}=h_p,
\end{equation}
and $\ca{F}$ is defined in terms of the Virasoro fusion matrix ${\bold F}$ \cite{Kusuki2018c,Kusuki2019} as
\begin{equation}
 \ca{F}_{\a_p, \a_H} \equiv \text{Res}\pa{-2 \pi i \  {\bold F}_{\a_p, \a} ; \a=\a_H  }.
\end{equation}
From the explicit form (\ref{eq:HHLLblock}), we can immediately show
\begin{equation}\label{eq:F1}
 \ca{F}_{\a_p, \a_H} 
   \left[
    \begin{array}{cc}
    \a_L   & \a_L   \\
     \a_H   &   \a_H \\
    \end{array}
  \right] 
\ar{\substack{\text{semiclassical} \\ \text{limit}  }}
	\d^{2 h_L} \pa{\fr{4}{\d}}^{h_p}.
\end{equation}
In a similar manner, the Regge limit of the semiclassical block can be related to the monodromy matrix as
\begin{equation}
\ca{F}^{LL}_{HH}(h_p|z) \ar{\substack{z \to 0 \\ \text{after} \\ (1-z) \to \ex{2\pi i}(1-z)}} 
 \ca{M}_{\a_p, 2\a_L}^{(+)}
   \left[
    \begin{array}{cc}
    \a_L   & \a_L   \\
     \a_H   &   \a_H \\
    \end{array}
  \right] .
\end{equation}
Therefore, we obtain
\begin{equation}\label{eq:M1}
 \ca{M}_{\a_p, 2\a_L}^{(+)}
   \left[
    \begin{array}{cc}
    \a_L   & \a_L   \\
     \a_H   &   \a_H \\
    \end{array}
  \right]
\ar{\substack{\text{semiclassical} \\ \text{limit}  }}
	\pa{\fr{2i}{\d}\sin \pi \d}^{-2h_L} \pa{-\fr{4i}{\d} \tan\fr{\pi \d}{2}}^{h_p}.
\end{equation}
Note that this is completely different from the monodromy matrix based on (\ref{eq:HHLLblock2}).
The dimension $h_p$ is order $O(1)$, therefore, the large $c$ limit does not change the hypergeometric function part of the HHLL block, unlike (\ref{eq:appHHLL}). As a result, we obtain
\begin{equation}
 \ca{M}_{\a_p, 2\a_L}^{(+)}
   \left[
    \begin{array}{cc}
    \a_L   & \a_L   \\
     \a_H   &   \a_H \\
    \end{array}
  \right]
\ar{\substack{\text{HHLL limit} \\ \text{and} \\ h_p \to 0 }}
	\pa{\fr{2i}{\d}\sin \pi \d}^{-2h_L} \pa{-\fr{2i}{\d} \sin\pi \d}^{h_p}.
\end{equation}

According to \cite{Fitzpatrick2014}, the LHHL block with heavy intermediate state can be given by just primary exchange.
Therefore, the following type of the fusion matrix is trivial,
\begin{equation}
 \ca{F}_{\a_H, \a_{L_1}+\a_{L_2}} 
   \left[
    \begin{array}{cc}
    \a_{L_1}   & \a_H   \\
     \a_{L_2}   &   \a_H \\
    \end{array}
  \right] 
\ar{\substack{\text{semiclassical} \\ \text{limit}  }}
	1.
\end{equation}

\section{Semiclassical 5-point Block} \label{app:5-pt.}

\subsection{Proof of (\ref{eq:F})}

In this Appendix, we show the detailed calculation of (\ref{eq:F}).
From the expression (\ref{eq:channel2}), we find that the Regge limit is given by

\begin{equation}\label{eq:AppF}
\begin{aligned}
 \ti{\ca{M}}_{0, 2\a_m}^{(-)}
   \left[
    \begin{array}{cc}
    \a_m   & \a_m   \\
     \a_{O}   &   \a_{O} \\
    \end{array}
  \right]
	\times (2i \e)^{h_{2\a_m}-2nmh_O}
 \parbox{\pow}{\usebox{\boxpo}},
\end{aligned}
\end{equation}
where $h_a=\a(Q-\a)$.
\footnote{
Here we assume $\a_{\text{min}}=\a_{2\a_m}$, which is naturally expected from the result in \cite{Kusuki2019}.
But this assumption is not necessary because we obtain the same conclusion (\ref{eq:5-pt.}) without fixing $\a_{\text{min}}$.
}
 If we take the limit $m \to 1$, then the monodromy matrix simply becomes one.
At this stage, what we need to evaluate the reflected entropy is the following asymptotics,
\newsavebox{\boxpq}
\sbox{\boxpq}{\includegraphics[width=190pt]{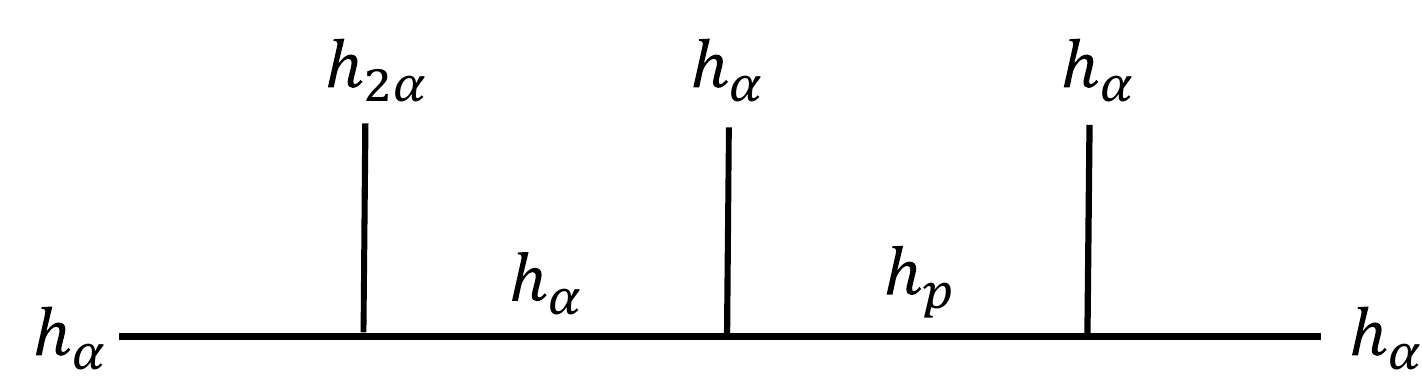}}
\newlength{\pqw}
\settowidth{\pqw}{\usebox{\boxpq}} 

\begin{equation}
 \parbox{\pqw}{\usebox{\boxpq}} \ar{\a \to 0} \ \ \  ?,
\end{equation} 
where we also take the large $c$ limit with $\fr{h_p}{c}$ fixed. It is important to note that the asymptotics of the 4-point semiclassical block is given by (see Appendix \ref{app:FM}),
\newsavebox{\boxpp}
\sbox{\boxpp}{\includegraphics[width=150pt]{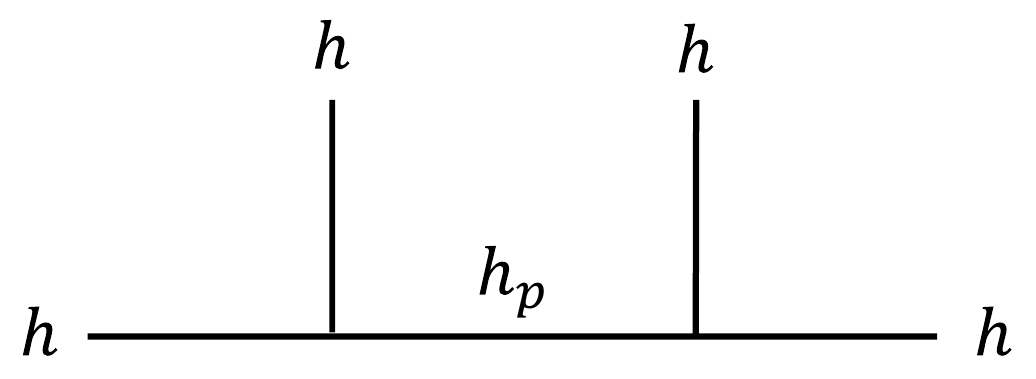}}
\newlength{\ppw}
\settowidth{\ppw}{\usebox{\boxpp}} 

\begin{equation}
 \parbox{\ppw}{\usebox{\boxpp}} \ar{h \to 0} z^{h_p} \pa{ \fr{1+\s{1-z}}{2}}^{-2h_p}=2^{2h_p}\pa{\fr{1+\s{1-z}}{1-\s{1-z}}}^{-h_p},
\end{equation}
which is used to reproduce the entanglement wedge cross section as in \cite{Tamaoka2019, Dutta2019}.
In fact, we can derive this semiclassical block with the intermediate state of order $c$
\footnote{
This approximated block is not the same as the regular part of the conformal block (i.e., the block with the heavy intermediate state) \cite{Zamolodchikov1987,Zamolodchikov1984}. The difference between these two blocks is that the former has the intermediate dimension of order $c$, whilst the later is calculated in the limit $h_p \gg c$.
}
 by the global block in the following way (instead of relying on (\ref{eq:HHLLblock});
\begin{equation}
\begin{aligned}
z^{h_p-2h} {}_2 F_1(h_p,h_p,2h_p;z) \ar{\substack{h_p \to \infty \\ \text{after} \\ h \to 0} } z^{h_p} \pa{ \fr{1+\s{1-z}}{2}}^{-2h_p},
\end{aligned}
\end{equation}
where the left-hand side is the well-known global block \cite{Dolan2000a,Dolan2003}. From this observation, we can deduce that the asymptotics of the 5-point block can be obtained by the 5-point global block, which has already calculated in \cite{Rosenhaus:2018zqn} as
\newsavebox{\boxpr}
\sbox{\boxpr}{\includegraphics[width=190pt]{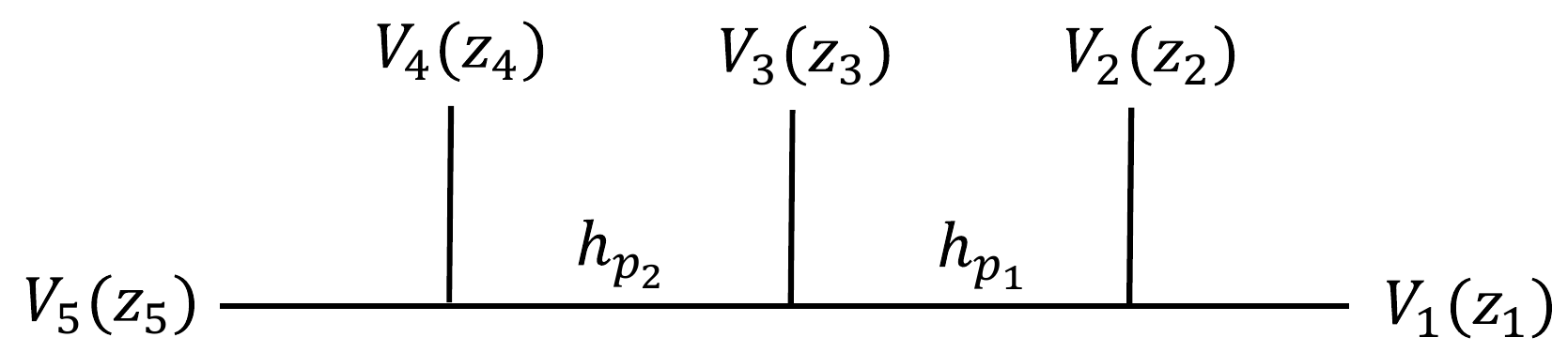}}
\newlength{\prw}
\settowidth{\prw}{\usebox{\boxpr}} 

\begin{equation}\label{eq:Rosenhaus}
\begin{aligned}
 \parbox{\prw}{\usebox{\boxpr}}
&=\ca{L}^{h_1, \cdots, h_5}(z_1, \cdots z_5)
\chi_1^{h_{p_1}} \chi_2^{h_{p_2}}  \\
& \times F_2  \left[
    \begin{array}{c}
      h_{p_1}+h_1-h_2, h_{p_2}+h_{p_1}-h_3, h_5+h_{p_2}-h_4   \\
        2h_{p_1}, 2h_{p_2} \\
    \end{array}
	; \chi_1, \chi_2
  \right]
,
\end{aligned}
\end{equation}
where $h_i$ $(i=1, \cdots 5)$ is the conformal dimension of the operator $V_i$, the cross ratio is defined by
$\chi_i \equiv \fr{z_{i,i+1} z_{i+2,i+3}}{z_{i,i+2} z_{i+1,i+3}} $ with $z_{i,j}=z_i-z_j$,
and the prefactor $\ca{L}$ is the leg factor as 
\begin{equation}
	\ca{L}^{h_1, \cdots, h_5}(z_1, \cdots z_5)
	\equiv  
	\pa{\fr{z_{23}}{z_{12}z_{13}}}^{h_1} \pa{\fr{z_{34}}{z_{35}z_{45}}}^{h_5} \prod_{i=1}^{3} \pa{\fr{z_{i,i+2}  }{  z_{i,i+1}z_{i+1,i+2}}}^{h_{i+1}}.
\end{equation}
The function $F_2$ is the Appell function defined as
\begin{equation}
 F_2  \left[
    \begin{array}{c}
     a_1, b,a_2   \\
        c_1, c_2 \\
    \end{array}
	; x_1, x_2
  \right]
	=
	\sum_{n_1,n_2=0}^{\infty} \fr{(a_1)_{n_1}  (b)_{n_1+n_2}  (a_2)_{n_2}   }{(c_1)_{n_1} (c_2)_{n_2} } \fr{x_1^{n_1}}{n_1 !} \fr{x_2^{n_2}}{n_2 !},
\end{equation}
where $(a)_n=\fr{\G(a+n)}{\G(a)}$ is the Pochhammer symbol and we define $(0)_n=\d_{n,0}$. By using this result, we obtain
\begin{equation}\label{eq:5-pt.}
 \parbox{\pqw}{\usebox{\boxpq}} \ar{\substack{h_p \to \infty \\ \text{after} \\ h_\a \to 0} } 
 {\chi_1}^{h_p} \pa{ \fr{1+\s{1-{\chi_1}}}{2}}^{-2h_p}.
\end{equation} 
Here we leave only the linear term $h_p$ in the log of the block, like (\ref{eq:HHLLblock}).
This approximated block is what we want (\ref{eq:F}), where the explicit form of the cross ratio $\chi_1$ is given by
\begin{equation}
\chi_1=\fr{  (-v_1+t)(-u_1+v_2)}{(-u_1+t)(-v_1+v_2)}.
\end{equation}

\subsection{Proof of (\ref{eq:F2})}

In this section, we show the asymptotics (\ref{eq:F2}).
The monodromy transformation in (\ref{eq:F2}) can be re-expressed as

\newsavebox{\boxpAe}
\sbox{\boxpAe}{\includegraphics[width=180pt]{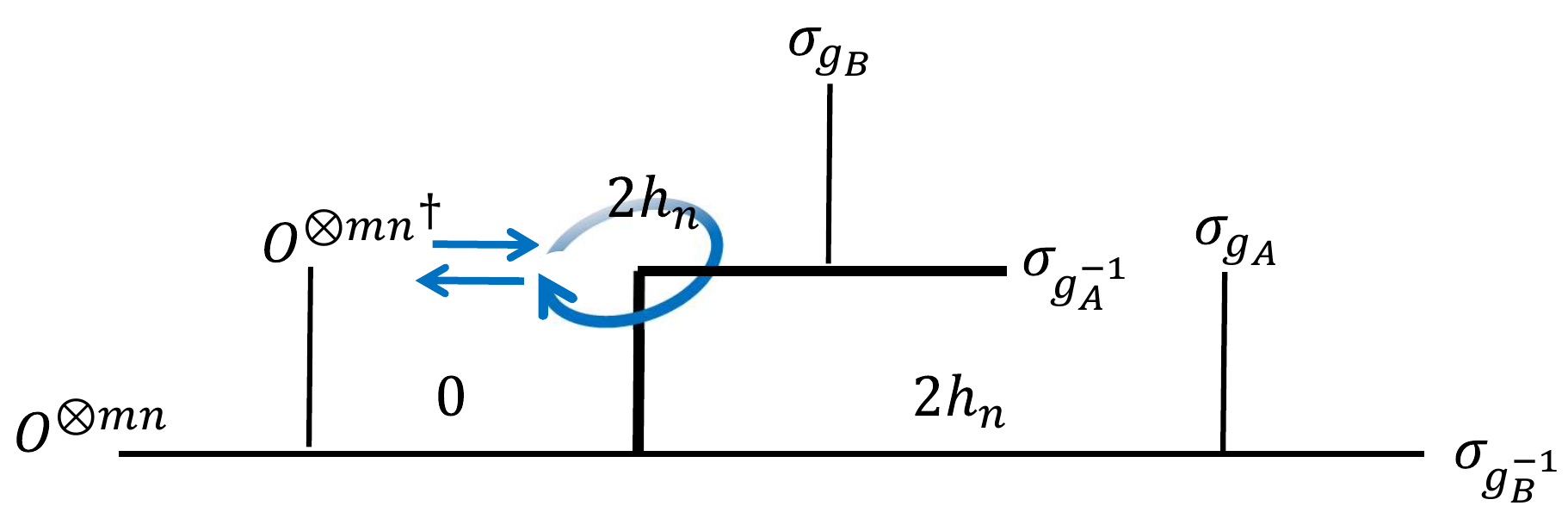}}
\newlength{\pAew}
\settowidth{\pAew}{\usebox{\boxpAe}} 

\begin{equation}\label{eq:monoB}
\parbox{\pzaw}{\usebox{\boxpza}}  = \parbox{\pAew}{\usebox{\boxpAe}}.
\end{equation}
This is just the monodromy transformation of $\dg{{O^{\otimes mn}}}$ around $\sigma_{g_B^{-1} g_A}$.
Let us recall that the orbifold block can be regarded as the square of the Virasoro block as explained in (\ref{eq:decoupleblock}).
Therefore, this monodromy effect comes from each Virasoro block (i.e., black and red in (\ref{eq:decoupleblock})) as

\newsavebox{\boxpAf}
\sbox{\boxpAf}{\includegraphics[width=190pt]{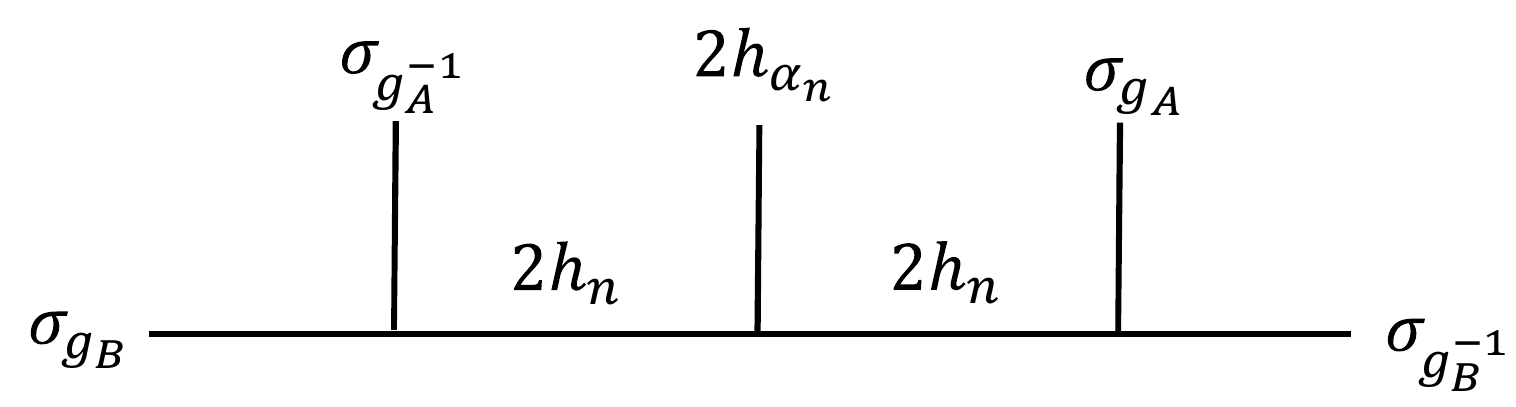}}
\newlength{\pAfw}
\settowidth{\pAfw}{\usebox{\boxpAf}} 

\begin{equation}\label{eq:B2}
 \ca{M}_{0, 2\a_n}^{(-)}
   \left[
    \begin{array}{cc}
    \a_n   & \a_n   \\
     \a_{O}   &   \a_{O} \\
    \end{array}
  \right]
{\red
 {\ca{M}}_{0, 2\a_n}^{(-)}
   \left[
    \begin{array}{cc}
    \a_n   & \a_n   \\
     \a_{O}   &   \a_{O} \\
    \end{array}
  \right]
}
	\times (2i \e)^{2h_{2\a_n}-4nh_O} 
	\times 
	\parbox{\pAfw}{\usebox{\boxpAf}},
\end{equation}
where we used the Regge limit of the block associated with $\bb{Z}_n$ symmetry \cite{Kusuki2019}.
The explict form of this monodromy matrix is (\ref{eq:M1}). To calculate the remaining 5-point conformal block, we can again make use of the global block (\ref{eq:Rosenhaus}). In fact, we can easily show 

\newsavebox{\boxpAg}
\sbox{\boxpAg}{\includegraphics[width=190pt]{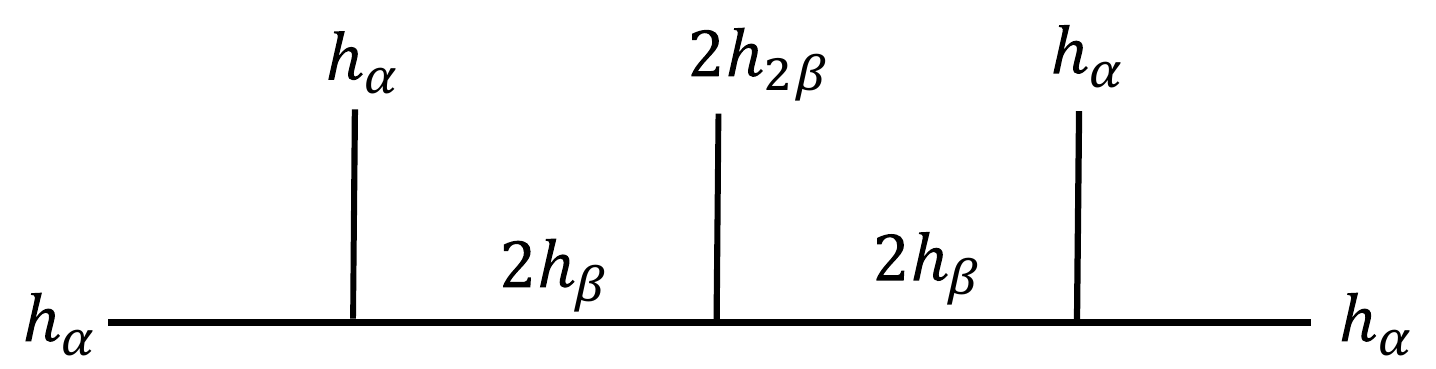}}
\newlength{\pAgw}
\settowidth{\pAgw}{\usebox{\boxpAg}} 

\begin{equation}
 \parbox{\pAgw}{\usebox{\boxpAg}} \ar{\substack{h_\a,h_\b \to 0} } 
\ca{L}^{h_\a, h_\a, 2h_{2\b}, h_\a, h_\a}(z_1, \cdots z_5)
\chi_1^{2h_{\b}} \chi_2^{2h_{\b}} ,
\end{equation} 
and substituting this result into (\ref{eq:B2}), we obtain (\ref{eq:F2}).

\subsection{Proof of (\ref{eq:REinH})}\label{app:squareblock}

The conformal blocks in (\ref{eq:CorrinH}) is given by the square of the Virasoro block as

\newsavebox{\boxpEa}
\sbox{\boxpEa}{\includegraphics[width=130pt]{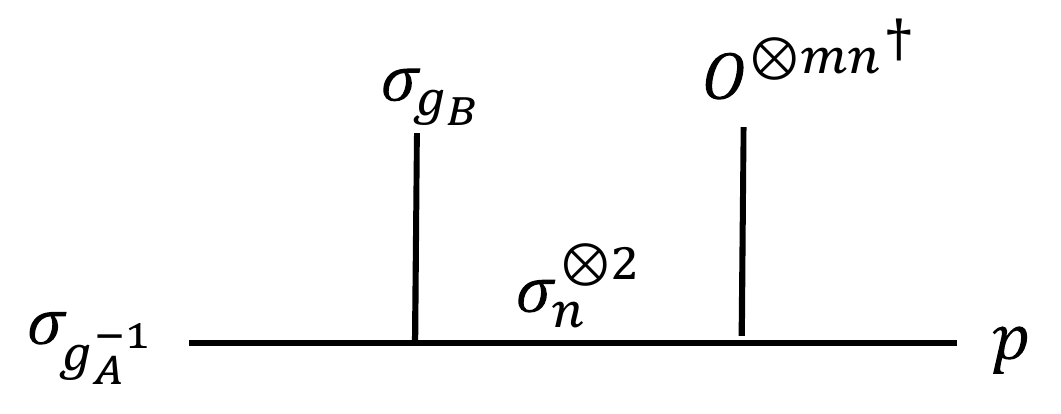}}
\newlength{\pEaw}
\settowidth{\pEaw}{\usebox{\boxpEa}} 

\newsavebox{\boxpEb}
\sbox{\boxpEb}{\includegraphics[width=130pt]{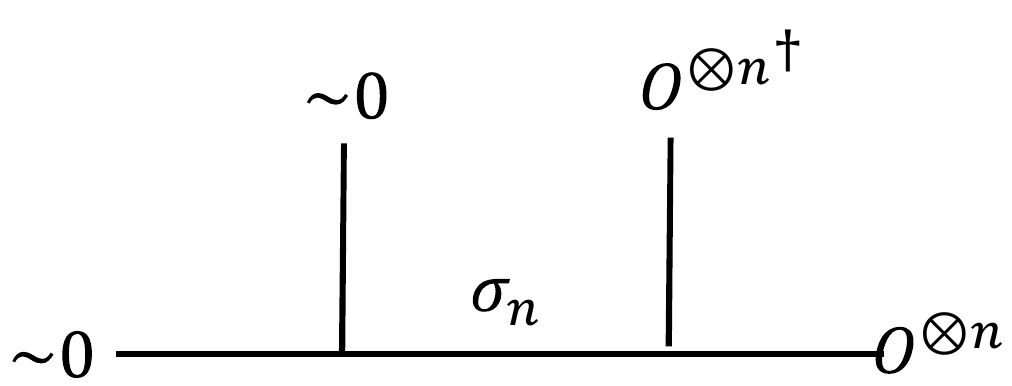}}
\newlength{\pEbw}
\settowidth{\pEbw}{\usebox{\boxpEb}} 

\begin{equation}
\begin{aligned}
 \parbox{\pEaw}{\usebox{\boxpEa}} =  \pa{\parbox{\pEbw}{\usebox{\boxpEb}}}^2,
\end{aligned}
\end{equation}
where $\sim 0$ means a state very close to the vacuum.
The detailed explanation of this squaring is shown in the main text (see \eqref{eq:square2}).
By using the HHLL approximation (\ref{eq:HHLLblock}) for the block in the parenthesis, we obtain (\ref{eq:REinH}).

\section{Heavy-Heavy-Light OPE Coefficient} \label{app:CnO}

The Heavy-Heavy-Light OPE coefficient can be calculated by the modular bootstrap equation for a 1-point function on a torus \cite{Kraus2016},

\begin{equation}
\begin{aligned}
\overline{\braket{O_H|O_L|O_H}} \sim 
\abs{
\hat{\g}^{\fr{h_L}{2}}  
\ex{-\fr{c-1}{6}\pi \pa{1-\s{1-\fr{24}{c-1}h_{\chi}}}\hat{\g}} 
\fr{1}{2\pi} \pa{1-\fr{24}{c-1}h_{\chi}}^{-\fr{h_L}{2}-\fr{1}{4}}
}^2
 \braket{\chi|O_L|\chi},
\end{aligned}
\end{equation}
where $\hat{\g}=\s{\fr{24}{c-1}h_H-1}$ and the operator $\chi$ is the lightest one with $\braket{\chi|O_L|\chi} \neq 0$.
The over-line means the average over all primary operators of fixed dimensions $h_H$, $\bar{h}_H$.
We take first the limit $c \to \infty$ with $\fr{h_H}{c}$ and $\fr{h_L}{c}$ fixed and then the limit $h_L \to 0$ as in the calculation of the reflected entropy,  this OPE coefficient is approximated by
\begin{equation}
\overline{\braket{O_H|O_L|O_H}} \sim \hat{\g}^{\fr{h_L}{2}}  \bar{\hat{\g}}^{\fr{\bar{h}_L}{2}}.
\end{equation}
If we consider $\braket{ {O_H}^{\otimes 2}|  {O_L}^{\otimes 2}  |{O_H}^{\otimes 2} }=  \braket{O_H|O_L|O_H}^2 $ (the square comes from the rule (\ref{eq:squaring}) and
set $O_L= \sigma_n$ and $O_H=O^{\otimes n}$, we obtain (\ref{eq:CnO}). 

Strictly speaking, this asymptotics holds only if $h_H$ is much larger than other parameters.
However, from the viewpoint of the holography, we expect that this result can be applied not only for $h_H \gg c, h_L$ but also $h_H>\fr{c}{12}$. This could be justified in the same way as the HKS method \cite{Hartman2014}, which is the justification of the Cardy formula for  $h>\fr{c}{12}$ in the large $c$ CFT (see \cite{Pal2019, Michel2019}).

\clearpage

\bibliographystyle{JHEP}
\bibliography{multi}

\end{document}